\documentstyle[epsfile,plotfile,twoside,submitp]{pio}

\newcommand{\ket}[1]{| #1 \rangle}
\newcommand{\bra}[1]{\langle #1 |}
\newcommand{\braket}[2]{\langle #1 | #2\rangle}
\newcommand{\ketbra}[2]{| #1 \rangle \langle #2 |}
\newcommand{\proj}[1]  {| #1 \rangle \langle #1 |}

\newcommand{\beq}{\begin{eqnarray}}
\newcommand{\eeq}{\end{eqnarray}}

\newcommand{\B}[2]{\ket{\Phi_{#1,#2}}}

\newcommand{\Bp}[2]{\proj{\Phi_{#1,#2}}}
\newcommand{\Bpbis}[2]{\Phi_{#1,#2}}

\newcommand{\BB}[2]{\ket{\Phi_{#1,#2}^*}}

\newcommand{\BBpbis}[2]{\Phi_{#1,#2}^T}

\newcommand{\Tr}{{\mathrm Tr}}

\renewcommand{\vec}[1]{\mathbf{#1}}

\newcommand{\e}{{\mathrm e}}

\begin{document}

\title{OPTICAL QUANTUM CLONING -- A REVIEW}

\author{Nicolas J. Cerf \\
{\it Centre for Quantum Information and Communication\\
Ecole Polytechnique\\
Universit\'{e} Libre de Bruxelles \\
1050 Bruxelles, Belgium}
\and 
Jarom\'{\i}r Fiur\'{a}\v{s}ek \\
{\it Department of Optics\\
Palack\'{y} University\\
77200 Olomouc, Czech Republic} 
}

\maketitle

\newpage

\vspace*{10mm}

\centerline{\bf ABSTRACT}

\vspace*{10mm}

After a brief introduction to the quantum no-cloning theorem
and its link with the linearity and causality of quantum mechanics,
the concept of quantum cloning machines is sketched, following,
whenever possible, the chronology of the main results. The important
classes of quantum cloning machines are reviewed, in particular
state-independent and state-dependent cloning machines. The 1-to-2 
cloning problem is then studied from a formal point of view, using the
isomorphism between completely positive maps and operators, which leads to 
the so-called double-Bell ansatz. This also yields an efficient numerical
approach to quantum cloning, based on semidefinite programming methods. The
derivation of the optimal N-to-M universal cloning machine in d dimensions
is then detailed, as well as the notion of asymmetric cloning machines.
In the second part of this review, the optical implementation of cloning
machines is considered. It is shown that the universal cloning of photons
can be achieved by parametric amplification of light or by symmetrization
via the Hong-Ou-Mandel effect. The various experimental demonstrations
of quantum cloning machines are reviewed. The cloning of orthogonally 
polarized photons is also considered, as well as the asymmetric and 
phase-covariant cloning of photons. Finally, the extension of quantum 
cloning to continuous variables is analyzed. The optimal cloning of 
coherent states of light by phase-insensitive amplification is explained,
as well as the experimental realization of continuous-variable quantum 
cloning with linear optics, measurement, and feed-forward operations.

\tableofcontents

\section{Introduction and history}
\subsection{The no-cloning theorem}

The history of quantum cloning can be traced back
to the controversial story of a paper by \cite{Herbert82} entitled
``FLASH -- A superluminal communicator based upon a new kind of measurement''.
In this paper, submitted early 1981 to {\it Foundations of Physics},
Herbert was discussing an idealized laser gain tube which would produce,
via {\it stimulated emission}, macroscopically distinguishable states of light 
for an incoming single photon in any polarization state. The claim was that
the noise in this process would, in principle, not prevent 
{\it perfectly} identifying the polarization state of the photon.
This process would supposedly open the way to faster-than-light communication, 
a possibility with which any physicist feels uncomfortable
since it violates causality.

Today, more than twenty years later, it is publicly known that 
GianCarlo Ghirardi and Asher Peres were requested to review this paper. 
The first of them recommended its rejection, 
based on the argument that the linear nature of quantum mechanics
must prevent such a process to exist, see \cite{vanderMerwe02}. 
The second referee wrote, see \cite{Peres02}, that he had realized the paper 
was wrong, but nevertheless recommended its publication because he expected 
that finding the error would raise a considerable interest! 
Herbert's paper was then published, 
and, funnily enough, the prediction of Peres happened to be true.

Soon afterwards, \cite{WZ82} published a paper in {\it Nature}, 
entitled ``A single quantum cannot be cloned'', which arrived 
at essentially the same conclusions as those drawn by Ghirardi
in his anonymous referee report dated of April 1981, which itself
was turned into a paper two years later, see \cite{Ghirardi83}.
Wootters and Zurek realized that, if one can build a ``cloning machine''  
that produces several clones of the horizontal- and vertical-polarization
states of an incoming photon, then circularly-polarized states 
cannot yield circularly-polarized clones. Instead, due to
the linearity of quantum mechanics, one gets a linear superposition
of vertically-polarized clones and horizontally-polarized clones.
Indeed, if the cloning machine is such that
\begin{eqnarray}
|H\rangle \; |C\rangle \to |H,H\rangle \; |C_H\rangle,   \qquad \qquad
|V\rangle \; |C\rangle \to |V,V\rangle \; |C_V\rangle,
\end{eqnarray}       
where $|H\rangle$ and $|V\rangle$ are horizontal- and vertical-polarization
states of the original photon, $|C\rangle$ is the initial state of 
the cloning machine, and $|C_H\rangle$ and $|C_V\rangle$ are the 
(arbitrary) final states of the cloning machine, 
then the left and right circularly-polarized states, 
$|L\rangle=2^{-1/2}(|H\rangle + i|V\rangle)$ and
$|R\rangle=2^{-1/2}(|H\rangle - i|V\rangle)$, are transformed as
\begin{eqnarray}
|L\rangle \; |C\rangle 
\to 2^{-1/2} (|H,H\rangle \; |C_H\rangle + i |V,V\rangle \; |C_V\rangle)
\ne |L,L\rangle \; |C_L\rangle,  \nonumber \\
|R\rangle \; |C\rangle 
\to 2^{-1/2} (|H,H\rangle \; |C_H\rangle - i |V,V\rangle \; |C_V\rangle)
\ne |R,R\rangle \; |C_R\rangle,
\end{eqnarray}
with $|C_L\rangle$ and $|C_R\rangle$ being (arbitrary) final states
of the cloning machine. As a consequence, the cloning of circularly-polarized
states fails, even in the special case where $|C_H\rangle=|C_V\rangle$.
This is the simplest explanation of what is known today as the
{\it quantum no-cloning theorem}.

Independently of this story, a related paper was published almost 
simultaneously by \cite{Dieks82} in {\it Physics Letters}, also 
showing that the ``FLASH'' proposal by Herbert was flawed. 
Here, the proof relies on the existence of EPR states, see \cite{EPR1935},
which give rise to quantum correlations between spatially
separated systems. If two photons are prepared in the EPR state
\begin{equation}
|{\mathrm EPR}\rangle = 2^{-1/2}(|H,V\rangle -|V,H\rangle)
\end{equation}
it is well known that measuring the linear polarization of one
of them in the horizontal-vertical basis allows one to  
immediately predict the outcome of a measurement of the linear polarization 
of the second one in the same basis, even if the measurement events 
are separated by a spacelike interval. For example, if the first photon is
found to be in the $|H\rangle$ state, then the second photon will necessarily
be observed in the $|V\rangle$ state. This property holds for any measurement
basis. It had been realized since the early times of quantum mechanics
that this property, called {\it quantum entanglement}, 
does not permit superluminal communication. 
Indeed, the statistics of any measurement performed
on one of the twin photons remains unchanged irrespectively of the
measurement (or, more generally, the operation) applied on the second one.
Dieks noticed, however, that if it was possible to clone perfectly
one of the twin photons when the other one has been measured,
then superluminal communication would become possible; hence, cloning
must be impossible.

Assume that Alice measures the first photon
either in the horizontal-vertical linear polarization basis 
or in the left-right circular polarization basis depending on whether
she wants to transmit a 0 or a 1 to Bob. In the former case, 
the second photon will be found by Bob to be in a balanced mixture of the
$|H\rangle$ and $|V\rangle$ states, while in the second case it will be
in a balanced mixture of the $|L\rangle$ and $|R\rangle$ states.
These two mixtures are indistinguishable (they are characterized
by the same density operator, proportional to the identity $I$), 
which is why quantum mechanics is said to
``coexist peacefully'' with special relativity. However,
if the second photon could be cloned perfectly, Bob would then get either 
a balanced mixture of $|H,H\rangle$ and $|V,V\rangle$, or a balanced mixture 
of $|L,L\rangle$ and $|R,R\rangle$. These mixtures being distinguishable,
Bob would have a way to infer Alice's bit instantaneously (with some error,
but which can be made arbitrarily small as the number of clones increases).
Dieks concluded from this paradox that such a cloning transformation 
cannot be consistent with quantum mechanics.

It appears that the quantum no-cloning theorem is thus one of these 
scientific results that has been rediscovered several times, 
at least by Dieks, Ghirardi, Wootters, and Zurek. 
Actually, it can be argued that it was already implicitly used by 
Stephen Wiesner in his famous paper entitled ``Conjugate coding'' written 
in the 70's but only published in \cite{Wiesner83}, which is sometimes
considered as the founding paper of quantum information theory.
In some sense, the no-cloning theorem was already intrinsically contained
in the roots of quantum mechanics and is thus trivial; but, on the other hand, 
its discovery has contributed to revisiting quantum mechanics 
in an information theoretic language, which has had a decisive influence 
on the dramatic development of quantum information science over the past
twenty years. 

\subsection{Beyond the no-cloning theorem}

Soon after the publication of the quantum no-cloning theorem, another paper 
appeared in {\it Nature}, written by \cite{Mandel83}. In this paper,
entitled ``Is a photon amplifier always polarization dependent?'',
Mandel drew attention to the physical origin of the impossibility of making
a perfect amplifying apparatus for light, namely {\it spontaneous emission}. 
He showed that, if the amplifier is a single 2-level atom with a dipole moment
$\mu$, then the amplification of an incoming photon of polarization 
vector $\epsilon$ depends on the scalar product between $\mu$ and $\epsilon$.
If the polarization vector $\epsilon$ of the incoming photon
is parallel to the dipole moment $\mu$,
then the state $|1\rangle_\epsilon$ will, after some
interaction time, evolve into a state containing 
the desired two-photon state $|2\rangle_\epsilon$
due to stimulated emission. On the contrary, if $\epsilon$ is orthogonal
to $\mu$, then the two-photon component of the resulting state
corresponds to $|1\rangle_\epsilon |1\rangle_{\bar \epsilon}$,
where ${\bar \epsilon}$ is a polarization vector orthogonal to $\epsilon$.
This is due to spontaneous emission, which spoils the amplification
since one of the two photons has the wrong polarization
${\bar \epsilon}$. In other words,
with such a simple one-atom amplifier, the final state depends on the
polarization of the incoming photon.

Interestingly, Mandel noticed that if we consider a more elaborate amplifier
made of two such atoms with orthogonal dipole moments ($\mu_1$ and $\mu_2$),
it becomes nevertheless possible to amplify the photon
independently of its polarization, although this process suffers 
from the unavoidable noise originating from spontaneous
emission. Assuming that the two atoms interact similarly
with the incoming photon, one understands intuitively that if one
atom amplifies the photon ``well'' (when $\epsilon$ is close to $\mu_1$),
then the second atom amplifies it ``poorly'' (because $\epsilon$ is then 
approximately orthogonal to $\mu_2$). The balance between these
two effects results in an amplification that does not depend on $\epsilon$.
By filtering out the resulting two-photon component, one gets
\begin{equation}
\ket{1}_\epsilon \; \ket{0}_{\bar \epsilon} \to 
{2\over 3} \, \ket{2}_\epsilon\bra{2} \otimes \ket{0}_{\bar \epsilon}\bra{0}
+{1\over 3} \, \ket{1}_\epsilon\bra{1} \otimes \ket{1}_{\bar \epsilon}\bra{1}
\label{MandelCloning}
\end{equation}
irrespective of $\epsilon$. In some sense, the perfect cloning of polarization
via stimulated emission works with probability 2/3, while
spontaneous emission blurs the polarization with probability 1/3.

Mandel's paper remained mostly unnoticed and, remarkably, 
one had to wait more than ten years before the notion of {\it quantum
cloning machine}, which was implicitly contained in this paper, became popular.
In a seminal paper, \cite{Buzek96} realized that, 
although {\it perfect} quantum cloning is ruled out 
by the no-cloning principle, 
some {\it imperfect} cloning may be possible. They found out that
a qubit (2-level quantum system) that is in an unknown state
can be approximately duplicated, resulting in two pretty good clones 
of the original state. This result holds in full generality,
regardless of the physical variable carrying the qubit, so it goes
much beyond the polarization-independent amplification of a single photon
considered before. This paper had a considerable impact
at that time because quantum information theory was born, and 
it had been realized how fruitful it is to investigate 
quantum mechanics using an information language.

Consider a qubit in the state 
$|\psi\rangle = \alpha |0\rangle+\beta |1\rangle$,
where $|0\rangle$ and $|1\rangle$ form an orthonormal basis of
the Hilbert space, while $\alpha$ and $\beta$ are arbitrary complex numbers
satisfying $|\alpha|^2+|\beta|^2=1$. \cite{Buzek96} addressed
the following formal problem: find a transformation acting on an
original qubit in state $|\psi\rangle$
together with an auxiliary system (commonly viewed as
the cloning machine itself) that produces two clones 
with the same fidelity and is state-independent, or {\it universal}. 
If the cloning machine is initially put in state $|C\rangle$, then
\begin{eqnarray}
|0\rangle \; |C\rangle \to |\Sigma_0\rangle,   \qquad \qquad
|1\rangle \; |C\rangle \to |\Sigma_1\rangle,
\end{eqnarray}       
with the final states $|\Sigma_0\rangle$ and $|\Sigma_1\rangle$ belonging to
the product Hilbert space ${\cal H}_A \otimes {\cal H}_B \otimes {\cal H}_C$,
where  ${\cal H}_A$ and ${\cal H}_B$ denote the spaces of the two clones
(called $A$ and $B$) and ${\cal H}_C$ denotes the space of the cloning 
machine $C$, see Fig.~\ref{fignicolasfigprogress1}.
By linearity, an arbitrary qubit state $|\psi\rangle$ is cloned as
\begin{eqnarray}
|\psi\rangle \; |C\rangle &\to& \alpha |\Sigma_0\rangle + \beta
|\Sigma_1\rangle \equiv |\Sigma\rangle  \; .
\end{eqnarray}       
The fidelity of the clones, which measures the overlap between the input state
and each clone, is given by
\begin{eqnarray}
f_A(\psi)=\langle \psi | {\mathrm Tr}_{BC}(\Sigma) | \psi \rangle \; , \qquad
f_B(\psi)=\langle \psi | {\mathrm Tr}_{AC}(\Sigma) | \psi \rangle \; ,
\end{eqnarray}
where Tr denotes the trace and $\Sigma\equiv |\Sigma\rangle\langle\Sigma|$
is a short-hand notation for the density operator of a pure state.
\cite{Buzek96} showed that, under the constraint that $f_A(\psi) = f_B(\psi)$
is independent of $\psi$, quantum mechanics permits the existence of a
cloning transformation which achieves a fidelity as high as
\begin{equation}
f^{\mathrm{univ}} = 5/6 \simeq 0.833
\end{equation}
\input #progressfig1
This transformation, which is called a {\it quantum cloning machine}, 
is given by
\begin{eqnarray}
|0\rangle \; |C\rangle &\to& |\Sigma_0\rangle \equiv
\sqrt{2/3} \; |00\rangle_{AB} \; |0\rangle_C
+ \sqrt{1/3} \; |\Psi^+\rangle_{AB} \; |1\rangle_C,
\nonumber\\
|1\rangle \; |C\rangle &\to& |\Sigma_1\rangle \equiv
\sqrt{2/3} \; |11\rangle_{AB} \; |1\rangle_C
+ \sqrt{1/3} \; |\Psi^+\rangle_{AB} \; |0\rangle_C,
\end{eqnarray}       
where $|\Psi^+\rangle = 2^{-1/2} (|01\rangle+|10\rangle)$ is one of the Bell
states, while $|0\rangle_C$ and $|1\rangle_C$ denote two orthogonal
states of the cloning machine. 
It is easy to check, by tracing
over the cloning machine, that the two clones of an input state $\ket{0}$
are left in the joint state
\begin{equation}
\rho_{AB}={\mathrm Tr}_{C}(\Sigma)={2\over 3} \; |00\rangle\langle 00|
 +{1 \over 3} |\Psi^+\rangle \langle\Psi^+|
\label{eq-buzek}
\end{equation}
which is equivalent to Eq.~(\ref{MandelCloning}) given
the bosonic statistics of photons. More generally, if the input state
is $|\psi\rangle$, the first term of the right-hand side 
of Eq.~(\ref{eq-buzek}) becomes a projector onto $|\psi\rangle^{\otimes 2}$,
while the second term is some ($\psi$-depending) maximally-entangled state. Therefore, by tracing over one of the clones,
the resulting state of the other clone is
\begin{equation}
\rho_A={\mathrm Tr}_{BC}(\Sigma)={2\over 3} \; |\psi\rangle\langle\psi|
 +{1 \over 6} \; I  \; ,
\qquad 
\rho_B={\mathrm Tr}_{AC}(\Sigma)={2\over 3} \; |\psi\rangle\langle\psi| 
 +{1 \over 6} \; I  \; ,
\label{eq-buzek-1-clone}
\end{equation}
with $I$ denoting the identity operator, confirming
that the two clones are left in the same state.
They can be viewed each as emerging from
a quantum {\it depolarizing} channel: 
they are found in the right state $|\psi\rangle$
with probability 2/3, while they are replaced by a random qubit
$I/2$ with probability 1/3.

Soon after the publication of this paper,
it was proved that this machine is actually the {\it optimal} universal 
cloning machine, that is, the highest fidelity of cloning permitted
by quantum mechanics is indeed 5/6, see \cite{Bruss98a}.

This discovery by \cite{Buzek96} triggered an immense interest 
and initiated an entire subfield of quantum information science
devoted to quantum cloning. 
In particular, further studies addressed cloning in dimensions
larger than 2, state-dependent cloning (considering a restricted set 
of input states),
the so-called $N$-to-$M$ cloning (where one produces $M$ identical clones
out of $N$ identical replicas of the original), the asymmetric cloning
(where the clones have unequal fidelities), the cloning of orthogonal 
qubit states, the cloning of continuous-variable states 
(such as coherent states), 
the economical cloning (where no ancillary space is necessary),
the probabilistic cloning (which is not deterministic, i.e., it 
does not succeed with probability 100\%), or even
the cloning of quantum entanglement (instead of quantum states). 
These numerous results will be reviewed in Section~\ref{section-overview}.

Aside from its utmost
importance for the foundations of quantum mechanics, the study of quantum
cloning has drawn a lot of interest probably also because it is closely
connected to quantum key distribution (QKD). Indeed, in many cases, the
cloning machine is known to be the most powerful eavesdropping strategy
against QKD protocols: the eavesdropper duplicates the quantum state
and sends one clone to the authorized party, while keeping the second
clone for later measurement. The characterization of cloning machines
is therefore crucial to assess the security of these QKD protocols (this
particular connection is out of the scope of the present review paper,
and will not be discussed any further).

\subsection{Quantum cloning without signaling}

Before entering the detailed study of quantum cloning, it is 
interesting to backtrack for a moment and further discuss
the proof of the quantum no-cloning theorem based on a pair of entangled
photons due to \cite{Dieks82}. 
As explained earlier, if Alice measures her photon 
either in the horizontal-vertical linear polarization 
basis or in the left-right circular polarization basis, and if Bob is
able to clone his photon perfectly, then he obtains two distinguishable 
2-photon mixtures, which apparently makes superluminal signaling possible.

A natural idea, due to \cite{Gisin98}, is to assume that Bob's cloning 
machine must necessarily introduce some intrinsic noise, and determine
the minimum amount of noise that must be added so that causality 
ceases to be violated. Remarkably, it so happens that the minimum noise
needed to comply with causality exactly coincides with that of the optimal
universal cloning machine. In other words, the upper bound on quantum 
cloning can be derived from simple principles.

As shown earlier, the two clones of the universal machine emerge each
from a quantum depolarizing channel, see \cite{Buzek96}. 
This channel can be interpreted
as giving rise to a shrinking of the vector representing the qubit state
in the Bloch sphere. Using the Bloch representation
\begin{equation}
\rho=(I+{\bf m}\cdot{\mathbf \sigma})/2,
\end{equation}
where ${\bf m}$ is a vector isomorphic to state $\rho$
and ${\mathbf \sigma}\equiv(\sigma_x,\sigma_y,\sigma_z)$
is the vector of Pauli matrices,
we see from Eq.~(\ref{eq-buzek-1-clone})
that a state associated with ${\bf m}$ yields two 
clones which are in a state associated to $2{\bf m}/3$, 
independently of the orientation of ${\bf m}$. Therefore, 
this universal cloning machine is sometimes also termed as {\it isotropic}.

Following \cite{Gisin98}, consider that the (pure) state of the original qubit 
is  associated with the (unit-norm) vector ${\bf m}$, 
and let us restrict our search to
cloning machines that are symmetric and isotropic, that is, 
the clones are in the states
\begin{equation}
\rho_{A}({\vec m})=\rho_{B}({\vec m})=
(1+\eta {\vec m}\cdot{\vec \sigma})/2,
\label{signal1}
\end{equation}
where $\eta$ is an unknown ``shrinking factor'' ($0\le \eta\le 1$). 
It is easy to check that $\eta$ is related to the fidelity by 
\begin{equation}
f_{A}({\vec m})=f_{B}({\vec m})=(1+\eta)/2.
\end{equation}
Using Eq.~(\ref{signal1}), the Hilbert-Schmidt decomposition 
of the joint state of the two clones can be written as
\begin{equation}
\rho_{AB}({\vec m})=\Big(I+\eta{\bf m}\cdot{\mathbf \sigma}\otimes I
+I \otimes \eta{\bf m}\cdot{\mathbf \sigma} + \sum _{j,k} t_{j,k}
\sigma_j\otimes \sigma_k \Big)/4,
\end{equation}
where the matrix $t_{j,k}$ measures the quantum correlations 
between the clones. \cite{Gisin98} went on deriving constraints
on $t_{j,k}$ that result from covariance and causality. The
covariance property (which will be explained in details later on) 
means, physically, that rotating the original qubit around, say, 
the $z$-axis before cloning must be equivalent to cloning the original
qubit and then rotating each of the two clones by the same amount
around the $z$-axis. 
Following Dieks' argument, the causality condition is taken into account
by imposing that
\begin{equation}
\rho_{AB}({\vec m}_1) + \rho_{AB}(-{\vec m}_1)=
\rho_{AB}({\vec m}_2) + \rho_{AB}(-{\vec m}_2)
\end{equation}
which expresses the fact that the two-clone states corresponding 
to two indistinguishable
mixtures of input states, $\{{\vec m}_1,-{\vec m}_1\}$ and 
$\{{\vec m}_2,-{\vec m}_2\}$, are themselves indistinguishable.
Putting all these conditions on $t_{j,k}$ together, one can show that
the maximum value of $\eta$ that preserves
the positivity of the two-clone state, $\rho_{AB}({\vec m})\ge 0$, 
is $\eta=2/3$; hence $f_A=f_B=5/6$. This provides an alternate proof of the optimality of the qubit universal cloner of \cite{Buzek96}.

To be complete, let us mention that such a use of the no-signaling condition
has been criticized in \cite{Bruss2000a}, the argument being that the linearity 
and trace-preservation properties of the cloning map (which, combined, 
imply the no-signaling condition) are not sufficient, strictly speaking, 
and need to be supplemented with the complete positivity condition 
in order to bound the cloning fidelity.
This simple technique, however, has proved to be successful 
to recover conditions on probabilistic cloning, see \cite{Hardy99},
on asymmetric universal cloning, see \cite{Ghosh99}, 
or even to find a new class of real cloning machines, see \cite{Navez2003}.

\subsection{Content of this review paper}

The rest of this review paper will be devoted to the study of quantum cloning
machines, as well as their optical realization. Let us sketch 
the content of the next Sections.  
Section~\ref{section-overview} will provide
an overview of the chronology of the main papers 
that have been written in this context, focusing on the results but
skipping the derivations. The numerous classes
of quantum cloning machines will be presented (universal cloners,
Pauli or Heisenberg cloners, phase-covariant cloners, Fourier-covariant cloners,
group-covariant cloners, real cloners, entanglement cloners, 
continuous-variable cloners, probabilistic cloners, or economical cloners).

In Section~\ref{section-cp-map},
we will consider the issue of quantum cloning from a formal point of view,
based on the description of the associated completely-positive (CP) map
and the notion of covariance. This study will be restricted 
to 1-to-2 cloning, and will focus on the isomorphism
between CP maps and operators. It will be shown that finding the optimal
cloning map reduces to a semidefinite programming problem, which can be
efficiently solved by numerical methods. It will also be shown that the unitary
realization of a cloning map based on the ``double-Bell' ansatz provides
a simple and efficient tool to investigate cloning analytically. Some
examples of $d$-dimensional 1-to-2 cloners will be provided.

In Section~\ref{section-universal-N-to-M}, this formal study will be 
extended to $N$-to-$M$ cloning machines in $d$~dimensions, 
but will be restricted the case of universal cloning. 
The derivation of the optimal cloning transformation
as well as the optimality proof will be detailed. In addition, the
extension to asymmetric cloning machines and the notion of universal-NOT
gate will be discussed. The reader who is mainly interested in the
optical realization of cloning machines and not so much in their
theoretical derivation may skip Sections~\ref{section-cp-map} and \ref{section-universal-N-to-M}, and immediately go to the next Sections.

In Section~\ref{section-cloning-photons}, 
the optical implementation of the universal quantum cloning machines
will be analyzed in details. The cloning experiments relying on 
stimulated parametric downconversion will be described first, followed by
those relying on the symmetrization that can be obtained 
with a Hong-Ou-Mandel interferometer. Next, the optical realization of
(universal) asymmetric cloning machines will be discussed,
as well as the (universal) cloning of a pair of orthogonal qubits.

In Section~\ref{section-phase-covariant}, the phase-covariant cloning machines
will be developed for qubits as well as $d$-dimensional systems, 
in a 1-to-2 or $N$-to-$M$ configuration. The experimental realization
of phase-covariant cloning for photonic qubits will be described.

In Section~\ref{section-continuous-variables}, the generalization
of quantum cloning to states belonging to an infinite-dimensional 
Hilbert space will be considered. In particular, the cloning of
coherent states of light by phase-sensitive amplification will be
explained, as well as the experimental realization of continuous-variable
cloning using linear optics, measurement, and feed-forward. The cloning
of a finite-width distribution of coherent states will be analyzed,
as well as the cloning of a pair of conjugate coherent states.
Finally, the conclusions will be drawn in Section~\ref{section-conclusions}.

\section{Overview of quantum cloning machines}

\label{section-overview}

\subsection{Universal cloning machines}

This Section will be devoted to a summary of the various cloning machines 
that have been introduced in the literature, following the chronology
as well as possible. Soon after the universal quantum cloning machine was
discovered by \cite{Buzek96}, the question arose whether
this machine was optimal. As already mentioned,
this cloning machine is required to be {\it symmetric}, that is,
the 2 clones must have equal fidelities $f_A(\psi) = f_B(\psi)$, $\forall\psi$.
In addition, it must be {\it universal}
(or state-independent), which means that all states are cloned 
with the same fidelity, independent of $\psi$.
It was proven by \cite{Bruss98a} that it is indeed the 
{\it optimal} symmetric universal duplicator for qubits,
so that $f=5/6$ is indeed the highest fidelity allowed by quantum mechanics
in this case. In the same paper, the concept of optimal {\it state-dependent} 
cloning machines was also introduced, that is, transformations 
that optimally duplicate only a particular subset of the input states.
Almost simultaneously, \cite{Gisin97}
introduced the concept of $N$-to-$M$ quantum cloning
machines, which transform $N$ identical replicas of an arbitrary state, 
$|\psi\rangle^{\otimes N}$, into $M>N$ identical clones.
They were able to prove for low  $N$ that the optimal universal $N$-to-$M$ cloning 
of qubits is characterized by the fidelity
\begin{equation}
f_{N\to M}^{\mathrm univ} = \frac{M(N+1)+N}{M(N+2)}.
\end{equation}
Incidentally, this confirms the optimality of the 1-to-2 universal
cloning machine with fidelity $f_{1\to 2}^{\mathrm univ}=5/6$.
The quantum network that realizes this $1$-to-$2$ universal cloning
of qubits was described by \cite{Buzek97a}, while it was
extended to $1$-to-$M$ universal cloning in \cite{Buzek98d}.
Note also that when the number of clones $M$ increases for fixed $N$,
the cloning fidelity decreases. This can simply be interpreted as a spreading
of quantum information over more clones. In the limit $M\to\infty$, 
the cloning transformation tends to a measurement, 
which confirms that the optimal
(state-independent) estimation of the state $|\psi\rangle^{\otimes N}$
of $N$ identical qubits has a fidelity
\begin{equation}
f_{N\to \infty}^{\mathrm univ} = \frac{N+1}{N+2}
\end{equation}
as originally derived in \cite{Massar95}.

Then, in early 1998, the extension of quantum cloning machines to higher-dimensional spaces was considered independently by \cite{Buzek98a,Cerf98,Werner98}.
The form of the optimal universal $1$-to-$2$ cloner in dimension $d$ was
conjectured by \cite{Buzek98a,Cerf98}, while the derivation and
full optimality proof of the universal $d$-dimensional $N$-to-$M$ cloner
was given by \cite{Werner98,Keyl99}. The optimal fidelity 
of the universal $1$-to-$2$ cloner of $d$-dimensional states
(or qu$d$its) was shown to be
\begin{equation}
f_{1\to 2}^{\mathrm univ}(d) = \frac{d+3}{2(d+1)}
\label{d-dim-univ}
\end{equation}
while, for arbitrary $N$ and $M>N$, it is
\begin{equation}
f_{N\to M}^{\mathrm univ}(d) = \frac{M(N+1)+(d-1)N}{M(N+d)}.
\end{equation}

In \cite{Cerf98}, the cloning of $d$-dimensional systems was actually
investigated in a more general setting: a large class of symmetric or asymmetric, universal or state-dependent, $1$-to-$2$ cloning machines 
was introduced in arbitrary dimension $d$. The optimality of this class
of cloners was only conjectured, but, in the special case of a symmetric 
and universal cloner, Eq.~(\ref{d-dim-univ}) was also derived.
For the set of asymmetric universal $1$-to-$2$ cloning machines, 
the balance between the fidelity of the two clones 
\begin{equation}
f_A^{\mathrm univ}(d)=\eta_A+(1-\eta_A)/d, \qquad \qquad
f_B^{\mathrm univ}(d)=\eta_B+(1-\eta_B)/d,
\label{univasymm2}
\end{equation}
was characterized by the simple relations
\begin{equation}
\eta_A=1-\alpha^2,\qquad \eta_B=1-\beta^2,\qquad
\alpha^2+\frac{2\, \alpha\beta}{d} + \beta^2 = 1 ,
\label{univasymm}
\end{equation}
where $\eta_A$ and $\eta_B$ are the ``shrinking'' factors associated
with the clones ($\eta$ is the probability that the input state
emerges unchanged at the output of the quantum depolarizing channel).
Here, $\alpha$ and $\beta$ are positive real variables. It is instructive
to notice that in the limit $d\to\infty$, the cloning of
quantum information resembles the distribution of a resource 
that can strictly not be shared: the probability that $\ket{\psi}$ is found
in one clone is complementary to the probability that it is found in
the second clone, that is, $\eta_A+\eta_B=1$.

The quantum circuit for the asymmetric universal cloning of qubits
was displayed in \cite{Buzek98c},
while it was generalized to $d$-dimensional symmetric or
asymmetric cloning in \cite{Braunstein01}.

Finally, even more general quantum cloning machines were obtained
in the special case of qubits ($d=2$) in an independent work 
by \cite{Niu98}. There, the $1$-to-$2$ asymmetric and state-dependent 
cloning of a qubit was investigated in full generality, and,
in particular, formulas (\ref{univasymm}) were recovered for $d=2$
without any assumption.

\subsection{Pauli and Heisenberg cloning machines}

The results of \cite{Cerf98} were later expanded
in \cite{Cerf2000a} for the case of qubits,
and \cite{Cerf2000b} for the case of $d$-dimensional systems.
The specificity of the approach to quantum cloning underlying these papers
is that one considers the cloning of a system that is initially 
maximally entangled with another system instead of the cloning of
a pure state. This second system acts as
a ``reference'' by keeping a memory of the original state after the
cloning has been achieved. The final state of the ``reference'', 
the two clones, and the cloning machine then fully characterizes 
the cloning transformation, as a consequence of the isomorphism 
between completely positive (CP) maps and operators (this will be explained in details in Section~\ref{section-isomorphism}).
By choosing an appropriate form for this final state, one generates
a large class of quantum cloning machines.

For qubits ($d=2$), this class corresponds to
the so-called {\it Pauli cloning machines}, whose 
clones emerge from two -- possibly distinct -- Pauli channels.
In a Pauli channel, the input qubit undergoes one of the three Pauli
rotations $\{\sigma_x,\sigma_y,\sigma_z\}$
or the identity $I$ with respective probabilities
$\{p_x,p_y,p_z,1-p_x-p_y-p_z\}$. For example,
it was shown that the whole class of {\it symmetric} Pauli cloning machines
corresponds to Pauli channels with probabilities $p_x=x^2$, $p_y=y^2$,
and $p_z=z^2$, with ${x,y,z}$ satisfying the condition
\begin{equation}
x^2+y^2+z^2+xy+xz+yz=1/2
\label{ellipsoid}
\end{equation}
The action of these Pauli cloners is easy to understand knowing that, if the
original qubit is in an eigenstate of $\sigma_x$, namely $(\ket{0}\pm\ket{1})/\sqrt{2}$, then it is rotated by an angle $\pi$ 
around the $y$-axis ($z$-axis) under $\sigma_y$ ($\sigma_z$)
while it is left unchanged (up to a sign)
by $\sigma_x$. Therefore, the cloning fidelity of the eigenstates
of $\sigma_x$ is $1-p_y-p_z$. Similarly,
the eigenstates of $\sigma _y$, namely $(\ket{0}\pm i \ket{1})/\sqrt{2}$,
are cloned with the fidelity $1-p_x-p_z$, while the
eigenstates of $\sigma_z$, namely $\ket{0}$ and $\ket{1}$, are
cloned with the fidelity $1-p_x-p_y$.
The universal $1$-to-$2$ symmetric cloning machine simply corresponds
to $p_x=p_y=p_z=1/12$. Note that these Pauli cloning machines
appear to be a special case of the state-dependent cloning transformations 
considered in \cite{Niu98}.

These considerations can be extended to $d$ dimensions in order to
obtain the set of so-called {\it Heisenberg cloning machines},
whose clones emerge from two -- possibly distinct -- Heisenberg channels. 
In a Heisenberg channel, the $d$-dimensional input state undergoes,
according to some probability distribution,
one of the $d^2$ error operators $E_{m,n}$ (with $0\le m,n\le d-1$)
that form the discrete Weyl-Heisenberg group.
It can be shown that the probability distribution of the $E_{m,n}$ errors 
for the first clone is dual, under a Fourier transform, to that of the
second clone. This corroborates the fact that if one clone is close-to-perfect
(its associated error distribution is peaked), then the second clone 
is very noisy (its associated error distribution is flat).
More precisely, this fidelity balance between the two clones can be shown 
to result from a {\it no-cloning uncertainty principle}, 
akin to Heisenberg principle, see also \cite{Cerf99}.
The quantum circuit realizing these Heisenberg cloning machines
was displayed in \cite{Braunstein01}.

Recently, the optimality of this entire class of (Pauli or Heisenberg)
quantum cloning machines has been rigorously proven by \cite{Chiribella2005} 
in the following sense: under some general invariance conditions, 
the cloners of this class 
coincide with all the {\it extremal} cloners. Therefore,
for a given (invariant) figure of merit, it is sufficient
to search the optimal cloner within this class to be guaranteed
that the found solution is the global optimal cloner.

\subsection{Phase- and Fourier-covariant cloning machines}

In 2000, an important class of state-dependent qubit cloning machines, 
named {\it phase-covariant cloning machines},
was introduced by \cite{Bruss2000b}. It is defined as a transformation
that clones all the {\it balanced} superpositions of basis states 
with the same (and highest) fidelity. These states 
\begin{equation}
\ket{\psi}=(\ket{0}+\e^{i\phi}\ket{1})/\sqrt{2}  \; ,
\end{equation}
with $\phi$ being an arbitrary phase,
are located on the equator of the Bloch sphere.
The optimal cloner also fulfills the
covariance condition with respect to the rotation of $\phi$, that is,
cloning the rotated original qubit is equivalent to cloning the original qubit
followed by a rotation of each of the clones. The optimal 
phase-covariant symmetric $1$-to-$2$ cloner was found to have
a fidelity
\begin{equation}
f_{1\to 2}^{\mathrm pc}(2) = \frac{1}{2} + \frac{1}{\sqrt{8}}
\simeq 0.854,
\end{equation}
which is higher than that of the corresponding universal cloner,
$f_{1\to 2}^{\mathrm univ}(2)=5/6$. In contrast, the resulting
fidelity for the states $\ket{0}$ and $\ket{1}$, corresponding to the poles
of the Bloch sphere, is equal to $3/4$, which is
lower than $f_{1\to 2}^{\mathrm univ}(2)$. In some sense, it is possible
to better clone some restricted set of states (the equator) 
at the expense of a worse cloning of some other states (near the poles).

Interestingly, this phase-covariant cloner
can be viewed simply as a special case of the Pauli cloners,
see \cite{Cerf2002b}. If we take $x=y=1/\sqrt{8}$ and $z=1/2-1/\sqrt{8}$, 
which satisfies Eq.~(\ref{ellipsoid}), we indeed recover the same cloner: 
the eigenstates of $\sigma_x$ are cloned with fidelity
$1-p_y-p_z=1/2+1/\sqrt{8}$, while the eigenstates of $\sigma _y$
are cloned with the fidelity $1-p_x-p_z=1/2+1/\sqrt{8}$. 
It was observed in \cite{Cerf2002b} that imposing these 4 states
lying symmetrically on the equator to be cloned with the same fidelity 
results in the phase-covariant cloner,
which actually gives the same fidelity for all states on the equator
(the deep reason for this equivalence was found in \cite{Chiribella2005}).
Finally, we verify that this Pauli cloner clones 
the eigenstates of $\sigma_z$ with a lower fidelity $1-p_x-p_y=3/4$.

One can summarize the results on qubit cloning machines
by noting that the eigenstates of the three Pauli matrices
play the role of three mutually unbiased (MU) bases for qubits
(MU bases are such that the modulus of the scalar
product of any two states taken from distinct bases is $1/\sqrt{d}$, with
$d$ being the dimension). One can thus define three generic classes
of qubit cloning machines, namely, the universal cloner (which can be obtained
by imposing the states of 3 MU bases to be cloned with the same 
and highest fidelity), 
the phase-covariant cloner (if the states of only 2 MU bases 
are cloned equally), and some particular Pauli cloner (if the states 
of all 3 MU bases are cloned with unequal fidelities).
The cloning of qubits having been essentially covered,
it became natural to turn to the state-dependent cloning of 
qutrits ($d=3$).

In \cite{Cerf2002b}, four kinds of Heisenberg
cloning machines were defined for qutrits, depending on whether
four, three, two or none of the MU bases are cloned
with the same fidelity. 
If none of the MU bases are cloned with equal fidelities, one has
a particular Heisenberg cloning machine.
On the contrary, if all four MU bases
are cloned with the same fidelity,
one recovers Eq.~(\ref{d-dim-univ}) for $d=3$ in the case of symmetric cloning, that is, 
the qutrit universal cloner with fidelity 
\begin{equation}
f_{1\to 2}^{\mathrm univ}(3) = 3/4 .
\end{equation}
If three MU bases are requested to be cloned with the same fidelity, 
one gets the so-called {\it double-phase covariant} 
qutrit cloner, with a fidelity
\begin{equation}
f_{1\to 2}^{\mathrm pc}(3) = \frac{5+\sqrt{17}}{12} \simeq 0.760
\label{eq-qutrit-pc}
\end{equation}
slightly higher than $f_{1\to 2}^{\mathrm univ}(3)$.
This cloner, which was independently derived in \cite{D'Ariano2001b},
has the property that it clones with the same and highest fidelity 
all the balanced superpositions
\begin{equation}
\ket{\psi}=\frac{1}{\sqrt{3}}(\ket{0}+\e^{i\phi_1}\ket{1}+\e^{i\phi_2}\ket{2})
\label{qutritbalanced}
\end{equation}
for arbitrary phases $\phi_1$ and $\phi_2$ (it is also
covariant with respect to both $\phi_1$- and $\phi_2$-rotations).
This can be understood by noting that if we complete the computational basis 
with any triplet of bases in order to make 4 MU bases, these
3 bases only consist of balanced superposition states. In analogy
with the qubit case, it then appears that imposing these 3 bases
to be cloned with the same (and highest) fidelity results in
a cloning machine that clones all states (\ref{qutritbalanced}) equally well,
that is, the double-phase-covariant cloner.

Finally, we may impose that two MU bases that are dual under a Fourier transform
are cloned with the same (and highest) fidelity,
the other two being also cloned with an equal (albeit lower) fidelity.
For example, the
computational basis $\{\ket{0},\ket{1},\ket{2}\}$
and the dual basis $|\ket{j}\rangle=3^{-1/2} \sum_{k=0}^2 \gamma^{jk} \ket{k}$
with $j=0,1,2$ and $\gamma=\e^{2\pi i/3}$ form such a pair of MU bases. 
We then get the so-called {\it Fourier-covariant} cloner for qutrits, 
see \cite{Cerf2002b}, with a fidelity
\begin{equation}
f_{1\to 2}^{\mathrm Fourier}(3) = \frac{1}{2}+\frac{1}{\sqrt{12}} \simeq 0.789
\end{equation}
which is even higher than $f_{1\to 2}^{\mathrm pc}(3)$
as expected since the considered set of input states is smaller than
for the double-phase-covariant cloner. This cloner is covariant
with respect to a Fourier transform, hence it clones two Fourier-conjugate
bases with the same fidelity.  

Note that, except in dimension 2,
any two MU bases cannot always be mapped onto any other two MU bases,
so that the Fourier-covariant cloner is not the unique transformation that
clones equally well two MU bases. Indeed, in \cite{Durt2003}, it was shown
that, in dimension 4, the cloner for two MU bases conjugate
under a Fourier transform differs from the cloner for two MU bases conjugate
under a double Hadamard transform. In the special case of qubits ($d=2$), 
however, all pairs of MU bases are unitarily equivalent, so that
the Fourier-covariant and phase-covariant cloners coincide,
$f_{1\to 2}^{\mathrm Fourier}(2) = f_{1\to 2}^{\mathrm pc}(2)$.

\subsection{Group-covariant cloning machines}

In \cite{D'Ariano2001b}, a general method for optimizing the 
{\it group-covariant cloners} was derived. More specifically, they
considered the optimal cloning transformations that are covariant under 
a proper subgroup $\Omega$ of the universal unitary group $U(d)$.
For example, the universal qubit cloner is covariant with respect
to $U(2)$, while the phase-covariant qubit cloner is 
covariant with respect to $U(1)$. They used this technique to
derive the symmetric double-phase-covariant cloner for qutrits 
corresponding to Eq.~(\ref{eq-qutrit-pc}), as well as 
the $1$-to-$3$ symmetric phase-covariant cloner
for qubits, associated with the fidelity
\begin{equation}
f_{1\to 3}^{\mathrm pc}(2) = \frac{5}{6} \simeq 0.833 .
\end{equation}
Owing to the complexity of the group-theoretical parametrization
of CP maps underlying this technique, its applicability seems rather 
limited. Nevertheless, using another method, \cite{Fan2001b}
were able to derive the optimal $1$-to-$M$ symmetric
phase-covariant cloning of qubits, yielding the fidelity
\begin{equation}
f_{1\to M}^{\mathrm pc}(2) = \left\{
\begin{array}{ll}
\frac{1}{2} +\frac{\sqrt{M(M+2)}}{4M}  & \qquad M {\mathrm~even},\\
\phantom{\Big(} \frac{1}{2}+\frac{M+1}{4M}
\phantom{\Big)} & \qquad M {\mathrm~odd}.
\end{array}
\right.
\label{phasecov1toM}
\end{equation}
More recently, \cite{D'Ariano2003} succeeded in applying
this theory of group-covariant cloning in order to confirm
Eq.~(\ref{phasecov1toM}), as well as to find
a general expression for $f_{N\to M}^{\mathrm pc}(2)$ and
the associated $N$-to-$M$ cloner. This expression, which was
partly conjectured in \cite{Fan2001b}, is quite complex,  
and depends on whether $N$ and $M$ have the same parity. It was 
noticed that, if the parities do not match, then the cloner 
that optimizes the fidelity of each of the clones does not
coincide with the optimal cloner with respect to the global 
fidelity (measuring how well the joint state of the clones
approximates $\ket{\psi}^{\otimes M}$, if $\ket{\psi}$ is the
state of the original). In the case of qutrits ($d=3$), 
\cite{D'Ariano2003} also found the optimal $1$-to-$M$ symmetric
phase-covariant cloner. The expression for its fidelity 
$f_{1\to M}^{\mathrm pc}(3)$ is rather complex, 
and depends on $M$ modulo 3.

\subsection{High-$d$ state-dependent cloning machines}

In parallel with this series of results on group-covariant cloning
involving several originals and clones but in low dimensions,
both the phase-covariant and Fourier-covariant $1$-to-$2$
cloning machines were extended to arbitrary dimensions $d$.
In \cite{Cerf2002a}, the $d$-dimensional symmetric
Fourier-covariant cloner was derived, and shown
to be characterized by the fidelity
\begin{equation}
f_{1\to 2}^{\mathrm Fourier}(d) = \frac{1}{2}+\frac{1}{\sqrt{4d}} .
\label{fidelity-fourier}
\end{equation}
It clones equally well two MU bases that are conjugate under a Fourier
transform, such as the computational basis
$\{\ket{0},\cdots \ket{d-1}\}$
and the dual basis $|\ket{j}\rangle=d^{-1/2} \sum_{k=0}^{d-1} \gamma^{jk} \ket{k}$
with $j=0,\cdots d-1$ and $\gamma=\e^{2\pi i/d}$.
The asymmetric Fourier-covariant cloners were also characterized
in the same paper.

Then, in \cite{Fan2003a}, the $d$-dimensional symmetric 
multi-phase-covariant cloner was derived, giving the fidelity
\begin{equation}
f_{1\to 2}^{\mathrm pc}(d) = \frac{1}{d} + 
\frac{d-2+\sqrt{d^2+4d-4)}}{4d} .
\label{fidelity-phase}
\end{equation}
It clones with the same (and highest) fidelity all balanced superpositions
of the states of the computational basis, with arbitrary phases.
This result was independently derived in \cite{Lamoureux2005,Rezakhani2005},
where it was also extended to asymmetric cloners. Note also that the
role of multi-phase-covariant cloners in the context of entanglement-based
QKD protocols was first studied in \cite{Durt-qutrits} for qutrits,
then in \cite{Durt-qudits} for $d$-dimensional systems.

\subsection{Cloning a pair of orthogonal qubits}

Another possible variant to the problem of cloning
was studied by \cite{Fiurasek2002}, who introduced universal cloning 
machines that transform 2 qubits that are in an antiparallel joint state 
$\ket{\psi} \ket{\psi^{\perp}}$ into $M$ clones of $\ket{\psi}$, with 
$\braket{\psi}{\psi^{\perp}}=0$. It was proven that, when $M$ is large
enough, such a cloner outperforms the standard $2$-to-$M$ cloner. 
One has the fidelity
\begin{equation}
f_{1,1\to M}^{\mathrm univ}(2) = \frac{1}{2}+
\frac{\sqrt{(M+2)/(3M)}}{2} 
\end{equation}
which is greater than $f_{2\to M}^{\mathrm univ}(2)$ for $M>6$.
In some sense, it is better to replace one of the two original states
$\ket{\psi}$ by its orthogonal state $\ket{\psi^{\perp}}$ if the goal is
to produce $M>6$ clones. This effect can be understood 
at the limit $M\to\infty$, 
\begin{equation}
f_{1,1\to \infty}^{\mathrm univ}(2) = \frac{1}{2}+
\frac{1}{2\sqrt{3}}  \simeq 0.789 ,
\end{equation}
that is, for the optimal measurement of a pair of antiparallel qubits. 
Indeed, it had been noticed earlier by \cite{Gisin99} that 
measuring $\ket{\psi}\ket{\psi^{\perp}}$ yields more information 
than measuring $\ket{\psi}^{\otimes 2}$, with
$f_{2\to \infty}^{\mathrm univ}(2) = 3/4$. An interpretation of this
property lies in the dimension of the Hilbert space spanned by $\ket{\psi}\ket{\psi^{\perp}}$, which is 4, while $\ket{\psi}^{\otimes 2}$
only spans the 3-dimensional symmetric subspace of 2 qubits, ${\cal H}^+$.

\subsection{Entanglement cloning machines}

Another problem, related to quantum cloning, has been investigated
in \cite{Lamoureux2004b}. There, it was shown that the amount of
entanglement contained in a two-qubit state cannot be cloned exactly,
in analogy with the impossibility of cloning the state itself. If
a cloning machine is devised that produces maximally-entangled clones
for maximally-entangled qubit pairs at the input, then it cannot
yield unentangled clones for all product states at the input.
Nevertheless, the approximate cloning of entanglement is very well possible.
In \cite{Lamoureux2004b}, a class of $1$-to-$2$ 
{\em entanglement cloning machines} was defined, which
are universal over the set of maximally-entangled two-qubit states. 
The symmetric cloner of this class
provides two clones of all maximally-entangled two-qubit states
with the optimal fidelity
\begin{equation}
f_{1\to 2}^{\mathrm entang}(2\times 2) = 
\frac{5+\sqrt{13}}{12}  \simeq 0.717
\end{equation}
corresponding to an entanglement of formation 0.285 e-bits.
In contrast, all product states are transformed into unentangled clones.
This was recently extended to the cloning of entanglement for 
$d\times d$-dimensional systems in \cite{Karpov2005}. 
The fidelity of the optimal symmetric entanglement cloner
that is universal over the set of maximally-entangled 
$d\times d$-dimensional states is
\begin{equation}
f_{1\to 2}^{\mathrm entang}(d\times d) = \frac{1}{4} \left[
\frac{d^2+1}{d^2-1} + \sqrt{ 1+\frac{4}{d^2}
\left( \frac{d^2-2}{d^2-1}  \right)^2  } 
\right] .
\end{equation}

\subsection{Real cloning machines}

Still another class of $d$-dimensional $1$-to-$2$ cloners was
introduced in \cite{Navez2003}, named {\it real cloning machines}.
It is defined as a transformation that clones all {\it real} superpositions
of the computational basis states with the same (and highest) fidelity.
The optimal $1$-to-$2$ symmetric real cloner in dimension $d$
was shown to have the fidelity
\begin{equation}
f_{1\to 2}^{\mathrm real}(d) = \frac{1}{2}+
\frac{2-d+\sqrt{d^2+4d+20}}{4(d+2)} .
\end{equation}
Note that in dimension $d=2$, the set of real states forms a circle
in the Bloch sphere which is unitarily equivalent to the equator, 
so that we have
$f_{1\to 2}^{\mathrm real}(2) = f_{1\to 2}^{\mathrm pc}(2)
=f_{1\to 2}^{\mathrm Fourier}(2) = 1/2+1/\sqrt{8}$. For any dimension
$d>2$, one has
\begin{equation}
f_{1\to 2}^{\mathrm univ}(d) < f_{1\to 2}^{\mathrm pc}(d) 
< f_{1\to 2}^{\mathrm real}(d) < f_{1\to 2}^{\mathrm Fourier}(d).
\end{equation}
Note that, for $d=4$, we have the identity
$f_{1\to 2}^{\mathrm real}(4)=f_{1\to 2}^{\mathrm entang}(2\times 2)$.
This comes from the fact that the set of maximally-entangled 
two-qubit states is isomorphic to the set of four-dimensional real states.

\subsection{Highly asymmetric cloning machines}

In \cite{Iblisdir2005} the concept of multipartite $N$-to-$M$ cloning machines
(with $M>2$) was introduced. 
These machines are highly asymmetric in the sense that
they produce $M$ clones of unequal fidelities. A very general group theoretical 
approach to the construction of the multipartite asymmetric cloning machines 
for qubits was then presented in \cite{Iblisdir2004a}. It was applied to
several particular examples such as the asymmetric $1\rightarrow N+1$ cloning machine, which produces two kinds of clones, one clone with fidelity $f^A$ 
and $N$ clones with fidelity $f^B$. The optimal fidelities read
\begin{equation}
f_{1\rightarrow N+1}^A=1-\frac{2}{3}x^2, \qquad f_{1\rightarrow N+1}^B= \frac{1}{2}+
\frac{1}{3N}(x^2+x\sqrt{(1-x^2)N(N+2)}),
\label{asymmetric1toN1}
\end{equation}
where $x\in(0,1)$ parametrizes the class of the optimal $1\rightarrow N+1$ asymmetric
cloners. Note that the formula (\ref{asymmetric1toN1}) holds only for $N>1$. 
It also was conjectured, based on exact analytical calculations for low $N$,
that the optimal $N \rightarrow N+1$ asymmetric cloner 
which produces, from $N$ replicas of a qubit, $N$ clones with fidelity $f^A$
and a single clone with fidelity $f^B$ achieves
\begin{equation}
f_{N\rightarrow N+1}^A=1-\frac{2}{N(N+2)}x^2, \qquad 
f_{N\rightarrow N+1}^B=1-\frac{1}{2}\left(\sqrt{\frac{N}{N+2}}x-\sqrt{1-x^2}\right)^2.
\end{equation}
The extension to $d$-dimensional systems was considered by \cite{Fiurasek2005} who
investigated the universal asymmetric quantum triplicator, which produces,
from a single replica of a qudit, three clones with three different fidelities 
$f^A$, $f^B$ and $f^C$. A simple parametric description of the class 
of the optimal universal highly-asymmetric triplicators was provided, 
extending Eqs.~(\ref{univasymm2}) and (\ref{univasymm}).
It was proved that the optimal fidelities can be expressed as
\begin{eqnarray}
f_{1\rightarrow 1+1+1}^{A}=
1-\frac{d-1}{d}\left[\beta^2+\gamma^2+\frac{2\beta\gamma}{d+1}\right], \nonumber \\
f_{1\rightarrow 1+1+1}^{B}=
1-\frac{d-1}{d}\left[\alpha^2+\gamma^2+\frac{2\alpha\gamma}{d+1}\right], \nonumber \\
f_{1\rightarrow1+1+1}^{C}=
1-\frac{d-1}{d}\left[\alpha^2+\beta^2+\frac{2\alpha\beta}{d+1}\right],
\end{eqnarray}
where the positive real parameters $\alpha$, $\beta$, $\gamma$ satisfy the
normalization condition
\begin{equation}
\alpha^2+\beta^2+\gamma^2+\frac{2}{d}(\alpha\beta+\alpha\gamma+\beta\gamma)=1.
\label{alfabetagammanorm}
\end{equation}

\subsection{Continuous-variable cloning machines}

Another interesting extension of quantum cloning,
which is often termed \break {\it continuous-variable quantum cloning}, concerns
the case of quantum systems lying in an infinite-dimensional Hilbert space.
In \cite{Cerf2000c}, the cloning of the set of coherent states
$\ket{\alpha}=\e^{\alpha a^{\dagger}} \ket{0}$ was investigated, 
with $\ket{0}$ denoting the vacuum state, $a^{\dagger}$ being 
the bosonic creation operator, and $\alpha=(x+ip)/\sqrt{2}$ being a c-number
which defines the position $(x,p)$ of $\ket{\alpha}$ in phase space. 
Here $x$ and $p$ are the so-called quadrature components.
A set of $1$-to-$2$ (symmetric or asymmetric)
cloning machines that are covariant with respect to the
Weyl group of displacements in phase space was derived.
The symmetric $1$-to-$2$ {\it Gaussian} cloner was found to have the fidelity
\begin{equation}
f_{1\to 2}^{\mathrm CV} = \frac{2}{3} \simeq 0.667
\label{cvclonerfid}
\end{equation}
and was conjectured to be optimal. It causes an independent 
Gaussian noise on $x$ and $p$, with a variance equal to one
shot-noise unit. Thus, the two clones are left 
in a thermal state (containing on average
$1/2$ thermal photon) which is displaced by $\alpha$. 
Let us also mention the independent derivation
of this $1$-to-$2$ Gaussian cloner 
as well as its extension to multiple clones ($M>2$) by \cite{Lindblad2000}.

\cite{Cerf2000d} later on derived an upper bound on the fidelity
of the symmetric $N$-to-$M$ Gaussian cloners, based on a link with
state estimation theory. Since it coincided with Eq. (\ref{cvclonerfid})
for $N=1$ and $M=2$, this proved that the above cloner 
is indeed the optimal cloner by means of a Gaussian operation. 
\cite{Cerf2001c} then showed that this $1$-to-$2$ Gaussian cloner
can be realized simply by use of an optical parametric amplifier of gain 2,
followed by a balanced beam splitter. The cloning noise then
originates from the vacuum fluctuations of the ancillary modes 
that are coupled to the input mode.

In \cite{Cochrane2004}, it was shown that if the ensemble of input
coherent states has a finite width, the $1$-to-$2$ Gaussian cloning
can be achieved with a higher fidelity. Clearly, if the task is to clone
a coherent state drawn from a distribution that is peaked around
the origin of phase space, the vacuum state is a very good approximation
of the original state, so cloning with a fidelity close to one is
possible. The above fidelity $f_{1\to 2}^{\mathrm CV}$ corresponds
to the opposite situation of an infinitely wide input distribution, 
that is, an arbitrary input coherent state.
If the input coherent state is distributed
according to a Gaussian distribution of zero mean
and given variance, \cite{Cochrane2004} gave a closed formula
for this fidelity as a function of the variance.

The optimal $N$-to-$M$ Gaussian cloning transformation that achieves 
the above-mentioned upper bound 
was obtained in \cite{Braunstein2001b,Fiurasek2001a}, yielding
\begin{equation}
f_{N\to M}^{\mathrm CV} = \frac{MN}{MN+M-N} .
\label{cvclonerfidNM}
\end{equation}
As for discrete-dimensional states, these cloners tend, 
at the limit $M\to\infty$, to the optimal measurement 
of $\ket{\alpha}^{\otimes N}$, with the fidelity
\begin{equation}
f_{N\to \infty}^{\mathrm CV} = \frac{N}{N+1}
\end{equation}
The optical realization of these symmetric $N$-to-$M$ cloners 
was also described there, while it was generalized 
to asymmetric $1$-to-$2$ cloners in \cite{Fiurasek2001a}.
In the latter case, the balance between the fidelities of the two clones
follows
\begin{equation}
f_A^{\mathrm CV} = \frac{1}{1+\sigma^2_A}, \qquad
f_B^{\mathrm CV} = \frac{1}{1+\sigma^2_B}, \qquad
\sigma_A \, \sigma_B = 1/2
\end{equation}
which corresponds to the no-cloning uncertainty relation derived in \cite{Cerf2000c}. Here, $\sigma^2_A$ and $\sigma^2_B$
are the variances of the added noise on clone A and $B$,
while one shot-noise unit is taken as 1/2.

Finally, in \cite{Cerf2001a}, a more general class of Gaussian
cloners was characterized, which transforms $N$ replicas of an arbitrary
coherent state $\ket{\alpha}$ and $N'$ replicas of its phase-conjugate
$\ket{\alpha^*}$ into  $M$ clones of $\ket{\alpha}$ and $M'$ clones
of $\ket{\alpha^*}$, with $N-N'=M-M'$. For well chosen ratios $N'/N$,
this cloner was shown to perform better than the $(N+N')$-to-$M$ cloner.
In addition, the special case of the {\it balanced} Gaussian cloner,
with $N=N'$ and $M=M'$ was shown to be optimal among all cloners in this class in the sense that it yields the highest fidelity for fixed $N+N'$ and
$M+M'$, namely
\begin{equation}
f_{N,N\to M,M}^{\mathrm CV} = \frac{4M^2N}{4M^2N+(M-N)^2}.
\end{equation}
Interestingly, in the limit $M\to\infty$, we have
\begin{equation}
f_{N,N\to \infty,\infty}^{\mathrm CV} = \frac{4N}{4N+1}
\end{equation}
which means that the optimal measurement of 
$\ket{\alpha}^{\otimes N} \ket{\alpha^*}^{\otimes N}$ gives
the same fidelity as the optimal measurement of $\ket{\alpha}^{\otimes 4N}$,
instead of $\ket{\alpha}^{\otimes 2N}$ as a simple counting of states seems
to imply. This advantage of phase conjugation was first noted 
in \cite{Cerf2001b}.

Coming back to the question of the symmetric $N$-to-$M$ cloning
of coherent states, \cite{Cerf2005} have recently investigated
the question of whether the above Gaussian cloners
really provide the absolute highest fidelity or, instead, 
transformations outside the realm of Gaussian operations need to be 
considered. Against all intuitions, it was shown that, provided $M$ is
finite, the cloning transformation that optimizes the single-clone
fidelity is slightly non-Gaussian. For example, the optimal
symmetric $1$-to-$2$ non-Gaussian cloner of coherent states 
was shown to have the fidelity
\begin{equation}
f_{1\to 2}^{\mathrm CV,~NG} = 0.683
\end{equation}
strictly larger than $f_{1\to 2}^{\mathrm CV} = 2/3 \simeq 0.667$.
In contrast, the optimal cloners of coherent states with respect to 
the global fidelity remain Gaussian. This discrepancy between optimal
cloners with respect to single-clone or global fidelities
is reminiscent to the situation for phase-covariant cloners
in finite-dimensional spaces.

For a review on continuous-variable quantum cloning,
see \cite{Cerf2003} and \cite{Braunstein2005}.

\subsection{Probabilistic cloning machines}

All the above listed cloning machines are deterministic, i.e., they always
produce (imperfect) clones. However, one can also consider probabilistic cloning
machines, which sometimes fail to generate the clones but, if they succeed, 
then the clones exhibit higher fidelities than those achieved 
by the best deterministic cloners. 
The concept of probabilistic cloning  was introduced by
\cite{Duan98a,Duan98b,Chefles98,Chefles99} who investigated cloning of a discrete finite set of pure states. 
They showed that a set of linearly independent states can be
perfectly copied with some probability $p$. In particular, an exact
1-to-2 cloning of two generally non-orthogonal 
pure states $|\psi_1\rangle$ and $|\psi_2\rangle$ is possible
with probability 
\begin{equation}
p_{1\rightarrow 2} = \frac{1}{1+|\langle \psi_1|\psi_2\rangle|}.
\end{equation}

The probabilistic cloning was then extended to infinite continuous sets of
input states in \cite{Fiurasek2004a}. It was shown that the optimal universal
cloning cannot be improved by using a probabilistic cloning strategy, due to a
very high underlying symmetry of the problem. Nevertheless, if one considers 
cloning of some restricted set of states, then probabilistic cloning may become
useful. A particular example is the optimal $N$-to-$M$ phase-covariant 
cloning of qubits, where the optimal probabilistic cloner achieves the
single-clone fidelity
\[
f_{N\rightarrow M}^{\mathrm{pc,~prob}} (2) 
= \frac{1}{2^M} \sum_{k=0}^N { M\choose
k+\left[\frac{1}{2}(M-N)\right]},
\]
where $[x]$ denotes the integer part of $x$. For $N>1$ the fidelity 
$f_{N\rightarrow M}^{\mathrm{pc,~prob}}(2)$ is larger than the fidelity 
$f_{N\rightarrow M}^{\mathrm{pc}}(2)$ of the optimal deterministic phase-covariant cloning.

\subsection{Economical cloning machines}

The 1-to-2 cloning transformation for $d$-dimensional systems
(qudits) can typically be expressed as a unitary operation on the Hilbert 
space of three qudits --- the input, a blank copy, and an ancilla.
The presence of an ancilla significantly affects the experimental implementation
of the cloning operation, which becomes more complicated and sensitive to
decoherence. These problems, which might drastically reduce the achieved 
cloning fidelity, may significantly be suppressed if an ``economical'' approach 
is followed, which avoids the ancilla. The 1-to-2 cloning is then realized 
as a unitary operation on two qudits only: the input and the blank copy. 
This is obviously much simpler to implement because it requires less qudits 
and two-qudit gates, and it requires to control the entanglement
of a pair of qudits only. It is thus likely to be much less sensitive to
noise and decoherence than its three-qudit counterpart.

To date, the only 1-to-2 cloning machine for which an economical 
realization is known is the phase-covariant qubit cloner due to \cite{Niu99}, 
which optimally clones all states on the equator of the Bloch sphere,
$|\psi\rangle_A=2^{-1/2}(|0\rangle_A+e^{i\phi}|1\rangle_A)$.
The qubit to be cloned is coupled to another qubit which becomes the second 
copy and is initially prepared in state $|0\rangle_B$. The unitary two-qubit
transformation reads
\[
|0\rangle_A|0\rangle_B \rightarrow |0\rangle_A |0\rangle_B, \qquad
|1\rangle_A|0\rangle_B \rightarrow \frac{1}{\sqrt{2}}
(|0\rangle_A |1\rangle_B+|1\rangle_A |0\rangle_B),
\]
for a symmetric phase-covariant cloner. It can easily be extended to an asymmetric setting as we will show in Eq.~(\ref{niu-griff-asymm}).

The possibility of economical realization of various 1-to-2
cloning machines for qudits has been analyzed in detail by \cite{Durt2005}.
It was shown there that economical universal cloning is not possible 
for any $d$.
It was also argued that the optimal 1-to-2 phase-covariant cloning 
of qudits does not admit economical implementation for any $d >2$, 
and this assertion was
rigorously proved for $d \leq 7$. A suboptimal economical phase-covariant 
cloner was nevertheless constructed, which does not require an ancilla and achieves the fidelity
\begin{equation}
f_{1\rightarrow 2}^{\mathrm pc,econ}(d)=\frac{1}{2 d^2}[d-1+(d-1+\sqrt{2})^2],
\end{equation}
which is only slightly below that of the optimal cloner. Similarly,
it was argued that the 1-to-2 Fourier-covariant cloning cannot be realized economically, albeit in dimension $d=2$ (in which case it is unitarily
equivalent to the phase-covariant cloner).

The concept of economical cloning can be extended to $N$-to-$M$ machines. 
As detailed in Section \ref{secPCqubits}, the paper by \cite{Fan2001b} 
implies that the optimal $N$-to-$M$ phase-covariant cloning of qubits ($d=2$), 
which maximizes the single-clone fidelity, admits an economical implementation for any $N$ and $M>N$. Moreover, 
the economical phase-covariant cloning of $d$-dimensional
systems (qudits) is also possible provided that $M=kd+N$, where $k$ is an integer, see \cite{Buscemi2005}.

\section{One-to-two quantum cloning as a CP map}
\label{section-cp-map}

\subsection{Isomorphism between CP maps and operators}
\label{section-isomorphism}
A very useful characterization of cloning relies on the isomorphism 
between completely positive maps ${\mathcal S}: {\mathcal H}_{\mathrm in}
\to {\mathcal H}_{\mathrm out}$ and positive semidefinite operators 
$S \geq 0$ acting on 
${\mathcal H}_{\mathrm in}\otimes {\mathcal H}_{\mathrm out}$,
where ${\mathcal H}_{\mathrm in}$ and ${\mathcal H}_{\mathrm out}$
denote, respectively, the input and output Hilbert spaces of ${\mathcal S}$, 
see \cite{Jamiolkowski72,Choi75}. To construct this isomorphism,
consider a maximally entangled
state on ${\mathcal H}_{\mathrm in}^{\otimes 2}$,
\begin{equation}
|\Phi^+\rangle=\frac{1}{\sqrt{d}}\sum_{j=0}^{d-1} |j\rangle|j\rangle,
\label{MaxEnt}
\end{equation}
where $d={\mathrm dim}({\mathcal H}_{\mathrm in})$.
If the map ${\mathcal S}$ is applied to the second subsystem of
$|\Phi^+\rangle$ while the first one is left unchanged, 
then the resulting (generally mixed) quantum state 
is isomorphic to ${\mathcal S}$ and reads
\begin{equation}\label{S3}
S= ({\mathcal I} \otimes {\mathcal S}) \; 
d \, \Phi^+,
\end{equation}
where $\Phi^+ \equiv |\Phi^+\rangle\langle \Phi^+|$
and ${\mathcal I}$ stands for the identity map,
while the prefactor $d$ is introduced for normalization purposes.
The map ${\mathcal S}$ can be characterized in terms of the state $S$
as follows,
\begin{equation}
\rho_{\mathrm out}= 
{\mathcal S}(\rho_{\mathrm in})=
{\mathrm Tr}_{\mathrm in}[(\rho_{\mathrm in}^T \otimes I_{\mathrm out}) S],
\end{equation}
where ``$\mathrm{in}$'' labels the input space, $I$ is the identity operator,
while $T$ denotes the transposition in the computational basis. 
A trace preserving map ${\mathcal S}$ implies that $S$ satisfies
the condition
\begin{equation}
{\mathrm Tr}_{\mathrm out} [S]=I_{\mathrm in},
\label{tracepreservation}
\end{equation}
while the complete positivity condition on ${\mathcal S}$
translates into $S\ge 0$.

In the following, we shall make this description specific to the
one-to-two quantum cloning machines, which produce two copies of a single
$d$-dimensional system (often called a qudit), see \cite{Fiurasek2001b}.
The output Hilbert space is endowed with a tensor product structure,
${\mathcal H}_{\mathrm out }={\mathcal H}_{A}\otimes{\mathcal H}_B$,
where the subscripts $A$ and $B$ label the two clones. For each particular
input state $|\psi\rangle$, the joint state of the clones is
\begin{eqnarray}
{\mathcal S}(\psi)=
{\mathrm Tr}_{\mathrm in} [(\psi_{\mathrm in}^T \otimes I_{AB}) S],
\label{cloningmap}
\end{eqnarray}
where $\psi\equiv |\psi\rangle\langle \psi|$. 
It will be useful in the following to note
that $\psi_{\mathrm in}^T$ is a rank-one projector
onto the state $|\psi^*_{\mathrm in}\rangle$, where $^*$ denotes the complex 
conjugation in the computational basis.

Using Eq.~(\ref{cloningmap}), the fidelity of the clones 
$A$ and $B$ is given by
\begin{eqnarray}
F_A({\mathcal S},\psi)&=& {\mathrm Tr}[(\psi_A\otimes I_B)
\, {\mathcal S}(\psi)]
~=~ {\mathrm Tr}[(\psi_{\mathrm in}^T \otimes \psi_A \otimes I_B) S],
\nonumber \\
F_B({\mathcal S},\psi)&=& {\mathrm Tr}[(I_A\otimes \psi_B)
\, {\mathcal S}(\psi)]
~=~ {\mathrm Tr}[(\psi_{\mathrm in}^T \otimes I_A \otimes \psi_B) S].
\label{FBEpsi}
\end{eqnarray}
The symmetric cloning machines are defined as the maps ${\mathcal S}$ 
verifying $F_A({\mathcal S},\psi)=F_B({\mathcal S},\psi)$, $\forall \psi$. Otherwise, the cloning machines are called asymmetric. 
When considering a universal cloning machine, 
we require that both $F_A({\mathcal S},\psi)$ and $F_B({\mathcal S},\psi)$ 
are independent of $\psi$, for all state $\psi$ in ${\mathcal H}_{\mathrm in}$. 
Other cloning machines, such as the phase-covariant, Fourier-covariant, or real
cloning machines will correspond to a constant fidelity over
a restricted set $R$ of input states $\psi$. 

\subsection{Covariance condition}
\label{sectiontwirling}

In what follows, we will always assume that the set $R$ of input states
$\psi$ is invariant under the action
of the group $G(\Omega)$ of unitaries $\{U_\omega | \omega \in \Omega\}$,
that is
\begin{equation}
U_\omega \, R \, U^{\dagger}_\omega = R, \qquad \forall \omega\in\Omega .
\end{equation}
The universal cloning machine is the special case
$U_\omega \in {\mathrm SU}(d)$.
A useful figure of merit to measure the quality of cloning
is the {\it global} fidelity,
\begin{eqnarray}
F({\mathcal S},\psi)&=& {\mathrm Tr}[(\psi_A\otimes \psi_B)\,
 {\mathcal S}(\psi)]
~=~ {\mathrm Tr}[(\psi_{\mathrm in}^T \otimes \psi_A \otimes \psi_B) S],
\end{eqnarray}
which measures how well the joint state of the two clones approximates
$\psi^{\otimes 2}$. When looking for a cloning machine
that optimally clones all the states of set $R$, one generally
defines the cloning fidelity of map ${\mathcal S}$
as the infimum of the global fidelity over all input states $\psi$,
\begin{equation}
F({\mathcal S})=\inf_{\psi\in {\mathrm R}} F({\mathcal S},\psi)
\end{equation}
It has been shown by \cite{Werner98} that, by using the so-called
{\it twirling} operation, there is no loss of generality 
in assuming the optimal cloning machine to be {\it covariant} 
with respect to the group $G(\Omega)$, hence the cloning fidelity
to be state-independent within the set $R$.  
The twirling operation consists in randomly applying a unitary $U_\omega$ 
to the input state and then undoing this by applying the reverse unitary $U^\dagger_\omega$ to each of the two clones with the probability density  $d\omega$ equal to the Haar measure on the group $G(\Omega)$.
This results in the twirled map
\begin{equation}
{\mathcal S}_{\mathrm twirl}[\psi]=\int_\Omega 
{\mathcal S}_\omega[\psi] \, d\omega
\end{equation}
with the rotated map ${\mathcal S}_\omega$ being defined as
\begin{equation}
{\mathcal S}_\omega[\psi]=
U_\omega^{\dagger \otimes 2} {\mathcal S}
\left[ U_\omega\psi U^\dagger_\omega \right]
U_\omega^{\otimes 2} .
\end{equation}
The core of the argument is that
\begin{eqnarray}
\lefteqn{  F({\mathcal S})=\inf_{\psi\in {\mathrm R}} F({\mathcal S},\psi)
\le \inf_{\phi\in {\mathrm R}}
\int_\Omega F\left({\mathcal S},U_\omega \phi U^{\dagger}_\omega\right)
\, d\omega    }\nonumber \\
&& = \inf_{\phi\in {\mathrm R}} \int_\Omega F\left({\mathcal S}_\omega,\phi\right)
\, d\omega 
= \inf_{\phi\in {\mathrm R}} F({\mathcal S}_{\mathrm twirl},\phi) 
= F({\mathcal S}_{\mathrm twirl}) ,
\end{eqnarray}
where we have used the invariance of $R$ under the unitaries $U_\omega$,
the invariance of the trace function under $U_\omega$, and
the linearity of the fidelity in ${\mathcal S}$. As a result, the operation
of twirling can only increase the cloning fidelity, so that the twirled map
${\mathcal S}_{\mathrm twirl}$ is at least as good as each of its
constituent maps ${\mathcal S}_\omega$.
Finally, as mentioned earlier,
we note that ${\mathcal S}_{\mathrm twirl}$ is {\it covariant} 
with respect to the group $G(\Omega)$, that is
\begin{equation}
{\mathcal{S}}_{\mathrm twirl}[U_\omega \psi U^{\dagger}_\omega]
= U_\omega^{\otimes 2} {\mathcal S}_{\mathrm twirl}(\psi)
U^{\dagger \otimes 2}_\omega \qquad \forall \omega\in\Omega, \forall \psi\in R .
\end{equation}
Physically, this covariance property means that rotating the original state
is exactly equivalent to rotating the two clones by the same amount. This 
also implies
that $F({\mathcal S}_{\mathrm twirl},\psi)$ does not depend on $\psi$.

In summary, we have shown that 
when looking for an optimal quantum cloning machine
(i.e., a machine that maximizes the worst-case global fidelity), 
it is sufficient to consider cloning maps 
that are covariant with respect to the group
under which the set of input states $R$ is invariant. The cloning
fidelity is therefore state-independent within the set $R$. 
In \cite{Keyl99}, it was proven that this reasoning also applies
more generally to the quantum cloning machines that maximize 
the {\it single-clone} fidelities ($F_A$ and $F_B$) 
instead of the global fidelity, provided that
{\it universal} cloning machines are considered. This, however, does not
hold for all quantum cloners (see, e.g., the case of 
phase-covariant or continuous-variable cloners treated later on).

\subsection{Cloning as a semidefinite programming problem}
\label{sectionsemidefinite}

Coming back to the characterization of the map ${\mathcal S}$
via its associated operator $S$, we can now use the fact that the
optimal cloning machine must have a state-independent fidelity
over the considered set of input states.
One can then turn to the average performance of the cloning machines, 
which is measured by the mean fidelities,
\begin{equation}
F_A({\mathcal S})=\int_{\psi} F_A({\mathcal S},\psi) \, d\psi, \qquad
F_B({\mathcal S})=\int_{\psi} F_B({\mathcal S},\psi) \, d\psi,
\label{FAB}
\end{equation}
where the measure $d\psi$ determines the kind of cloning machines we are
dealing with. In particular, universal cloning machines 
correspond to choosing $d\psi$ to be the invariant measure 
on the factor space $SU(d)/SU(d-1)$ induced by the Haar
measure on the group $SU(d)$. The fidelities (\ref{FAB})
can be expressed as linear functions of the operator $S$,
\begin{equation}
F_{A}=\mathrm{Tr}[S\, R_{A}], \qquad F_{B}=\mathrm{Tr}[S\, R_B],
\end{equation}
where we have defined the positive semidefinite operators
\begin{equation}
R_{A}=\int_{\psi} \psi_{\mathrm{in}}^T \otimes \psi_A \otimes I_B \, d \psi,
\qquad
R_{B}=\int_{\psi} \psi_{\mathrm{in}}^T \otimes I_A \otimes \psi_B \, d \psi.
\label{RBEdefinition}
\end{equation}
We will see in the next Sections how these operators can be calculated
for different kinds of cloning machines. Note that 
for a {\it symmetric} cloning machine, 
one should simply maximize the average of the mean fidelities,
\begin{equation}
F({\mathcal S})={1\over 2}[F_A({\mathcal S})+F_B({\mathcal S})]={\mathrm Tr}[S\, R],
\end{equation}
with $R=(R_A+R_B)/2$. This can be justified by an argument similar to
the one used for the twirling operation. By averaging over the permutation 
between the two clones, one obtains a map whose global fidelity
can only be better than that of the original map. Therefore, we can
restrict ourselves to cloning transformations that are covariant
with respect to the interchange of the clones, hence satisfying
$F_A({\mathcal S})=F_B({\mathcal S})$.

From this argument based on twirling and permutation, we conclude
that maximizing the mean fidelity, averaged over the two clones
(with equal weights), should yield a cloning map which has 
a state-independent and clone-independent fidelity. The asymmetric
cloners can also be obtained with the same maximization but
by putting different weights in front of $F_A$ and $F_B$.

An interesting point to note is that finding the optimal 
cloning map ${\mathcal S}$ reduces to a {\it semidefinite programming} 
problem, namely finding the operator $S$
verifying $S\ge 0$ and ${\mathrm Tr}_{AB}[S]=I_{\mathrm in}$
that maximizes $\mathrm{Tr}[S\, R]$, with $R$ depending on the
considered cloning machine (\cite{Audenaert2002}). There exist very 
efficient numerical methods for solving semidefinite programs, 
see e.g. \cite{Vandenberghe1996}. Even more importantly, it can be shown 
with the help of Lagrange duality lemma that the optimal cloning trace-preserving CP map, which maximizes ${\mathrm Tr}[SR]$, must satisfy
\begin{eqnarray}
(R-\lambda_{\mathrm{in}}\otimes I_{AB})S=0, 
\label{Soptimality} \\
\lambda_{\mathrm{in}}\otimes I_{AB}-R \geq 0,
\label{lambdaoptimality}
\end{eqnarray}
where $\lambda \geq 0$ is a positive semidefinite operator whose matrix elements
represent the Lagrange multipliers accounting for the trace-preservation constraint
${\mathrm Tr}_{AB}[S]=I_{\mathrm in}$. Note that $\lambda$ can be expressed in terms
of the optimal CP map, $\lambda={\mathrm Tr}_{AB}[S\, R]$. If both equations
(\ref{Soptimality}) and (\ref{lambdaoptimality})  are
satisfied, then $S$ is the optimal CP map 
maximizing ${\mathrm Tr}[S\, R]$, a property which is
useful to prove and check the optimality of a given map $S$ that is
conjectured to be optimal. 

The proof that Eqs.~(\ref{Soptimality})  and
(\ref{lambdaoptimality}) 
imply optimality is rather simple and we briefly sketch it here. 
Suppose that (\ref{lambdaoptimality}) is satisfied, then it holds 
for any trace-preserving CP map that 
${\mathrm Tr}[S(\lambda\otimes I -R)] \geq 0$
and ${\mathrm Tr}[\lambda\otimes I S]={\mathrm Tr}\lambda$, due to the
trace-preservation condition. It follows that the fidelity is upper bounded by the
trace of the Lagrange multiplier, $\mathrm{Tr}[RS] \leq \mathrm{Tr}[\lambda]$, and that the optimal map 
which satisfies (\ref{Soptimality}) saturates this bound.

Very often a simpler method is sufficient to prove the optimality, namely, the
fidelity can be bounded by the maximum eigenvalue $r_{\mathrm max}$ of $R$. 
Since $R \leq r_{\mathrm max}I$, we immediately have
\begin{equation}
F(S) \leq  d \, r_{\mathrm max},
\label{fidelityupperboundmaxeig}
\end{equation}
where $d={\mathrm dim}({\mathcal H}_{\mathrm in})$.
If there exists a CP map $S$ which saturates (\ref{fidelityupperboundmaxeig}),
then this transformation is optimal. Note, however, that in
certain cases such as the cloning of a pair of orthogonal qubits, 
the bound on fidelity (\ref{fidelityupperboundmaxeig}) is not tight and cannot be saturated.

\subsection{Double-Bell ansatz}

Let us now consider the {\it unitary realization} of the cloning map
${\mathcal S}$. We know that any CP map can be realized physically
by supplementing the input system with an ancilla (hence, extending
the Hilbert space) and acting with a unitary operator 
in this extended space. Here, the
ancilla can be viewed as the cloning machine itself, and it must
be traced over after applying the unitary operator. The resulting
map can be written as
\begin{equation}
{\mathcal S}(\psi_{\mathrm in})=
{\mathrm Tr}_{{\mathrm in},C}[(\psi_{\mathrm in}^T \otimes I_{ABC}) \Sigma],
\end{equation}
where $C$ denotes the cloning machine
and $\Sigma$ is the operator that is isomorphic to this extended map
${\mathcal H}_{\mathrm in } \to 
{\mathcal H}_{A}\otimes{\mathcal H}_B\otimes{\mathcal H}_C$. Since
this extended map is some unitary operation $U_{ABC}$ in the extended
space ${\mathcal H}_{A}\otimes{\mathcal H}_B\otimes{\mathcal H}_C$, 
the operator $\Sigma$ must be some (unnormalized) rank-one projector or  
pure state in the joint space ${\mathcal H}_{\mathrm in }\otimes 
{\mathcal H}_{A}\otimes{\mathcal H}_B\otimes{\mathcal H}_C$. We thus have
\begin{equation}
\Sigma= |\sigma\rangle\langle\sigma|, \qquad {\mathrm with~~}
|\sigma\rangle_{{\mathrm in},ABC} 
= \sqrt{d} \, (I_{\mathrm in} \otimes U_{ABC}) \, 
|\Phi^+\rangle_{{\mathrm in},A} |0\rangle_{B,C}
\end{equation}
where $|0\rangle_{B,C}$ is the (arbitrary) initial state of the blank copy $B$
and cloning machine $C$, and $|\Phi^+\rangle$
is defined as in Eq.~(\ref{MaxEnt}). The prefactor $\sqrt{d}$,
with $d={\mathrm dim}({\mathcal H}_{\mathrm in})$, is introduced
for normalization purposes, and the cloning map corresponds to
Eq.~(\ref{cloningmap}) with $S={\mathrm Tr}_C \Sigma$.

\input #progressfig2

Physically, the state $|\sigma\rangle$ has a very simple interpretation, see Fig.
\ref{fignicolasfigprogress2}.
If we start with two qudits prepared in a maximally entangled state
$|\Phi^+\rangle$ and process one of them in the quantum cloning machine
while the other one is left unchanged (kept as a reference), then
$|\sigma\rangle$ is the joint state of this ``reference'' qudit 
(denoted as ``in'' since it keeps a memory of the input state), 
the clones $A$ and $B$, as well as the cloning machine $C$.
Remember that if we project the reference qudit 
onto the state $|\psi^*\rangle$, then, 
in the absence of cloning, qudit $A$ is found in state $|\psi\rangle$. 
By causality, it is irrelevant whether this projection onto $|\psi^*\rangle$
is done before or after the cloning machine has been applied 
on qudit $A$. Therefore, projecting the reference qudit
of state $|\sigma\rangle$ onto $|\psi^*\rangle$ yields 
the joint state of $A$, $B$, and $C$ that would have been obtained 
by cloning the state $|\psi\rangle$, namely
\begin{equation}
|\psi\rangle \to 
\ket{\psi_{\mathrm out}}_{ABC} =
 ~_{\mathrm in}\langle\psi^*|\sigma\rangle_{{\mathrm in},ABC} .
\end{equation}
We can say that $|\sigma\rangle$ 
fully encodes the information about the cloning of any state.

It was suggested in \cite{Cerf98,Cerf2000a,Cerf2000b}
that a generic form for state $|\sigma\rangle$
involving a superposition of double-Bell states
may encompass most of the interesting quantum cloning machines,
including the universal or state-dependent -- symmetric as
well as asymmetric -- cloners. This so-called {\it double-Bell ansatz}
corresponds to taking $|\sigma\rangle=\sqrt{d} \, |{\cal A}\rangle$, with
\begin{eqnarray}
|{\cal A}\rangle_{{\mathrm in},A;B,C}= \sum_{m,n=0}^{d-1} a_{m,n} \,
\B{m}{n}_{{\mathrm in},A} \,  \BB{m}{n}_{B,C} ,
\label{eq_ansatz}
\end{eqnarray}
where it is assumed that it is sufficient to use a Hilbert space
for the cloning machine $C$ which has the same dimension $d$ as that 
of the input or the clones. Note that the Schmidt decomposition
of $|{\cal A}\rangle$ for the partition ``in'' vs. $ABC$
implies that ${\mathrm dim}({\mathcal H}_C)\ge d$.
In Eq.~(\ref{eq_ansatz}), the $a_{m,n}$'s are complex amplitudes
which satisfy the normalization condition $\sum_{m,n=0}^{d-1}|a_{m,n}|^2=1$,
while $\B{m}{n}$ denote Bell states in $d$ dimensions. As the latter states
play an important role in what follows, we will first discuss them in details,
as well as some useful related properties. Note also that $|{\cal A}\rangle$ 
is a quantum state of norm one, while $|\sigma\rangle$ 
has a norm equal to $d$.

\subsubsection{Useful properties of $d$-dimensional Bell states}

A standard generalization of the Bell states in $d$ dimensions is
\beq
\B{m}{n}=\frac{1}{\sqrt{d}} \sum_{j=0}^{d-1} \gamma^{nj}\ket{j}_1 \ket{j+m}_2 ,
\eeq
where 1 and 2 denote two $d$-dimensional systems. These states
form a set of $d^2$ maximally-entangled states of systems
1 and 2, where $m,n\in \{0,1,\cdots d-1\}$ and $\gamma=\e^{2\pi i /d}$ stands
for the $d$-th root of unity. Note that, in what follows,
the "bra" and "ket" labels are always taken modulo $d$. 
In the case of qubits ($d=2$), we recover the standard Bell states
\beq
&&\B{0}{0}=\ket{\Phi^+} = (\ket{00}+\ket{11})/\sqrt{2}, \quad
\B{0}{1}= \ket{\Phi^-} = (\ket{00}-\ket{11})/\sqrt{2}, \nonumber\\
&&\B{1}{0}=\ket{\Psi^+} = (\ket{01}+\ket{10})/\sqrt{2}, \quad
\B{1}{1}=\ket{\Psi^-} = (\ket{01}-\ket{10})/\sqrt{2}. \nonumber\\
\eeq
Taking the partial trace of any state  $\Bpbis{m}{n}\equiv\Bp{m}{n}$
over one of the two systems (1 or 2) results in the maximally mixed state,
\beq
\Tr_1(\Bpbis{m}{n})=\Tr_2(\Bpbis{m}{n})= I/d ,  \qquad \forall m,n
\eeq
so that the
states $\B{m}{n}$ are indeed maximally entangled. It is easy to check that the
states $\B{m}{n}$ form a complete orthonormal basis in the considered $d^2$-dimensional Hilbert space. The resolution of identity reads
\beq
\lefteqn{ \sum_{m,n=0}^{d-1}  \Bpbis{m}{n} 
= \frac{1}{d} \sum_{m,n} \sum_{j,j'} \gamma^{n(j-j')}  \,
\ketbra{j}{j'} \otimes \ketbra{j+m}{j'+m} } \nonumber\\
&& = \sum_{m} \sum_{j,j'} \delta_{j,j'}  \,
\ketbra{j}{j'} \otimes \ketbra{j+m}{j'+m} 
= I_{12},
\eeq
where we have used the identity
$\frac{1}{d} \sum_{n=0}^{d-1} \gamma^{nj} = \delta_{j,0}$.

Let us focus on the Bell state with $m=n=0$, that is
\beq
\B{0}{0}=\frac{1}{\sqrt{d}} \sum_{j=0}^{d-1} \ket{j}\ket{j}
\eeq
which is another notation for the state $|\Phi^+\rangle$
as defined in Eq.~(\ref{MaxEnt}). This state is particularly useful
because it satisfies the relation
\beq
(U^* \otimes U) \B{0}{0} = \B{0}{0}
\label{eq_U*U}
\eeq
for any unitary transformation $U$, as can readily be checked by
using the unitarity condition $U U^\dagger=I$ and the completeness relation 
$\sum_j\ketbra{j}{j}=I$. 
Note that the symbol * denotes 
the complex conjugation operation in the computational
basis $\{\ket{j}\}$; thus $\ket{j^*}=\ket{j}$.
The identity (\ref{eq_U*U}), or equivalently
\beq
(I \otimes U) \B{0}{0} = (U^T \otimes I) \B{0}{0}
\eeq
corresponds to the following useful property: 
if the joint system 12 is prepared in the
state $\B{0}{0}$ and system 1 is projected onto $\ket{\psi^*}$, then
the resulting state of system 2 is $\ket{\psi}$. 
Indeed, taking $\ket{\psi}=U\ket{0}$, 
we have $\bra{\psi^*}=\bra{0}U^T$, so that
\beq
\lefteqn{ (\proj{\psi^*}\otimes I) \B{0}{0}
= (\ket{\psi^*}\bra{0}\otimes I)(U^T\otimes I)\B{0}{0} } \nonumber\\
&=& (\ket{\psi^*}\bra{0}\otimes I)(I \otimes U)\B{0}{0} 
=d^{-1/2} \ket{\psi^*} (U\ket{0})
=d^{-1/2} \ket{\psi^*} \ket{\psi}.
\eeq
Interestingly, it makes no difference whether 1 prepares and sends 
the state $\ket{\psi}$ to 2, or whether 1 projects its part of a shared
entangled state $\B{0}{0}$ onto $\ket{\psi^*}$ so to create $\ket{\psi}$
at a distance on 2.

In what follows, we will also need the discrete group of Weyl-Heisenberg
operators (also called {\it error operators}), namely
\beq
E_{m,n}=\sum_{j=0}^{d-1} \gamma^{jn}\ketbra{j+m}{j}
\eeq
with $m,n\in \{0,1,\cdots d-1\}$, generalizing the Pauli matrices 
for more than two dimensions. For qubits ($d=2$), we have
\beq
&&E_{0,0}= I , \quad ~
E_{0,1}= \sigma_z, \nonumber\\
&&E_{1,0}= \sigma_x, \quad
E_{1,1}= \sigma_x \sigma_z = -i \sigma_y .\nonumber\\
\eeq
In arbitrary dimension, the error operator $E_{m,n}$ shifts the
state by $m$ units (modulo $d$) in the computational basis and multiplies it by
a phase so as to shift its Fourier transform by $n$ units (modulo $d$). Indeed,
in the computational basis $\{\ket{j}\}$, we have $E_{m,0}\ket{j}=\ket{j+m}$,
while in the dual basis $\{ |\ket{j}\rangle
=(1/\sqrt{d})\sum_{k=0}^{d-1}\gamma^{jk}\ket{k} \}$, 
we have $E_{0,n}|\ket{j}\rangle = |\ket{j+n}\rangle$.
The error operators fulfill the following properties,
\beq
&& E_{m,n}^* = E_{m,-n}, \qquad \qquad
E_{m,n}^T = \gamma^{-mn} E_{-m,n}, \\
&& E_{m,n}^{\dagger} = \gamma^{mn} E_{-m,-n}, \qquad
E_{m,n}E_{\mu,\nu} = \gamma^{n\mu} E_{m+\mu,n+\nu}.
\eeq
Interestingly, the Bell states can be transformed into each other by applying
an error operator locally (on one of the two systems, leaving the 
other one unchanged), 
\beq
\B{m}{n} = (I \otimes E_{m,n}) \B{0}{0} = 
(E_{m,n}^T \otimes I) \B{0}{0}.
\label{eq_local_transf_bell}
\eeq
This also implies that the Bell states are invariant (up to a phase)
under correlated error operators $(E_{\mu,\nu}^* \otimes E_{\mu,\nu})$.
We can check this by calculating
\beq
\lefteqn{  (E_{\mu,\nu}^* \otimes E_{\mu,\nu}) \, \B{m}{n} 
 = (I \otimes E_{\mu,\nu} E_{m,n}) \, (E_{\mu,\nu}^* \otimes I)
\, \B{0}{0}  } \nonumber\\
&&= (I \otimes E_{\mu,\nu} E_{m,n}) \, 
(I \otimes E_{\mu,\nu}^{\dagger} ) \, 
\B{0}{0}  \nonumber \\
&&= \gamma^{m\nu -n\mu} (I \otimes E_{m,n}) \, \B{0}{0}  
= \gamma^{m\nu -n\mu} \, \B{m}{n},
\label{eq_invariant_bell}
\eeq
where we have used property (\ref{eq_local_transf_bell}) as well as
\beq
\lefteqn{  E_{\mu,\nu} E_{m,n} E_{\mu,\nu}^{\dagger} 
= \gamma^{\mu\nu} \, E_{\mu,\nu} E_{m,n} E_{-\mu,-\nu}
= \gamma^{\mu\nu-n\mu} \, E_{\mu,\nu} E_{m-\mu,n-\nu}  } \hspace{1cm}\nonumber\\
&& = \gamma^{\mu\nu-n\mu+\nu(m-\mu)} \, E_{m,n} 
= \gamma^{m\nu - n\mu} \, E_{m,n} .
\eeq

\subsection {Heisenberg cloning machines}
\label{sub-section-Heisenberg}

Coming back to the double-Bell ansatz, Eq.~(\ref{eq_ansatz}),
the quantum cloning machine is thus completely
characterized by the $d \times d$ matrix ${\vec a}=\{a_{m,n}\}$.
The form (\ref{eq_ansatz}) is particularly interesting because, 
when tracing over B and C, the systems ``in'' and A are left 
in a mixed state that is diagonal in the Bell basis,
\beq
\rho_{{\mathrm in},A}= \sum_{m,n=0}^{d-1} |a_{m,n}|^2 \; \Bpbis{m}{n}
\eeq
with $\Bpbis{m}{n}\equiv\Bp{m}{n}$.
Since the original system is maximally entangled with the reference system ``in'' (the initial state
being $\B{0}{0}$), this implies that clone A undergoes 
the error $E_{m,n}$ with probability $|a_{m,n}|^2$. It emerges from
a Heisenberg channel characterized by the probability
distribution $|a_{m,n}|^2$.

An important property of state $|{\cal A}\rangle$ is that, 
when interchanging clones A and B, it can be re-expressed 
as a superposition of double-Bell states
albeit with different amplitudes,
\beq
|{\cal A}\rangle_{{\mathrm in},B;A,C}= 
\sum_{m,n=0}^{d-1} b_{m,n} \B{m}{n}_{{\mathrm in},B} \BB{m}{n}_{A,C}
\eeq
with
\beq
b_{m,n}=\frac{1}{d}\sum_{x,y=0}^{d-1} \gamma^{nx-my} a_{x,y} .
\label{fouriertransform}
\eeq
Again, when tracing over A and C,
systems ``in'' and B are left in a Bell-diagonal mixed state,
\beq
\rho_{{\mathrm in},B}=\sum_{m,n=0}^{d-1} |b_{m,n}|^2 \; \Bpbis{m}{n}
\eeq
implying that clone B undergoes 
the error $E_{m,n}$ with probability $|b_{m,n}|^2$
(it emerges from another Heisenberg channel).
Remarkably, Equation~(\ref{fouriertransform}) implies that
the matrix ${\vec b}=\{b_{m,n}\}$ is related to
${\vec a}=\{a_{m,n}\}$ by a (bivariate and $d$-dimensional) 
discrete Fourier transform, ${\vec b}={\cal F}[{\vec a}]$.
So, the cloning map can be characterized equivalently by the matrix $\vec{a}$
(characterizing  the noise of clone A) or its Fourier transform $\vec{b}$
(characterizing the noise of clone B), and we see that
the complementarity between these two clones simply originates from
a Fourier transform: the more noisy is clone A, the less noisy is clone B.
This leads to a no-cloning uncertainty relation, 
see \cite{Cerf99,Cerf2000b}.

Finally, we can use the ansatz (\ref{eq_ansatz})
to express the map associated with an arbitrary Heisenberg cloner 
in the simple form
\beq
\ket{\psi}&\to& \ket{\psi_{\mathrm out}} = 
\sum_{m,n=0}^{d-1} a_{m,n}
\Tr_{\mathrm in} \Big[\left( \psi^T_{\mathrm in} \otimes I_{ABC}\right)
\Big( (\Bpbis{m}{n})_{{\mathrm in},A} \otimes (\BBpbis{m}{n})_{BC} \Big) \Big]
\nonumber\\
&=&\sum_{m,n=0}^{d-1} a_{m,n} \; E_{m,n} \ket{\psi}_A \otimes \BB{m}{n}_{BC}.
\eeq
Incidentally, we note here that by measuring the clone $B$ together
with the cloning machine $C$ in the Bell basis, we get a pair 
of indices $(m,n)$ which can be used to undo the noise on clone $A$
simply by applying $E_{m,n}^{\dagger}$. This process, which bears some
analogy with quantum teleportation, will be exploited in Section~\ref{secUCOasymmetry} in order to convert a symmetric cloner
into an asymmetric cloner.

\subsubsection{Covariance w.r.t. the Weyl-Heisenberg group}

It can be proven that the Heisenberg cloning machines are covariant
with respect to the discrete Weyl-Heisenberg group of error operators
$\{E_{m,n}\}$. Recall that $E_{\mu,0}$ corresponds
to a cyclic relabeling of the computational basis states, while 
$E_{0,\nu}$ corresponds to a cyclic relabeling of the dual
basis states; $E_{\mu,\nu}=E_{\mu,0}E_{0,\nu}$ 
simply corresponds to a sequence of these cyclic permutations. Thus,
the Heisenberg cloners are covariant with respect to cyclic permutations 
of the basis states in the computational and dual basis.

Using Eq. (\ref{cloningmap}), it can be easily shown that
the covariance condition of the cloning map
${\mathcal{S}}$ with respect to the unitary operator $U$, namely 
\begin{equation}
{\mathcal{S}}[U \psi U^{\dagger}]
= U^{\otimes 2} {\mathcal S}(\psi)
U^{\dagger \otimes 2} \qquad \forall \psi\in R
\end{equation}
translates into the condition
\begin{equation}
(U^*\otimes U^{\otimes 2}) \, S \,(U^T \otimes U^{\dagger \otimes 2}) = S
\Longleftrightarrow [S,U^*\otimes U^{\otimes 2}]=0
\end{equation}
on the operator $S$ that is isomorphic to ${\mathcal{S}}$. 
We may also impose that when the original is transformed
according to the unitary $U$, the cloning machine is transformed 
according to the unitary $U^*$. This condition, named
{\it strong covariance}, can be expressed as
a constraint on state $|\sigma\rangle$ or $|{\cal A}\rangle$, namely
\begin{equation}
(U^* \otimes U^{\otimes 2} \otimes U^*) 
|{\cal A}\rangle_{{\mathrm in},A,B,C} = |{\cal A}\rangle_{{\mathrm in},A,B,C}.
\label{strongcov}
\end{equation}
It was shown recently that, provided that the set of input states
is invariant with respect to the Weyl-Heisenberg group, the class
of strongly-covariant cloning maps is equivalent to the 
class of {\it extremal} covariant maps, see \cite{Chiribella2005}. 
Thus, substituting covariance with strong covariance 
greatly simplifies the search for optimal cloners since, given that the
covariant cloners form a convex set, it is sufficient to search 
among extremal cloners.

The strong covariance of the Heisenberg cloners can be checked by
using condition (\ref{strongcov}) with $U=E_{\mu,\nu}$ for all $\mu$ and $\nu$.
This equation indeed holds for each component of ${\cal A}$, namely
\beq
\lefteqn{ (E_{\mu,\nu}^*\otimes E_{\mu,\nu} \otimes E_{\mu,\nu}
 \otimes E_{\mu,\nu}^*) \B{m}{n}\BB{m}{n} } \nonumber\\
&=& \gamma^{m\nu -n\mu} \, \B{m}{n} \, \gamma^{-(m\nu - n\mu)} \, \BB{m}{n}
\nonumber\\
&=& \B{m}{n}\BB{m}{n},
\eeq
where we have used Eq. (\ref{eq_invariant_bell}). Thus, the Heisenberg
cloning machines defined by the ansatz state $|{\cal A}\rangle$ 
for an arbitrary matrix $\vec{a}$ have the nice property that they keep
the same form when making a cyclic permutation of the basis states 
(in both the computational and dual bases).

This covariance property also implies that the 
reduced cloning maps are {\it unital}. It is trivial to prove that
applying an error operator $E_{m,n}$ chosen at random (uniformly
among the $d^2$ possibilities) on an arbitrary state
$\rho$ always gives a maximally disordered state,
\beq
\frac{1}{d^2} \sum_{m,n=0}^{d-1} E_{m,n} \, \rho \, E_{m,n}^{\dagger}  =
\frac{I}{d}.
\label{eq_disordered_errors}
\eeq
Consider an arbitrary input state of the cloner $\psi$,
the covariance and the linearity of the reduced cloning map 
${\cal S}_A$ or ${\cal S}_B$ imply that
\beq
  {\cal S}_{A,B}\left(\frac{1}{d^2} \sum_{m,n=0}^{d-1} E_{m,n} \, 
\psi \, E_{m,n}^{\dagger}\right) 
= \frac{1}{d^2} \sum_{m,n=0}^{d-1} E_{m,n} \, {\cal S_{A,B}}(\psi) \, 
E_{m,n}^{\dagger}
\eeq
so that, using Eq. (\ref{eq_disordered_errors}), we verify 
that the reduced cloning maps ${\cal S_{A,B}}$ are indeed unital
\beq
{\cal S}_{A,B}[I/d ]=I/d .
\eeq

\subsection{Three special cases of Heisenberg cloners}

\subsubsection{Universal cloners}

Let us now discuss several interesting special cases 
of Heisenberg cloning machines.
The first example is the \emph{universal} (or isotropic)
cloning machine, where the
channel underlying each output is a quantum depolarizing channel. This
implies that all the probabilities $p_{m,n}=|a_{m,n}|^2$ except
$p_{0,0}$ must be equal. The same holds for the probabilities $q_{m,n}=|b_{m,n}|^2$ associated with the second clone. 
These conditions put very strong
constraints on the matrix $a$, whose elements can be thus parametrized by two
real coefficients $v$ and $x$,
\begin{eqnarray}
a_{m,n}&=&(v-x) \,\delta_{n,0} \,\delta_{m,0}+x \nonumber \\
&=&\left ( \begin{array}{cccc}
v & x &\cdots  & x\\
x & x &\cdots  & x\\
\vdots & \vdots & \ddots & \vdots \\
x & x & \cdots & x
\end{array} \right)
\end{eqnarray}
The Fourier transform yields the matrix elements of $b$, namely
\begin{equation}
b_{m,n}=(v'-x')\,\delta_{n,0} \,\delta_{m,0}+x',
\end{equation}
with
\begin{equation}
x'=\frac{v-x}{d}, \qquad v'=\frac{v+(d^2-1)x}{d}.
\end{equation}
The cloning is a trace-preserving operation so the condition 
${\mathrm Tr}[\rho_{\mathrm in,A}]=1$ must be satisfied, which provides the normalization constraint,
\begin{equation}
v^2+(d^2-1)x^2=1.
\end{equation}
If the input state is $\ket{0}$, the error operators $E_{0,n}$, $\forall n$,
leave it unchanged up to a phase, while all the other $E_{m,n}$'s
produce a state that is orthogonal to it. Therefore, 
the fidelities of the two clones for any input state 
can be expressed as
\begin{equation}
F_A=v^2+(d-1)x^2, \qquad F_B=v'^2+(d-1)x'^2.
\label{twofidelities}
\end{equation}
Note that there is only a single free parameter $x$ which controls the
asymmetry of the cloner. We can also characterize the cloner
as a function of the fidelity $F_A$ of the first clone, namely,
\begin{equation}
x^2=\frac{1-F_A}{d(d-1)}, \qquad v^2=\frac{(d+1)F_A-1}{d}.
\end{equation}
The symmetric cloner is obtained by putting $x=x'$, which results in
\begin{equation}
x^2=\frac{1}{2d(d+1)}, \qquad v^2=\frac{d+1}{2d}
\end{equation}
and is associated with the fidelity given in Eq.~(\ref{d-dim-univ})

Note that, as rigorously proved recently for any $d$
in \cite{Fiurasek2005}, this isotropic Heisenberg cloner represents 
the optimal asymmetric cloning machine which, for a fixed fidelity $F_A$ of 
the first clone, maximizes the fidelity of the second clone $F_B$. 
Note also that the optimality of the Heisenberg cloners, based the double-Bell ansatz, was explained in \cite{Chiribella2005} 
as a consequence of the extremality of these cloners.
It is worth stressing that by exploiting this double-Bell ansatz, 
these machines can be derived almost without any effort 
as they follow immediately from the general isotropy 
and trace preservation conditions. 

\subsubsection{Fourier-covariant cloners}

As a second example, we shall consider the {\it Fourier-covariant}
machine, which clones equally well two mutually unbiased bases, the computational basis $\{|k\rangle\}$ 
and the dual basis 
\begin{equation}
||l\rangle\rangle=\frac{1}{\sqrt{d}}\sum_{k=0}^{d-1}e^{2\pi i (kl/d)} |k\rangle.
\end{equation}
The cloner copies equally well the states of both bases if the
matrix $a$ has the form,
\begin{eqnarray}
a_{m,n}&=&(v-2x+y)\,\delta_{m,0}\,\delta_{n,0}
+(x-y)\,(\delta_{m,0}+\delta_{n,0})+y
\nonumber \\
&=&\left ( \begin{array}{cccc}
v & x &\cdots  & x\\
x & y &\cdots  & y\\
\vdots & \vdots & \ddots & \vdots \\
x & y & \cdots & y
\end{array} \right)
\end{eqnarray}
where the parameters satisfy the trace preservation condition
\begin{equation}
v^2+2(d-1)x^2+(d-1)^2y^2=1 
\end{equation}
The matrix $b$ then has a similar form with $v$, $x$, and $y$ being replaced by 
\begin{eqnarray}
v'&=&\frac{1}{d}[v+2(d-1)x+(d-1)^2y], \nonumber \\
x'&=&\frac{1}{d}[v+(d-2)x+(1-d)y], \\
y'&=&\frac{1}{d}[v-2x+y]. \nonumber
\end{eqnarray}
The fidelities of the two clones are again given by Eq.~(\ref{twofidelities})
but, here, we have two free parameters, say, $x$ and $y$. To eliminate
one of them, one has to maximize Bob's fidelity $F_B$ for a given value of Alice's fidelity $F_A$ (using the normalization relation), which is a simple
constrained optimization problem. The resulting optimal asymmetric cloner 
is characterized by
\begin{equation}
v=F_A, \qquad x=\sqrt{\frac{F_A(1-F_A)}{d-1}}, \qquad y=\frac{1-F_A}{d-1}.
\end{equation}
which depends on the single parameter $F_A$.
The symmetric Fourier-covariant cloner can again be obtained
by setting $x=x'$ and $y=y'$ , which gives Eq.~(\ref{fidelity-fourier})
for the fidelity.


\subsubsection{Phase-covariant cloners}

As a third example, consider the {\it phase-covariant} machine,
which optimally clones all balanced superpositions of the form
\begin{equation}
\ket{\psi}=d^{-1/2} [\ket{0}+\e^{i\phi_1} \ket{1}
+\cdots \e^{i\phi_{d-1}}\ket{d-1} ]
\end{equation}
where the $\phi_i$'s are arbitrary phases.
Here, it can be easily shown that the matrix $a$
must take the form
\begin{eqnarray}
a_{m,n}&=&(v-y)\,\delta_{m,0}\,\delta_{n,0}+(y-x)\,\delta_{m,0}+x.
\nonumber \\
&=&\left ( \begin{array}{cccc}
v & y &\cdots  & y\\
x & \multicolumn{2}{c}{\cdots}  & x\\
\vdots & \multicolumn{2}{c}{\ddots} & \vdots \\
x & \multicolumn{2}{c}{\cdots} & x
\end{array} \right)
\end{eqnarray}
while the trace preservation condition is
\begin{equation}
v^2+(d-1)y^2+d(d-1)x^2=1
\end{equation}
The matrix $b$ has the same form, albeit with $v$, $x$, and $y$ being replaced
by
\begin{eqnarray}
v'&=&\frac{1}{d}[v+d(d-1)x+(d-1)y], \nonumber \\
x'&=&\frac{1}{d}[v-y] , \\
y'&=&\frac{1}{d}[v-dx+(d-1)y]. \nonumber
\end{eqnarray}
If the input state is $||0\rangle\rangle$ (all the states of the dual basis
are balanced superpositions), the error operators $E_{m,0}$, $\forall m$,
leave it unchanged up to a phase, while all the other $E_{m,n}$'s
produce a state that is orthogonal to it. Therefore, the
fidelities of the two clones are again given by Eq.~(\ref{twofidelities}),
and we have two free parameters, say, $x$ and $y$. We can eliminate 
one of them by maximizing $F_B$ for a given $F_A$, which yields the optimal
phase-covariant cloner. In the special case
of a symmetric cloner, we have $x=x'$ and $y=y'$
resulting in the fidelity given by Eq.~(\ref{fidelity-phase}).

\section{$N$-to-$M$ universal quantum cloning}
\label{section-universal-N-to-M}

\subsection{Optimal cloning transformation}
\label{secUCoptimalmap}

In this Section we will focus on the universal (state-independent) cloning.
An ideal universal $N \rightarrow M$ quantum cloning machine would be a device 
that prepares $M$ exact clones of an arbitrary  state $\psi \in \mathcal{H}$ 
from  $N$ copies of $\psi$. The input Hilbert space of the cloning
transformation is the symmetric subspace ${\mathcal{H}}_{+}^{\otimes N}$ of 
$N$ qudits, and $d=\dim \mathcal{H}$ shall denote the dimension of the Hilbert
space of the input states in what follows.
As already explained above, exact deterministic quantum cloning
is forbidden by the linearity of quantum mechanics and only approximate copying 
with fidelity less than unity is possible.


As noted before, two different kinds of cloning fidelities are considered in the literature. 
The global fidelity compares the global state of $M$ clones with the ideal
output $\psi^{\otimes M}$. Let $\mathcal{S}$ denote the cloning CP map. Then the
global fidelity of cloning of the state $\psi$ can be expressed as
$F_{N \rightarrow M}^{\mathrm univ,G}({\mathcal{S}},\psi)=\mathrm{Tr}[\psi^{\otimes M}
{\mathcal{S}}(\psi^{\otimes N})]$. Generally, the fidelity of the cloning can
depend on $\psi$ and one may define the cloning fidelity as the infimum of
$F_{N \rightarrow M}^{\mathrm univ,G}({\mathcal{S}},\psi)$ over all input states $\psi$,
\begin{equation}
F_{N \rightarrow M}^{\mathrm univ,G}({\mathcal{S}})=\inf_{\psi} \mathrm{Tr}[\psi^{\otimes M}
{\mathcal{S}}(\psi^{N})].
\end{equation}
The single-clone fidelity quantifies how well each clone resembles the
desired output $\psi$. For $k$-th clone we can write
$F_{N \rightarrow M}^{\mathrm univ,SC}(S,\psi,k)=\mathrm{Tr}[\psi \mathrm{Tr}_{k}^\prime{\mathcal{S}}(\psi^{\otimes N})]$
where $\mathrm{Tr}_k^\prime$ denotes trace over all $M$ qudits except for the $k$-th
qudit. When judging the performance of the cloning machine, we should take the
infimum of $F_{N \rightarrow M}^{\mathrm univ,SC}({\mathcal{S}},\psi, k)$ over all input states and all $M$
clones and define
\begin{equation}
F_{N \rightarrow M}^{\mathrm univ,SC}({\mathcal{S}})=\inf_k \inf_{\psi} \mathrm{Tr}[\psi \mathrm{Tr}_{k}^\prime
{\mathcal{S}}(\psi^{N})],
\end{equation}
where $k\in\{1,\cdots,M\}$.

The \emph{universal} cloning machine should clone equally well all 
quantum states so the fidelity should not depend on $\psi$. 
Any transformation $\mathcal{S}$ can be converted 
into a universal cloning transformation whose fidelity is state independent
by twirling operation consists of applying randomly a unitary $U(\Omega)$ 
to each input $\psi$ and then undoing this by applying a unitary 
$U^\dagger(\Omega)$ to each clone with the probability density  $d\Omega$ 
equal to the Haar measure on $\mathrm{SU}(2)$, see also Sec. \ref{sectiontwirling}. The effective map 
\begin{equation}
{\mathcal{S}}_{\mathrm{twirl}}(\psi)=\int_\Omega U^{\dagger
\otimes M}(\Omega) {\mathcal{S}}[(U(\Omega)\psi U^\dagger(\Omega))^{\otimes N}]
U^{\otimes M}(\Omega) \, d\Omega
\end{equation} 
is covariant, i.e.,
${\mathcal{S}}_{\mathrm{twirl}}[(U\psi U^{\dagger})^{\otimes N}]
=U^{\otimes M} {\mathcal{S}}_{\mathrm{twirl}}(\psi^{\otimes N})U^{\dagger \otimes M}$
and, consequently, $F_{N \rightarrow M}^{\mathrm univ,G}$ does not depend on $\psi$.
To guarantee the independence of the single-clone fidelity 
on the clone index $k$, it is also necessary to randomly permute the $M$ clones
after the twirling.  The important feature of the twirling operation and the
permutations is that they do not modify the \emph{mean fidelity} calculated as
the average of $F_{N \rightarrow M}^{\mathrm univ,G}({\mathcal{S}},\psi)$ or 
$\frac{1}{M}\sum_{k=1}^M F_{N \rightarrow M}^{\mathrm univ,SC}({\mathcal{S}},\psi,k)$ over all
input states $|\psi\rangle=U(\Omega)|\psi_0\rangle$ with the measure $d\Omega$. 

The universal cloning has been studied extensively by many authors
(\cite{Buzek96,Gisin97,Hillery97,Buzek98a,Buzek98c,Cerf98,Werner98,Niu98,Keyl99,Cerf99,Cerf2000a,Cerf2000b}).
The task of cloning can be rephrased as diluting the quantum information 
carried by the $N$ input qudits into $M$ output qudits. The universal 
cloning should not prefer any direction in the Hilbert space and should 
be isotropic. As shown by \cite{Werner98}, the optimal universal cloning operation
${\mathcal{S}}_{\mathrm{opt}}$ can be expressed as follows,
\begin{equation}
{\mathcal{S}}_{\mathrm{opt}}(\psi^{\otimes N})=\frac{D(N,d)}{D(M,d)}
\Pi_{M,d}^{+}(\psi^{\otimes N} \otimes I^{\otimes (M-N)})\Pi_{M,d}^{+},
\label{UCmapS}
\end{equation}
where $\Pi_{M,d}^+$ is the  projector onto the fully symmetric (Bose) subspace 
of $M$ qudits and 
\begin{equation}
D(M,d)={d+M-1 \choose M}
\end{equation}
is the dimension of this subspace. We can see from Eq. (\ref{UCmapS}) 
that the optimal cloning formally consists in attaching $M-N$ blank copies prepared 
in maximally mixed state $I/d$ to the input state $\psi^{\otimes N}$ 
and then projecting the whole state of $M$ qudits onto the symmetric subspace of $M$
qudits, ${\mathcal{H}}_{+}^{\otimes M}$.
With proper normalization as given in Eq. (\ref{UCmapS}) ${\mathcal{S}}_{\mathrm{opt}}$
is a trace preserving completely positive map and can be therefore realized 
deterministically.

The maximal global cloning fidelity achieved by the optimal cloner
(\ref{UCmapS}) reads
\begin{equation}
F_{N \rightarrow M}^{\mathrm univ,G}=\frac{D(N,d)}{D(M,d)}= \frac{(d+N-1)! \, M!}{(d+M-1)!\, N!}.
\label{}
\end{equation}
The density matrix of each output clone is a convex 
mixture of the input state $\psi$ and the maximally mixed state $I/d$,
\begin{equation}
\rho=\gamma \psi +\frac{1}{d}(1-\gamma) I.
\label{UCsingleclonerho}
\end{equation}
This expression  reveals the high isotropy of the universal symmetric
quantum cloning  which is fully characterized by a single parameter, namely
the shrinking factor $\gamma(N,M)$,
\begin{equation}
\gamma(N,M)=\frac{N}{N+d}\frac{M+d}{M}.
\end{equation}
The single-clone fidelity can be immediately determined from Eq.
(\ref{UCsingleclonerho}) and we get
\begin{equation}
F_{N \rightarrow M}^{\mathrm{univ,SC}}= \frac{MN+M+N(d-1)}{M(N+d)}.
\label{UCsingleclonefidelity}
\end{equation}

\subsubsection{Connection with quantum state estimation}

There is a close relationship between optimal quantum cloning and optimal
quantum state estimation. As shown by \cite{Bruss98b}, 
in the limit of  infinite number of clones, $M\rightarrow \infty$, the
single-clone fidelity $F_{N \rightarrow M}^{\mathrm{univ,SC}}$ becomes equal to the fidelity 
of the optimal estimation of the state $\psi$ from $N$ copies (\cite{Massar95,Bruss99a,Hayashi2004})
\begin{equation}
F_{N \rightarrow \infty}=\frac{N+1}{N+d}.
\label{UCestimationfidelity}
\end{equation}
Consequently, in the limit $M\rightarrow \infty$ the optimal cloning becomes equivalent to
the optimal state estimation from $N$ copies of $\psi$ followed by
preparation of infinitely many copies of the estimated state. 
This relationship between optimal universal cloning  and optimal state 
estimation can be explored to prove the optimality of the cloning
transformation (\ref{UCmapS}). It follows from the symmetry, isotropy and linearity of
universal quantum cloning that  the single-qudit outputs must have the form
(\ref{UCsingleclonerho})
and that for concatenated 
universal cloners the shrinking factors multiply. Since the concatenation of the
optimal  $N \rightarrow M$ and $M\rightarrow L$ cloners cannot be better than the optimal
$N\rightarrow L$ cloner, we get
\begin{equation}
\gamma(N,L) \geq \gamma(N,M)\gamma(M,L).
\label{UCshrinkingfactorsinequality}
\end{equation}
We take the limit $L\rightarrow \infty$ and take into account that the
shrinking factor corresponding to the fidelity (\ref{UCestimationfidelity}) 
reads $\gamma(N,\infty)=N/(N+d)$. 
In this way, we get from the inequality (\ref{UCshrinkingfactorsinequality}) 
an upper bound on $\gamma(N,M)$,
 \begin{equation}
 \gamma(N,M) \leq \frac{\gamma(N,\infty)}{\gamma(M,\infty)}= \frac{N}{N+d}\frac{M+d}{M},
 \end{equation}
 which is saturated by the optimal universal cloning transformation
 (\ref{UCmapS}).

\subsubsection{Unitary realization and quantum circuit}
 
So far the optimal cloning transformation was presented in the form of the rather
abstract CP map (\ref{UCmapS}). It holds that every trace preserving CP map admits
a unitary realization with the use of an ancilla system. The unitary
realization of cloning requires $2(M-N)$ ancilla qudits; $M-N$ blank copies 
and $M-N$ additional ancillas. For the sake of
presentation simplicity we will consider here the $N\rightarrow M$ cloning of qubits 
(\cite{Gisin97}). The unitary  cloning transformation can be expressed 
in a covariant form,
\begin{equation}
U|N\psi\rangle_{\mathrm{in}}|R\rangle_{\mathrm{anc}}=\sum_{j=0}^M \alpha_j |(M-j)\psi,
j\psi^{\perp}\rangle_{\mathrm{clones}}
|(M-N-j)\psi^{\perp},j\psi\rangle_{\mathrm{anc}}.
\label{UCqubitmap}
\end{equation}
Here $|k\psi,(N-k)\psi^\perp\rangle$ denotes a symmetric state of $N$ qubits
with $k$ qubits in state $|\psi\rangle$ and $N-k$ qubits in an orthogonal state
$|\psi^\perp\rangle$, $\langle \psi|\psi^\perp\rangle=0$, $|R\rangle_{\mathrm{anc}}$
denotes the initial state of the ancilla qubits and
\begin{equation}
\alpha_j=(-1)^j {M-j \choose N}^{1/2} {M+1 \choose M-N }^{-1/2}.
\end{equation}
The generalization of the formula (\ref{UCqubitmap}) to qudits with arbitrary $d$ was
obtained by \cite{Fan2001a}. In Sec. \ref{secUCOdownconversion} we shall show that 
the transformation (\ref{UCqubitmap}) naturally arises in stimulated amplification 
of light when the qubits are represented by the polarization states of single photons.

\input #progressfig3

Quantum information theory teaches us that an arbitrary unitary operation $U$ can be
implemented as a sequence of single-qubit rotations and a two-qubit controlled-NOT
gates, $U_{\mathrm CNOT}=|j\rangle_c|k\rangle_t=|j\rangle_c|k\oplus j\rangle_t$ , where
$\oplus$ denotes addition modulo 2, $c$ is the control qubit and $t$ is the
target qubit. The quantum network for the optimal universal 
$1\rightarrow M$ cloning of qubits (\cite{Buzek97a,Buzek98d}) is depicted in 
Fig. \ref{fignetwork}. First, the $2(M-1)$ ancilla qubits 
$a_1,\cdots,a_{M-1}$ and $b_1,\cdots,b_{M-1}$ are prepared 
in an entangled state
\begin{equation}
|\Phi\rangle_{ab}=\frac{1}{M}\sqrt{\frac{2}{M+1}} \sum_{k=0}^{M-1}
\left(e_k|M-1,k\rangle_{a}+f_k|M-1,k-1\rangle_{a}\right) |M-1,k\rangle_b,
\label{UCancillapsi}
\end{equation}
where $e_k=M-k$ and $f_k=\sqrt{k(M-k)}$, and $|M-1,k\rangle$ denotes a symmetric
state of $M-1$ qubits with $k$ qubits in state $|1\rangle$ and $M-1-k$ qubits in
state $|0\rangle$. The state (\ref{UCancillapsi}) can be generated by a sequence
of single-qubit rotations and C-NOT gates starting from any initial pure state
of the ancilla. The cloning itself consists of a sequence of $2(M-1)$ C-NOT
gates where the qubit $a_0$ that contains the state to be copied serves as a
control qubit and the ancillas are target qubits. This is followed by another
sequence of $2(M-1)$ C-NOT gates where now the qubit $a_0$ is target and the
ancilla qubits are controls. The $M$ clones are stored in the qubits
$a_0,\cdots, a_M-1$ while the qubits $b_j$ represent the ancillas.

\subsection{Optimality proof for $1\rightarrow M$ cloning of qubits}
\label{secUCproof}

The optimality of the $1\rightarrow 2$ symmetric cloning machine for qubits
was first proved by \cite{Bruss98a}. 
The optimality of the cloning transformation (\ref{UCmapS})
for arbitrary number of inputs $N$, outputs $M$  and dimension $d$ 
was proved by \cite{Werner98} for the global fidelity and later by \cite{Keyl99} for
the single-clone fidelity using powerful techniques of group theory. 
Here we shall present a simple optimality proof for the class of 
$1\rightarrow M$ universal symmetric cloning machines for qubits. This proof
follows the general concept outlined in Sec. \ref{sectionsemidefinite} where 
it was shown that the
fidelity is upper bounded by the maximum eigenvalue of a certain positive
semidefinite operator. This optimality proof when single-clone fidelity is used
as a figure of merit is due to \cite{Gisin97} and it has also been extended to the
global fidelity (\cite{Fiurasek2001b}). The advantage of this approach is that it can be
easily generalized to asymmetric cloning, as discussed in the next section.

Consider the maximization of the single-clone fidelity and let
us assume that the output Hilbert space of the cloning map $\mathcal{S}$ is the
symmetric subspace, since the desired outputs $|\psi\rangle^{\otimes M} \in
{\mathcal{H}}_+^{\otimes M}$. Then all the clones have the same fidelity by
construction and we can express the operator $S$ that is 
isomorphic to the CP map $\mathcal{S}$ as follows,
\begin{equation}
S=\sum_{i,j=0}^1 \sum_{k,l=0}^M S_{ik,jl} |i\rangle_{\mathrm{in}} \langle j| \otimes
|M,k\rangle_{\mathrm{out}}\langle
M,l|.
\end{equation}
The mean single-clone fidelity can be calculated by averaging over the surface
of the Poincar\'e sphere,
\begin{equation}
F_{N \rightarrow M}^{\mathrm{univ,SC}}=\int_\psi \mathrm{Tr} [(\psi^T \otimes \psi)
 Tr_{out}^\prime (S)] d\psi,
\end{equation}
where $\mathrm{Tr}_{\mathrm{out}}'$ denotes tracing over all output qubits except for the first 
one, and 
\begin{equation}
\int_\psi d\psi \equiv \frac{1}{4\pi} \int_0^{2\pi}\int_0^\pi\sin\vartheta d\vartheta d\phi,
\qquad |\psi\rangle=\cos\frac{\vartheta}{2} |0\rangle+
e^{i\phi}\sin\frac{\vartheta}{2} |1\rangle.
\end{equation}
After the tracing and integration, we find that the mean single-clone fidelity
is a linear function of $S$, $F_{N \rightarrow M}^{\mathrm{univ,SC}}=\mathrm{Tr}[SR_{SC}]$, where the positive
semidefinite operator $R_{SC}$ reads
\begin{eqnarray*}
R_{SC}&=&\frac{1}{6M} \sum_{k=0}^M \left[(2M-k)|0\rangle\langle 0|
+(M+k)|1\rangle\langle 1| \right] \otimes |M,k\rangle\langle M,k|
\nonumber \\
&&+\frac{1}{6M}\sum_{k=0}^{M-1} \sqrt{(M-k)(k+1)} \, |1\rangle\langle 0| \otimes
|M,k+1\rangle\langle M,k|+\mathrm{h.c.}
\end{eqnarray*}
According to Eq. (\ref{fidelityupperboundmaxeig}) the fidelity 
$F_{N \rightarrow M}^{\mathrm{univ,SC}}$ 
is upper bounded by the maximum
eigenvalue of $R_{SC}$, $F_{N \rightarrow M}^{\mathrm{univ,SC}}\leq 2 r_{\mathrm{SC,max}}$.
The matrix $R_{SC}$ has a block diagonal structure and it is easy to show that
all the eigenstates of $R_{SC}$ have the form
$\alpha|0\rangle|M,k\rangle+\beta|1\rangle|M,k+1\rangle$. The calculation of the
eigenvalues of $R_{SC}$ thus reduces to finding roots of quadratic polynomials and one
finds that $R_{SC}$ has only three different eigenvalues, $r_1=(2M+1)/(6M)$,
$r_2=1/3$, and $r_3=1/6$. This provides an  upper bound 
$F_{N \rightarrow M}^{\mathrm{univ,SC}}\leq (2M+1)/(3M)$ which is saturated 
by the cloning machine (\ref{UCmapS}).
This proves that the machine (\ref{UCmapS}) is optimal.

Similar chain of arguments can be used to demonstrate the optimality of the
machine (\ref{UCmapS}) when the global fidelity is the figure of merit. The mean global
fidelity can be written as $F_{N \rightarrow M}^{\mathrm univ,G}=\mathrm{Tr}[SR_G]$, where 
\begin{equation}
R_{G}=\int_\psi \psi^T_{\mathrm{in}}\otimes \psi_{\mathrm{out}}^{\otimes M}d\psi.
\end{equation}
With the help of Schur's lemma this integral can be easily evaluated and one
obtains $R_G=\frac{1}{M+2}\Pi_{M+1}^{+,T_1}$, where  $\Pi_{M+1}^+$ is a
projector onto the symmetric subspace of $M+1$ qubits and $T_1$ denotes partial
transposition with respect to the first qubit. Again, the matrix $R_G$ is block
diagonal and its eigenvalues can be easily determined analytically. One finds
that $r_{G,\mathrm{max}}=1/(M+1)$ which implies 
$F_{1 \rightarrow M}^{\mathrm univ,G}\leq 2/(M+1)$ and this bound is
achieved by the cloner (\ref{UCmapS}).

\subsection{Universal asymmetric quantum cloning}
\label{secUCasymmetry}

The quantum cloning machines serve as universal distributors of quantum information among
several parties. The symmetric cloner divides the information equally among all $M$ copies
but it is also possible to distribute the information unequally. A lot of attention has
been devoted to the universal  asymmetric $1\rightarrow 2$ cloning machines for qubits 
(\cite{Cerf98,Buzek98b,Niu98,Cerf99,Cerf2000a}) and
qudits (\cite{Cerf2000b,Cerf2002a}), which produce two clones $A$ and $B$ with 
different fidelities $F_A$ and $F_B$. The optimal asymmetric cloner can be defined 
as a machine that for a given fixed fidelity of the first clone $F_A$ maximizes the fidelity
of the second clone $F_B$. Such machines can find applications, e.g., in eavesdropping 
on quantum key distribution protocols, where they allow one to investigate the trade-off 
between the information gained by the eavesdropper and the disturbance observed 
at the receiver's station.
 
In terms of the cloning CP map $S$, the mean fidelities of the two clones 
can be expressed as $F_{A}=\mathrm{Tr}[S R_{A}]$ and  $F_{B}=\mathrm{Tr}[S R_B],$
where the positive semidefinite operators $R_j$ are given by
\begin{equation}
R_{A}=\int_{\psi} \psi_{\mathrm{in}}^T \otimes \psi_A \otimes I_B \, d \psi, 
\quad
R_{B}=\int_{\psi} \psi_{\mathrm{in}}^T \otimes I_A \otimes \psi_B \, d \psi.
\end{equation}
The optimal asymmetric cloning machine should maximize a convex mixture 
of the mean fidelities $F_{A}$ and $F_{B}$
(\cite{Fiurasek2003b,Lamoureux2004b,Fiurasek2005}),
\begin{equation}
F=pF_{A}+(1-p)F_{B}=\mathrm{Tr}[SR],
\label{convexmixture}
\end{equation}
where $R=p R_A+(1-p)R_{B}$ and $p$ is a parameter that controls the asymmetry of
the cloner.  The maximization of $F$ for a given value of $p$ can be
equivalently rephrased as a maximization of $F_{B}$ for a fixed value of $F_A$.
After some algebra, we find
\begin{equation}
R=\frac{1}{d(d+1)}[I_{\mathrm{in},AB}+ d p\Phi_{\mathrm{in},A}^+\otimes I_{B} 
+d(1-p)\Phi_{\mathrm{in},B}^{+} \otimes I_A].
\end{equation}
The maximum eigenvalue of $R$ is $d$-fold degenerate and the corresponding 
eigenvector is
\begin{equation}
 |r_{\mathrm{max}};k\rangle=\alpha\, |\Phi^+\rangle_{AR}|k\rangle_{B}
+\beta\, |\Phi^+\rangle_{BR}|k\rangle_{A},
\label{UCasymeigenstates}
\end{equation}
where 
the coefficients $\alpha,\beta \geq 0$ are some functions of $d$ and $p$.  
By properly normalizing the eigenstates (\ref{UCasymeigenstates}) we get
\begin{equation}
\alpha^2+\beta^2+\frac{2\alpha\beta}{d}=1.
\label{UCasymnormalization}
\end{equation}

The operator $S$ isomorphic to the optimal cloning CP map $\mathcal{S}$ 
is proportional to the projector onto the subspace spanned by the eigenstates
(\ref{UCasymeigenstates}). The unitary realization of this map requires a single ancilla
qudit $C$ and can be written in a covariant way,
\begin{equation}
|\psi\rangle \rightarrow 
\alpha \, |\psi\rangle_{A}|\Phi^+\rangle_{BC}
+\beta \, |\psi\rangle_{B}|\Phi^+\rangle_{AC}.
\label{UCasymunitary}
\end{equation}
From this expression we can evaluate the fidelities of the two clones,
\begin{equation}
F_A=1-\frac{d-1}{d}\beta^2, \qquad F_B=1-\frac{d-1}{d}\alpha^2.
\label{UCasymfidelityAB}
\end{equation}
Note that the parameters  $\alpha^2$ and $\beta^2$ are the so-called 
{\it depolarizing fractions} as discussed in \cite{Cerf98,Cerf2000a,Cerf2000b}. 
The one-parametric class of the
optimal universal asymmetric $1\rightarrow 2$ cloning machines is characterized by the 
Eqs. (\ref{UCasymunitary}) and (\ref{UCasymfidelityAB})  together with the normalization condition (\ref{UCasymnormalization}).

\subsection{Universal NOT gate}
\label{secUCunot}

The process of optimal quantum cloning is closely connected to another 
impossible operation in quantum mechanics, the so-called universal NOT gate for qubits. 
The hypothetical universal NOT gate would perfectly  reverse any spin-$\frac{1}{2}$ state.
This device should thus produce from the input qubit $|\psi\rangle$ an orthogonal state
$|\psi_\perp\rangle$. However, this is impossible, because the transformation $|\psi\rangle\rightarrow
|\psi_\perp\rangle$ is anti-unitary. More generally 
one can consider an extended scenario where $N$ copies of the state $|\psi\rangle$ 
are available and the task is to prepare a single copy of the flipped 
spin $|\psi_\perp\rangle$. The best approximation to this forbidden
operation was found in \cite{Gisin99,Buzek99,Buzek2000}.

The optimal universal NOT gate $S_{\mathrm{UNOT}}$ can be made covariant by twirling so that 
the fidelity $F=\langle\psi_\perp|{\mathcal{S}}_{\mathrm{UNOT}}(\psi^{\otimes N})|\psi_\perp\rangle$ 
does not depend on $|\psi\rangle$ and can be written as  
$F_{\mathrm{UNOT}}=\mathrm{Tr} [S_{\mathrm{UNOT}} R_{\mathrm{UNOT}}]$, where
\begin{equation}
R_{\mathrm{UNOT}}=\int_{\psi} [\psi^{\otimes N}]^{T} \otimes \psi_\perp d\psi 
=\frac{1}{N+2}(U^{\otimes N} \otimes I) \Pi_{+,N+1}
(U^{\dagger \otimes N}\otimes I)
\end{equation}
see \cite{Fiurasek2001b}.
The unitary operation $U$ provides the link between the states
$|\psi^\ast\rangle$ and $|\psi_\perp\rangle,$ $|\psi^\ast\rangle=U|\psi_\perp\rangle$,
 $U|0\rangle=-|1\rangle$, $U|1\rangle=|0\rangle$. The fidelity of the U-NOT gate is bounded
by the maximum eigenvalue of $R_{\mathrm{UNOT}}$. Since this operator is proportional to a
projector, we immediately find $r_{\mathrm{UNOT, max}}=1/(N+2)$. The dimension of the input
Hilbert space ${\mathcal{H}}_+^{\otimes N}$ is $d=N+1$ and we obtain,
\begin{equation}
F_{\mathrm{UNOT}}=\frac{N+1}{N+2}.
\label{UCUNOTFidelity}
\end{equation}
Remarkably, this fidelity coincides with the optimal fidelity of the estimation of the state
$|\psi\rangle$ from $N$ copies. If we possess an estimate of $|\psi\rangle$ then we can also
produce an estimate of $|\psi_\perp\rangle$  with the same fidelity, simply by flipping the
estimated spin. This implies that the optimal U-NOT gate can be realized by performing the
optimal estimation of the state $|\psi\rangle$ followed by the preparation of the flipped
estimated state. In this way we can generate arbitrary many approximate copies of
$|\psi_\perp\rangle$, all with the same fidelity (\ref{UCUNOTFidelity}).
Remarkably, the optimal cloning transformation (\ref{UCqubitmap}) simultaneously also 
implements the optimal approximate U-NOT gate. This machine produces $M-N$ approximate 
anti-clones, which are stored in the ancilla, 
see \cite{Buzek99,DeMartini2002}. 

It is possible to generalize the concept of U-NOT gate to qudits, by noting that the
state $|\psi_\perp\rangle$ is unitarily equivalent to the state $|\psi^\ast\rangle$. 
The complex conjugation is well defined for any dimension $d$, and one can look for the 
transformation that optimally approximates the (generalized) transposition map 
$\psi^{\otimes N}\rightarrow \psi^\ast \equiv \psi^T$. Using similar reasoning 
as before, one can prove that the  maximal fidelity of the approximate transposition
is equal to the fidelity (\ref{UCestimationfidelity}) of the optimal state estimation 
from $N$ copies, see \cite{Fiurasek2004a}.


\section{Universal cloning of photons}
\label{section-cloning-photons}

\input #progressfig4

\subsection{Amplification of light}
\label{secUCOdownconversion}

In quantum optics, single photons are very often used as carriers of quantum
information. Photons represent ideal flying qubits; they can be transmitted over long
distances via low-loss optical fibers and their interaction with the environment is very
weak so they do not suffer from a significant decoherence. Quantum bits can be encoded
into single photons in various ways. One natural option is to exploit the polarization
degrees of freedom and to represent a qubit as a superposition of vertically ($|V\rangle$)
and horizontally ($|H\rangle$) polarized photon, 
$|\psi\rangle=\alpha|H\rangle+\beta|V\rangle$. 
Another possibility is to use the so-called time-bin encoding where the photon can be 
located in one of $d$ different time slots
(\cite{Marcikic2002,Riedmatten2004a}). Such encoding has the advantage that it is
not restricted to qubits and the photon can thus represent a $d$-dimensional system with
arbitrary $d$ (\cite{Riedmatten2004b}). Arbitrary time-bin qudits can be prepared using unbalanced Mach-Zehnder
interferometer. Time-bin encoding has been
advantageously used for long-distance quantum key distribution.

The cloning of the quantum states of single photons requires that the number of output
photons be higher than the number of input photons. This simple fact immediately
leads to the insight that 
the optimal copying of photons can be performed by means of
amplification of light (\cite{DeMartini2000,Simon2000a,Simon2000b}). This is very 
natural because the goal of quantum cloning is to ``amplify'' the quantum information 
carried by the photons. There are several physical mechanisms that can be used for 
cloning such as parametric down-conversion or amplification
of light in atomic media. In all cases, the cloning is achieved due to the process of
stimulated emission, which means that the medium emits preferably photons in the same
quantum state as that of the input photons injected into the medium.

Most of the quantum cloning experiments based on the stimulated  amplification of light 
were carried out using the process of stimulated parametric down-conversion. Consider a
nonlinear crystal with second order nonlinearity $\chi^{(2)}$. In such crystal, a single
``blue'' pump photon with frequency $\omega_{P}$ can be converted into two ``red''
photons with frequencies $\omega_S$ and $\omega_I$ such that
$\omega_S+\omega_I=\omega_P$, which expresses the energy conservation.
The two downconverted photons are referred to as signal (S) and idler (I),
respectively, for historical reasons. An efficient down-conversion requires the
conservation of momentum, which translates into the phase-matching condition
${\bf k}_{S}+{\bf k}_I={\bf k}_P$, where ${\bf k}_j$ stands for the wave-vector of the j-th
photon and $|{\bf k}_j|=n_j \omega_j/c$, where $n_j$ is the refraction index at 
frequency $\omega_j$. Efficient phase matching in the nonlinear crystal can be achieved
by exploring the birefringence and using different polarizations for the pump, 
signal and idler beams. We can distinguish two different kinds of phase matching. In Type
I matching, the pump beam is, say, vertically polarized, and both signal and idler are
horizontally polarized. On the other hand, in the Type II matching, the signal and idler
photons are orthogonally polarized. Besides their polarization states, the signal and
idler beams can also be distinguished spatially. So in non-degenerate Type-II
downconversion we deal with modes $A_H$ and $A_V$ for the signal beam and $B_H$ and
$B_V$ for the idler beam. It is possible to arrange the configuration of the pump beam
and nonlinear crystal and to select only certain directions in the output beams such that
the effective Hamiltonian which describes this process reads
\begin{equation}
H= i\kappa (a_V^\dagger b_H^\dagger-a_H^\dagger b_V^\dagger) +\mathrm{h.c.}.
\label{UCsingletHamiltonian}
\end{equation}
This Hamiltonian is obtained in the limit of strong coherent pumping and the coupling
constant $\kappa$ is proportional to the pump beam amplitude $\alpha_P$ and to the
second-order nonlinearity $\chi^{(2)}$, while $a_{j}^\dagger$ is the creation operator for the
$j$-th mode. An essential feature of the Hamiltonian (\ref{UCsingletHamiltonian}) is
that it is invariant with respect to the simultaneous identical transformation of the
polarization basis of signal and idler photons. Mathematically, we have
$(U\otimes U) H (U^\dagger \otimes U^\dagger)=H$, where $U a_V U^\dagger=
u_{VV}a_V+u_{VH}a_H$, $U a_H U^\dagger=
u_{HV}a_V+u_{HH}a_H$, the matrix $u_{ij}$ is unitary and identical transformation rules
hold also for $b_V$ and $b_H$. This covariance property guarantees that
the cloning process is universal and the cloning fidelity is the same for all input
states. It therefore suffices to consider only one particular input state.

A sketch of the cloning of polarization states of photons via stimulated downconversion
is shown in Fig. \ref{figdownconversion_scheme}. The signal mode is initially prepared 
in $N$-photon state $|\psi\rangle_a^{\otimes N}$. This can be in practice achieved
e.g. by means of spontaneous parametric downconversion in crystal C$_1$ and 
conditioning on observing $N$ photons in the output idler mode with photodetector PD.
 After the passage through the crystal C$_2$, $M-N$ photon pairs can be generated with
certain probability. If this happens, then $M$ clones are present in the mode A while
the mode B contains $M-N$ anti-clones. Note that the cloning is only
probabilistic and we cannot predict \emph{a-priori} the number of clones that will be
generated. The particular $N \rightarrow M$ cloning event can be selected only
\emph{a-posteriori} by accepting only the events when $M$ photons were detected 
in mode A.

Let us start with a simple example of $1\rightarrow 2$ cloning
which will illustrate all the main features.  In this case, the input state is given by 
a single photon in mode A and a vacuum in mode B. As already explained, without loss of
generality we can assume that the photon is vertically polarized and we have
$|\psi_{\mathrm{in}}\rangle=|1\rangle_{a_V}|0\rangle_{a_H}|0\rangle_{b_V}|0\rangle_{b_H}$.
The generation of the second clone requires that a single photon pair is emitted in the
nonlinear crystal. In the first order perturbation theory, the output state is given by
\begin{equation}
H |1\rangle_{aV}|0\rangle_{a_H}|0\rangle_{b_V}|0\rangle_{b_H} \propto 
\sqrt{2} |2\rangle_{a_V}|0\rangle_{a_H}|0\rangle_{b_V}|1\rangle_{b_H}
-|1\rangle_{a_V}|1\rangle_{a_H}|1\rangle_{b_V}|0\rangle_{b_H}.
\end{equation}
Notice the prefactor $\sqrt{2}$ which arises because the emission of the second
vertically polarized photon in mode A is stimulated by the presence of a vertically
polarized photon in this spatial mode. This cloning is optimal since it yields the
maximum fidelity. We can immediately see that the global fidelity is $2/3$. To determine
the single-clone fidelity we note that with probability $2/3$ both photons in mode A
are vertically polarized and with probability $1/3$ only one photon is vertically
polarized. So the probability that one randomly chosen photon in spatial mode A is
vertically polarized is $\frac{2}{3} \times 1 + \frac{1}{3} \times \frac{1}{2} =
\frac{5}{6}$, which is the maximal
single-clone fidelity for $1\rightarrow 2$ cloning of qubits.

\input #progressfig5

Several experiments on cloning via parametric down-conversion 
have been reported, see
\cite{DeMartini2000,Linares2002,Pelliccia2003,Sias2003,DeMartini2004}.
The experimental setup used by \cite{Linares2002} is
shown in Fig. \ref{figlinarescloningfig1}. A nonlinear BBO crystal is pumped by a second harmonic of a
Ti:sapphire laser which emits 120 fs long pulses. A tiny part of the coherent master laser beam was
split on the first beam splitter $BS$ and used as a seed for the down-conversion. With
probability $p \ll 1$, the beam contained exactly one photon. 
This beam was fed to the BBO
crystal and the output is analyzed using a sequence of wave-plates, polarizing beam
splitters (PBS) and single-photon photo-detectors. The probability of a pair
generation in the crystal $p_2 \ll p $  which guarantees that the dominant event leading
to two photons in mode $a$ and one photon in mode $b$ is when a single photon was in the
weak coherent beam $a$  and a single pair was emitted in the crystal. The
conditioning on observing a click of the trigger detector D1 is important since it
eliminates the events when there were two photons in mode $a$ and no pair was generated
in the crystal. 

\input #progressfig6

In the experiment, one measures the number of coincidence clicks of the photodetectors D2
and D3 as a function of the time delay between the input photon beam in mode $a$ and the 
pump beam. If those two beams do not overlap in the BBO crystal, then there is no
stimulated down-conversion and the polarization of the second photon emitted in mode $a$
is fully random. If the two beams overlap, then stimulated amplification sets on and the
second photon is emitted preferably with the same polarization as the input photon.
Optimal cloning is achieved when the overlap is perfect. A detector setting with PBS was
used to measure the number of orthogonally polarized photon pairs $N(1,1)$. To detect the number
of pairs with the same polarization $N(2,0)$, PBS was replaced by a polarizer followed by 
an ordinary beam splitter. The observed coincidence rates as a function of the time delay
are shown in Fig. \ref{figlinarescloningfig3} for three different polarizations. 
We can see that $N(1,1)$ does not depend on the delay as expected, while $N(2,0)$ 
decreases with increasing delay. The average experimental cloning fidelity determined 
from these data reads $F\approx 0.81$ which is very close to the theoretical 
maximum $5/6 \approx 0.833$.

An improved experimental setup involving double passage of the pump beam through the
nonlinear crystal has been developed (\cite{Pelliccia2003,DeMartini2004}), see 
Fig. \ref{figdemartiniUNOTfig1}.
In this setup, the photon to be cloned is generated during the first passage of the pump
pulse through the crystal. Since the signal and idler beams are entangled, projecting
idler beam onto state $|\psi\rangle$ prepares the signal in state $|\psi_\perp\rangle$.
The click of the trigger detector $D_T$ heralds the preparation of a single photon in
mode labeled $-k_1$ in Fig.~\ref{figdemartiniUNOTfig1}. This photon is then cloned by sending it again 
through the nonlinear BBO crystal. The delay between the pump and signal is controlled by
moving the mirror M$_P$. In this experiment, the states of both clones and the
anti-clone were analyzed simultaneously and it was demonstrated that this device
accomplishes jointly the optimal $1\rightarrow 2$ cloning and also the optimal
universal-NOT gate for qubits. The attained fidelities were $F_{\mathrm CLON}=0.81$ and
$F_{\mathrm UNOT}=0.62$.

The stimulated down conversion can be used to probabilistically 
implement any $N \rightarrow M$ cloning of qubits (\cite{Simon2000a}) and even qudits
(\cite{Kempe2000,Fan2002b}).
The unitary transformation induced by the Hamiltonian $H$ can written in a factorized form as
follows,
\begin{equation}
e^{-iHt}=e^{\lambda(a_V^\dagger b_H^\dagger-a_H^\dagger b_V^\dagger)}
(1-\lambda^2)^{n_{\mathrm{tot}}/2+1}
e^{-\lambda(a_V b_H-a_H b_V)}
\label{UCunitaryfactorization}
\end{equation}
where  $\lambda=\tanh(\kappa t)$, $t$ is an effective interaction time, and 
 $n_{\mathrm{tot}}=a_V^\dagger a_V+a_H^\dagger a_H
+b_V^\dagger b_V+b_H^\dagger b_H$ is the total number of photons in spatial modes $a$
and $b$. Since the Hamiltonian $H$ is covariant it is enough to consider the input state
$|\psi_{\mathrm{in}}\rangle=|N\rangle_{aV}|0\rangle_{aH}|0\rangle_{bV}|0\rangle_{bH}$.
With the help of the factorization (\ref{UCunitaryfactorization}) we find that the corresponding output state reads
\begin{equation}
e^{-iHt}|\psi_{\mathrm{in}}\rangle= (1-\lambda^2)^{N/2+1}\sum_{M=N}^\infty
\lambda^{M-N}|\Psi_M\rangle,
\label{UCdcoutput}
\end{equation}
where 
\begin{equation}
|\Psi_M\rangle=\sum_{k=0}^{M-N} (-1)^k \sqrt{{M-k \choose N}} \,
|M-k\rangle_{aV}|k\rangle_{aH}|k\rangle_{bV}|M-N-k\rangle_{bH}.
\label{UCdcpsiM}
\end{equation}
We can see that the output state (\ref{UCdcoutput}) is a weighted superposition of states 
$|\Psi_{M}\rangle$ with different numbers of clones $M$. The state  $|\Psi_{M}\rangle$ 
and hence the  fidelity of $N\rightarrow M$  cloning is independent of the coupling strength $\lambda$ 
and only the probability of generating exactly $M$ clones depends on $\lambda$.
One can also immediately see that the state  (\ref{UCdcpsiM}) coincides (up to an irrelevant overall
normalization factor) with the outcome of the optimal cloning transformation
(\ref{UCqubitmap}), hence
the universal cloning via parametric down-conversion is optimal.

\input #progressfig7

We now extend the concept of cloning via amplification to qudits represented
by a single photon in  $d$ different spatial modes or 
time-bins. The use of time-bin encoding seems to be particularly advantageous since only a
single nonlinear Type-I matched crystal is required, and the pump beam should 
consist of a sequence of $d$ pulses.  We associate creation operators $a_j$ and $b_j$ 
with $j$-th time-bin of signal and idler beams, respectively. 
The Hamiltonian governing the evolution of this system can be expressed as
\begin{equation}
H_d=i\kappa \sum_{j=1}^d (a_j^\dagger b_j^\dagger -a_j b_j).
\label{UCHamiltonianqudit}
\end{equation}
This Hamiltonian is invariant with respect to simultaneous unitary transformations 
of the signal and idler modes, 
$(U \otimes U^{\ast}) H (U^{\dagger} \otimes U^{T})= H$, where $U\in SU(d)$.
This covariance property guarantees that the cloning is universal and the cloning
fidelity does not depend on the input state so it suffices to consider 
the  input state $|\psi_{\mathrm{in},d}\rangle=|N\rangle_{a_1}|0\rangle_{a_2}\cdots|0\rangle_{a_d}
|0\rangle_{b_1}|0\rangle_{b_2}\cdots|0\rangle_{b_d}$. The unitary operation $\exp(-iHt)$
can be again factorized similarly as in Eq. (\ref{UCunitaryfactorization}) and  we get
$
e^{-iH_d t} |\psi_{\mathrm{in},d}\rangle=(1-\lambda^2)^{N/2+d}\sum_{M=N}^\infty \lambda^{M-N} |\Psi_{M,d}\rangle
$ where the state containing $M$ clones reads
\begin{equation}
|\Psi_{M,d}\rangle= \sum_{\bf m} \sqrt{{N+m_1 \choose N}}
|N+m_1\rangle_{a_1}|m_2\rangle_{a_2} \ldots |m_d\rangle_{a_d}
|m_1\rangle_{b_1}|m_2\rangle_{b_2} \ldots |m_d\rangle_{b_d}.
\label{UCdcquditoutput}
\end{equation}
In this formula,  $\sum_{\bf m}$ indicates summation over all vectors 
${\bf m}=(m_1,\ldots,m_d)$
satisfying  $\sum_{j=1}^d m_j=M-N$. The optimality of this cloning transformation can
be proved by explicit evaluation of the fidelity.  It can be shown that 
there are ${M-N+d-2-m \choose d-2}$  different terms in Eq. (\ref{UCdcquditoutput}) 
with $N+m$ photons in mode $a_1$, each with weight ${N+m \choose m}$. 
The average single-clone fidelity can be thus expressed as,
\begin{equation}
F=\frac{1}{\mathcal{N}} \sum_{m=0}^{M-N} {N+m\choose N} {M-N+d-2-m \choose d-2} \frac{N+m}{M},
\label{UCdcquditfidelity}
\end{equation}
where the normalization factor is given by
\begin{equation}
{\mathcal{N}}\equiv \sum_{m=0}^{M-N} {N+m\choose N} {M-N+d-2-m \choose d-2} =
{M+d-1\choose N+d-1}.
\label{UCdcquditnormalization}
\end{equation}
The summation in Eq. (\ref{UCdcquditfidelity}) can be performed with the help 
of the identity given  in Eq. (\ref{UCdcquditnormalization}) and one recovers 
the optimal fidelity (\ref{UCsingleclonefidelity}). 


Instead of parametric down-conversion it is also possible to amplify the light 
by sending it through an inverted atomic medium (\cite{Simon2000a,Kempe2000,Fan2002b}). 
The atoms should possess $d$ different ground states $|g_j\rangle$ and 
an excited state $|e\rangle$.  We assume that each atomic transition 
$|e\rangle \rightarrow |g_j\rangle$ is strongly coupled to a single optical 
mode $a_j$ and the qudits are represented by single photons in those $d$ modes.
The universality of the cloning requires that the coupling strength $\kappa$
must be the same for all $d$ transitions $|e\rangle \rightarrow |g_j\rangle$.
In the interaction picture and in the rotating wave approximation, the interaction
of light with atoms is governed by the Jaynes-Cummings Hamiltonian, 
\begin{equation}
H_{JC}=\kappa \sum_{k=1}^L \sum_{j=1}^d  a_j^\dagger |g_{jk}\rangle\langle e_k|
+\mathrm{h.c.},
\label{UCHamiltonianatoms}
\end{equation}
where $L$ is the number of atoms and $|g_{jk}\rangle$ stands for the ground 
state $|g_j\rangle$ of $k$-th atom. This Hamiltonian satisfies the covariance property
$U\otimes U^\ast H U^\dagger \otimes U^T = H$, where 
$U a_j^\dagger U^\dagger=u_{jk}a_k^\dagger$,
$U|e_k\rangle=|e_k\rangle$, and $U|g_{jk}\rangle=\sum_{l=1}^d u_{jl}|g_{lk}\rangle$.

Suppose that all $L=M-N$ atoms are initially prepared in the excited state
and that all $N$ input photons are in mode $A_1$. The joint atoms-photons input state reads 
$|\psi_{\mathrm{in},LA}\rangle=|N\rangle_{a_1}|0\rangle_{a_2}\ldots|0\rangle_{a_d}|e_{1}\rangle
\ldots|e_L\rangle$. If each atom emits a photon during the passage of the light through 
the atoms then  $M$ clones are generated and all  atoms end up in ground states.

In the regime of weak coupling we can express 
the output state conditional on all atoms being in some ground state 
using the $L$-th order perturbation theory,
\begin{equation}
|\psi_{\mathrm{out}}\rangle \propto  
(\sum_{j=1}^d  a_j^\dagger b_j^\dagger c)^L \, |\psi_{\mathrm{in},LA}\rangle,
\end{equation}
where the operator $b_j^\dagger c$  is defined as
$b_j^\dagger c=\sum_{k=1}^L|g_{jk}\rangle\langle e_k|$. 
 Note that with this notation, 
the Hamiltonian (\ref{UCHamiltonianatoms}) becomes similar to the down-conversion 
Hamiltonian (\ref{UCHamiltonianqudit}). 
Since the atoms are supposed to be identical, the photons emitted by them do not carry any
information about which atom emitted which photon. Consequently, if all atoms emit
photons, then the atoms relax to symmetric ground state. Suppose that $m_j$ photons were
emitted to mode $a_j$, $j=1,\ldots,d$. The corresponding symmetrized ground atomic state
reads
\begin{equation}
|g_{\bf{m}}\rangle=C_{L \bf{m}}^{-1}\sum_{\pi(\bf{k})}|g_{1 k_1}\rangle 
\ldots |g_{1 k_{m_1}}\rangle |g_{2 k_{m_1+1}}\rangle \ldots
|g_{2 k_{m_1+m_2}}\rangle \ldots |g_{d k_{L-m_d+1}}\rangle  \ldots |g_{d k_L}\rangle,
\end{equation}
where  $\sum_{\pi(\bf{k})}$  denotes  summation over all $L!$ values of the 
subscripts $k_l$, $l=1,\ldots,L$, which can be obtained as permutations of 
$\{1,\ldots,L\}$. The normalization coefficient 
\begin{equation}
C_{L \bf{m}}^2=m_1! \, m_2! \cdots m_d ! \, L!
\end{equation}
is chosen such that $\langle g_{\bf{m}}|g_{\bf{m}}\rangle=1$.
After some algebra, one finds that the output state (\ref{}) can be expressed as follows,
\begin{eqnarray*}
|\psi_{\mathrm{out}}\rangle &\propto & \sum_{\bf{m}}\frac{C_{L\bf{m}} \, L! }{m_1! m_2! \ldots m_d!}
a_1^{\dagger m_1} a_2^{\dagger m_2} \ldots a_d^{\dagger m_d} 
|N\rangle_{a_1}|0\rangle_{a_2}\ldots|0\rangle_{a_d}|g_{\bf{m}}\rangle
\nonumber \\
&\propto& \sum_{\bf{m}} \sqrt{N+m_1 \choose N} \,
|N+m_1\rangle_{a_1}|m_2\rangle_{a_2}\ldots|m_d\rangle_{a_d}|g_{\bf{m}}\rangle.
\end{eqnarray*}
Since this state is fully equivalent to the state (\ref{UCdcquditoutput}), the cloning is optimal. 
Although this result was obtained within the framework of a perturbation theory, a
detailed analysis reveals that it holds for any interaction strength. It can be also shown
that if only $M'-N<L$ atoms emit photons and the rest of the  atoms remain in the excited
state, then $M'$ optimal photonic clones are generated (\cite{Fan2002b}). 

A proof-of principle experiment on cloning via stimulated emission
was reported by \cite{Fasel2002}. In that work, a commercially available polarization
insensitive Erbium doped fiber amplifier was utilized. The amplifier was injected with a 
weak vertically polarized coherent signal with mean photon number $\bar{n}_{\mathrm{in}}$.
After the amplification, the output mean numbers $\bar{n}_V$ and $\bar{n}_H$ 
of vertically and horizontally polarized photons were measured. The fidelity of the
amplification process can be simply defined as $F=\bar{n}_V/(\bar{n}_V+\bar{n}_H)$. 
The output mean intensities depend linearly on the input intensity (\cite{Shimoda57}),
\begin{equation}
\bar{n}_V=G \bar{n}_{\mathrm{in}}+ \frac{1}{Q}(G-1), \qquad \bar{n}_H=\frac{1}{Q}(G-1).
\label{UCamplification}
\end{equation} 
Here $G$ is the gain of the amplifier and $Q$ is a factor depending on the properties of
the amplification process. The term $G \bar{n}_{\mathrm{in}}$ represents the amplified 
injected input signal while $(G-1)/Q$ represents the noise arising due to spontaneous 
emission. For quantum-noise limited amplification, $Q=1$.
From Eqs. (\ref{UCamplification}) we can express $G$ in terms of $\bar{n}_{\mathrm{in}}$, 
$\bar{n}_{\mathrm{out}}=\bar{n}_V+\bar{n}_H$ 
and Q, and we find $G=(Q \bar{n}_{\mathrm{out}}+2)/(Q \bar{n}_{\mathrm{in}}+2)$. 
On inserting this expression into the formula for the fidelity, we obtain
\begin{equation}
F=\frac{Q \bar{n}_{\mathrm{out}}\bar{n}_{\mathrm{in}}+\bar{n}_{\mathrm{out}}
+\bar{n}_{\mathrm{in}}}{Q \bar{n}_{\mathrm{out}}\bar{n}_{\mathrm{in}}+2\bar{n}_{\mathrm{out}}}.
\label{UCamplificationfidelity}
\end{equation}
If we formally replace $\bar{n}_{\mathrm{in}}$ with $N$ (the number of input copies) 
and $\bar{n}_{\mathrm{out}}$ with $M$ (the number of output clones), then for $Q=1$ the 
formula (\ref{UCamplificationfidelity}) becomes the optimal fidelity of $N \rightarrow M$ 
cloning of qubits. Experimentally, $G=1.3$ and  $Q=0.8$ was observed which is quite 
close to the optimal value $Q=1$. For instance, the fidelity of $1 \rightarrow 2$ cloning 
for $Q=0.8$ inferred from (\ref{UCamplificationfidelity}) reads $F=0.821$ which
is only slightly lower than the optimal fidelity $F=5/6 \approx 0.833.$

\subsection{Symmetrization}
\label{secUCOsymmetrization}

We have seen in Section \ref{secUCoptimalmap} that the optimal universal 
$N \rightarrow M$ quantum cloning can be accomplished by symmetrizing the state 
of $N$ input copies and $M-N$ maximally mixed states. Since photons are bosons, 
the projection onto symmetric subspace can be easily carried out with the use 
of linear optics, namely by mixing the $M$ photons on an
array of $M-1$ beam splitters and selecting only the events when all photons 
are collected in a single spatial mode. 

\input #progressfig8

Let us first illustrate this method on the example of $1\rightarrow 2$ cloning of
polarization states of photons (\cite{Ricci2004,Irvine2004,Sciarrino2004b}). The setup is 
schematically illustrated in Fig. \ref{figsymmetrization}(a). The photon in mode A whose state is to be 
cloned is combined on a balanced beam splitter BS$_1$ with a blank copy photon prepared 
in a maximally mixed state. Only the cases when both photons leave the beam splitter 
in the left output mode are post-selected and the
two clones are spatially separated by an auxiliary balanced beam splitter BS$_2$. At the
heart of cloning via symmetrization is the Hong-Ou-Mandel effect (\cite{Hong87}). If two photons with
identical polarization state interfere on a balanced beam splitter, then they both end up 
in the same spatial mode and one does not observe any coincidences of one photon in
mode A and one in mode B. So for the input $|\psi\rangle_A|\psi\rangle_B$, there is
probability $1/2$ of having two photons in the left output port and probability $1/2$ of splitting
them in the two output modes A' and B'. Altogether, the conditional transformation reads 
\begin{equation}
|\psi\rangle_{A}|\psi\rangle_{B}\rightarrow 
\frac{1}{2}|\psi\rangle_{A'}|\psi\rangle_{B'},
\label{UCsymparallel}
\end{equation}
On the other hand, if the two photons are initially in orthogonal polarization states,
$|\psi\rangle_A|\psi_\perp\rangle_B$, then they are distinguishable and do not interfere
on BS$_1$. With probability $1/4$, the  photon in mode A is reflected and the photon in mode
$B$ is transmitted and they are both in the left output. Again, there is probability $1/2$  that
the two photons will be divided on a balanced beam splitter BS$_2$. Since the photon in
state $|\psi\rangle$ can be either reflected or transmitted on BS$_2$, the final state of
photons in modes $A'$ and $B'$ is a balanced superposition of these two possibilities,
namely a symmetric state, 
 \begin{equation}
|\psi\rangle_{A}|\psi_{\perp}\rangle_{B} \rightarrow \frac{1}{4}(
|\psi\rangle_{A'}|\psi_{\perp}\rangle_{B'}+|\psi_{\perp}\rangle_{A'}
|\psi\rangle_{B'}).
\label{UCsymorthogonal}
\end{equation}
Since the projector onto the symmetric subspace acts as 
$\Pi_{+}|\psi\rangle|\psi\rangle=|\psi\rangle|\psi\rangle$ and
$\Pi_{+}|\psi\rangle|\psi_\perp\rangle=\frac{1}{2}(|\psi\rangle|\psi_\perp\rangle
+|\psi_\perp\rangle|\psi\rangle)$, it immediately follows from 
Eqs.~(\ref{UCsymparallel}) and (\ref{UCsymorthogonal}) that the
setup shown in Fig.~\ref{figsymmetrization}(a) implements with probability $1/4$ the projection onto the 
symmetric subspace followed by a spatial separation of the two photons.

The maximally mixed polarization state in mode $B$ can be obtained for instance by
preparing the blank copy photon in state $|V\rangle$ or $|H\rangle$ with probability 
$1/2$ each. Another, more intriguing, option is to send into the port B one part of the
maximally entangled two-photon singlet state
$|\Psi^{-}\rangle=\frac{1}{\sqrt{2}}(|\psi\rangle_B|\psi_\perp\rangle_C-|\psi_\perp\rangle_B\psi\rangle_C).$
In this case, if the symmetrization succeeds, then we obtain in the spatial mode $C$
the optimal anti-clone of $|\psi\rangle$ i.e. a state that has a fidelity $2/3$ with
$|\psi_\perp\rangle$.
 
The optimal $1\rightarrow 2$ cloning based on symmetrization has been experimentally demonstrated 
by two groups (\cite{Ricci2004,Irvine2004}). In both experiments, the input photon whose 
state was cloned was obtained from a weak coherent beam, and it was combined on 
a balanced beam splitter with one photon from a maximally entangled  singlet 
state generated in a nonlinear crystal by means of spontaneous parametric down-conversion. 
The triple coincidence events were selected where there were two clones and one anticlone
present and the intensity of the weak coherent beam was adjusted such that the dominant
contribution to the triple coincidence events originated from the cases when there was a
single photon in the coherent beam and a single entangled photon pair was generated in the
nonlinear crystal. The observed mean cloning fidelities in these two experiments were 
$F=0.82$ (\cite{Ricci2004}) and $F=0.81$  (\cite{Irvine2004}), respectively. 
The simpler setup depicted in Fig. \ref{figsymmetrization}(a) and involving only two photons was also implemented
experimentally (\cite{Sciarrino2004a}). A single photon pair was generated in a nonlinear
crystal. One photon representing the input was prepared in the state
$|\psi\rangle$ using wave plates  while the other photon was randomly prepared in the
state $|V\rangle$ or $|H\rangle$. This experiment is much simpler than the previous one,
because only two-photon coincidence events were observed instead of tree-photon
coincidences. This resulted in much higher rate of cloning, and also in better visibility
and mean cloning fidelity $F=0.826$ very close to the theoretical maximum $F=0.833$.

An extension of the symmetrization procedure to $M$ photons is illustrated in Fig.
\ref{figsymmetrization}(b).
The photons are combined on an array of $M-1$ beam splitters $BS_j$ and
the symmetrization succeeds if all $M$  photons are bunched in the same 
spatial mode (\cite{Sciarrino2004b}). To confirm this we 
can split the output signal into $M$ different spatial modes 
using another array of $M-1$ beam splitters BS$_j^\prime$ and postselect only 
events when each of $M$ photodetectors PD registers one photon.

We now demonstrate that the array of beam splitters accomplishes the desired projection
onto the symmetric subspace. The symmetric  two-mode $L$-photon states  $|L,k\rangle$ 
with $L-k$ photons polarized vertically  and $k$ photons
polarized horizontally form a basis in the symmetric
space ${\mathcal{H}}_{+}^{\otimes L}$. We prove our claim by induction. Consider the $L$th
beam splitter BS$_L$ in the scheme in Fig. \ref{figsymmetrization}(b).  The state impinging from the left is a
symmetric $L$-photon state while a single photon impinges on BS$_L$ from the
bottom. The beam splitter BS$_L$ need not be balanced but its transmittance $t$ and
reflectance $r$ should be independent of the polarization. 
In the Heisenberg picture, the mixing of the modes on the beam splitter is described by
 linear input output canonical transformations of the creation operators,
\begin{equation}
a_{V,\mathrm{out}}^\dagger= r a_{V,\mathrm{in}}^\dagger +t b_{V,\mathrm{in}}^\dagger, 
\qquad b_{V,\mathrm{out}}^\dagger= r b_{V,\mathrm{in}}^\dagger -t a_{V,\mathrm{in}}^\dagger,
\label{UCsymbeamsplitter}
\end{equation}
and similar formulas hold also for horizontal polarization. The state transformation 
on a beam splitter can be most easily determined by expressing all states in terms of
the creation operators acting on the vacuum,  
\begin{equation}
|L,k\rangle=\frac{1}{\sqrt{k! (L-k)!}}\,  a_{H,\mathrm{in}}^{\dagger k}a_{V,\mathrm{in}}^{\dagger L-k}|\mathrm{vac}\rangle,
\quad |V\rangle=b_{V,in}^{\dagger} |\mathrm{vac}\rangle, \quad
 |H\rangle=b_{H,in}^{\dagger}|\mathrm{vac}\rangle.
 \label{UCsymcreation}
\end{equation}
From Eqs. (\ref{UCsymbeamsplitter}) we express the ``in'' operators as linear 
combinations of the ``out'' operators and substitute into the formulas (\ref{UCsymcreation}). 
Using this technique it is easy to show that if all $L+1$ photons bunch in 
the right output mode then the following conditional transformation takes place:
\begin{eqnarray}
|L,k\rangle|V\rangle &\rightarrow & t_L^L r_L \sqrt{L+1-k}|L+1,k\rangle,
\nonumber \nonumber \\
|L,k\rangle|H\rangle &\rightarrow & t_L^L r_L \sqrt{k+1}|L+1,k+1\rangle.
\label{MkVH}
\end{eqnarray}
Consider now the projection of the states $|L,k\rangle|V\rangle$ and 
$|L,k\rangle|H\rangle$ onto the symmetric subspace of $L+1$ photonic qubits. 
One finds  that
\begin{eqnarray}
\Pi_{+,L+1}|L,k\rangle|V\rangle=\sqrt{\frac{L+1-k}{L+1}}|L+1,k\rangle,
\nonumber \nonumber \\
\Pi_{+,L+1}|L,k\rangle|H\rangle=\sqrt{\frac{k+1}{L+1}}|L+1,k+1\rangle.
\label{MkVHproj}
\end{eqnarray}
The transformations (\ref{MkVH}) and (\ref{MkVHproj}) are equivalent up to 
state-independent prefactor $\sqrt{L+1} \, t_L^L r_L$ which proves that 
the array of $M-1$ beam splitters BS in Fig. \ref{figsymmetrization}(b) projects the
input states onto  the symmetric subspace of $M$ qubits. 
The probability of success of the projection 
can be determined by comparing the coefficients in Eqs. (\ref{MkVH}) 
and (\ref{MkVHproj}) and we find
\begin{equation}\label{PSNdef}
P=P_S M! \prod_{j=1}^{M-1} T_j^j (1-T_j),
\end{equation}
where $T_j=t_j^2$ and $P_S= \mathrm{Tr[\Pi_{M}^{+}\rho_{in}]}$ is the overlap of 
the input $M$-photon state $\rho_{\mathrm{in}}$ with projector onto the symmetric subspace. 
The optimal transmittance $T_j$ of the $j$th beam splitter
leading to maximal $P$ can be obtained by maximizing $T_j^j(1-T_j)$ which yields
$T_{j,\mathrm{opt}}=j/(j+1).$
Note that $T_{j,\mathrm{opt}}$ does not depend on the input 
$N$-photon state. On inserting the optimal $T_j$ into Eq. (\ref{PSNdef}) we get
\begin{equation}
P_{\mathrm{opt}}=P_S \frac{M!}{M^{M}}.
\end{equation}
Recently, the optimal universal $1\rightarrow 3$ and $2\rightarrow 3$ cloning of
polarization states of photons via symmetrization has been demonstrated
experimentally in \cite{Masullo2004}. The three photons used in the
experiment consisted of a pair of photons generated in the process of
spontaneous parametric downconversion and a single photon in very weak
coherent beam. These three photons were combined on two beam splitters
and only the events where all photons bunched in a single spatial mode
were chosen by post selection.  Wave plates, a polarizing beam
splitter, an array of beam splitters and photodetectors were employed
to analyze the clones. The experimentally observed fidelity of the
$1 \rightarrow 3$ cloning was $F_{1\rightarrow 3}^{\mathrm exp}=0.758$ which is very close to the
theoretical maximum $F_{1\rightarrow 3}^{\mathrm th}=7/9 \approx 0.778$. The observed fidelity
of $2\rightarrow 3$ cloning, $F_{2\rightarrow 3}^{\mathrm exp}=0.894$, was
also close to the optimum value $F_{2\rightarrow 3}^{\mathrm th}=11/12
\approx 0.917$.

The symmetrization on a beam splitter can be naturally extended to qudits. Symmetrization
of two photonic qudits represented by a state of a photon in $d$ different spatial modes would
require an array of $d$ balanced beam splitters, each mixing the $j$-th mode of the first
and second qudits. It may be more advantageous to work with time-bin qudits, where the
symmetrization would require only one balanced beam splitter where the two photons would
interfere. Similarly as before, only the events when the two photons bunch and leave 
the beam splitter in the same spatial port have to be post-selected.

\subsection{Universal asymmetric cloning of photons}
\label{secUCOasymmetry}

So far, we have presented various optical implementations of symmetric cloning machines. 
In this section we will consider optimal $1\rightarrow 2$ asymmetric cloning of qubits. 
We will describe two methods, both based on the interference of photons on unbalanced beam splitters. 
The first approach, introduced in Section~\ref{sub-section-Heisenberg},
is to start from the output of the optimal symmetric cloner
and convert it into an output of the optimal asymmetric cloner, which is given by
\begin{equation}
|\Psi\rangle=
\frac{1}{\sqrt{2-2p+2p^2}}\left[|\psi\rangle_A|\psi\rangle_B|\psi_\perp\rangle_C-
p|\psi\rangle_A|\psi_\perp\rangle_B|\psi\rangle_C 
-(1-p)|\psi_\perp\rangle_A|\psi\rangle_B|\psi\rangle_C\right].
\label{UCasymmetric}
\end{equation}
Here $p\in [0,1]$ is an asymmetry parameter and the fidelities of the clones in qubits $A$
and $B$ read
\begin{equation}
F_A= 1-\frac{(1-p)^2}{2(1-p+p^2)}, \qquad F_B= 1-\frac{p^2}{2(1-p+p^2)}.
\end{equation}
The symmetric cloner is recovered when  $p=1/2$ and we have
\begin{equation}
|\Psi\rangle_{\mathrm{sym}}=
\frac{1}{\sqrt{6}}\left[2|\psi\rangle_A|\psi\rangle_B|\psi_\perp\rangle_C-
|\psi\rangle_A|\psi_\perp\rangle_B|\psi\rangle_C 
-|\psi_\perp\rangle_A|\psi\rangle_B|\psi\rangle_C\right].
\end{equation}
Suppose first that the second clone (qubit B) and the anti-clone (qubit C) 
are projected on the singlet state  $|\Psi^{-}\rangle$. We obtain
\begin{equation}
I_A\otimes \Pi_{BC}^{-} |\Psi\rangle_{\mathrm sym}=\frac{\sqrt{3}}{2} \, |\psi\rangle_A |\Psi^{-}\rangle_{BC},
\label{UCsymreversal}
\end{equation}
where $\Pi_{BC}^{-}=|\Psi^{-}\rangle\langle \Psi^{-}|$.
The original input state $|\psi \rangle$ is perfectly recovered in the qubit $A$.  
The projection on singlet forms a part of the Bell measurement, i.e. a
measurement in the basis of four maximally entangled Bell states. There is an
interesting analogy between Eq. (\ref{UCsymreversal}) and the process of 
quantum teleportation (\cite{Bennett93,Bouwmeester97,Boschi98,Marcikic2003}). In quantum
teleportation, two parties, Alice and Bob, share a maximally entangled singlet state.
Alice wants to send to Bob an unknown state $|\psi\rangle$ using only classical
communication and shared entanglement as a resource. To achieve this, Alice carries out a
Bell measurement on the state $|\psi\rangle$ and her part of the shared singlet state. 
She communicates the measurement outcome (two classical bits) to Bob, who accordingly
applies one of four different unitary transformations to his part of the singlet. In this
way, the state is teleported from Alice to Bob.

The cloning can be deterministically reversed by performing Bell measurement on 
one of the clones and the anti-clone and applying an appropriate correcting unitary to
the first clone (\cite{Bruss2001a}). In the Bell measurement, the singlet is detected with
probability $3/4$ while each of the triplet Bell states is detected with probability
$1/12$, independently of the input state.  

\input #progressfig9

This full reversal of cloning can be generalized to a partial reversal which converts the 
symmetric cloner to asymmetric one (\cite{Filip2004a}). The idea is to apply to qubits 
B and C a filter  $\Pi_{BC}^{-}+a \Pi_{BC}^{+}$, where $a\in[0,1]$ controls the asymmetry. If
$a=0$, then we get projection on singlet and full reversal, while for $a=1$ the two qubits
are multiplied by identity and nothing happens. Let us now consider arbitrary $a$.
 The state after filtering,
\begin{equation}
|\Psi_{\mathrm{proj}}\rangle=I_A \otimes (\Pi_{BC}^{-}+a \Pi_{BC}^{+})|\Psi_{\mathrm{sym}}\rangle, 
\label{UCasymfiltration}
\end{equation}
can be after normalization expressed as follows, 
\begin{eqnarray}
|\Psi_{\mathrm{proj}}\rangle &= &\frac{1}{\sqrt{6(a^2+3)}}\left[
(3+a)|\psi\rangle_A|\psi\rangle_B|\psi_\perp\rangle_C-
(3-a)|\psi\rangle_A|\psi_\perp\rangle_B|\psi\rangle_C \right. \nonumber \\
 && \left.-2a|\psi_\perp\rangle_A|\psi\rangle_B|\psi\rangle_C\right]
\end{eqnarray}
We can immediately see that this state coincides with the outcome of the optimal
asymmetric cloner (\ref{UCasymmetric}) and  $p=(3-a)/(3+a)$.

For optical polarization qubits, the filtration (\ref{UCasymfiltration})  can be implemented by letting the two
photons interfere on an unbalanced beam splitter and post-selecting only the events when a
single photon is detected in each output  port. There are two ways how the photons can
exit the beam splitter in different spatial modes. Either both photons are reflected or
they are both transmitted. The unitarity dictates that these two alternatives acquire a
mutual phase shift $\pi$. If the two photons are in the same state $|\psi\rangle$, then
these two alternatives interfere destructively, while if the photons are in orthogonal
polarization states then there is no interference. The resulting conditional
transformation reads 
\begin{equation}
|\psi\psi\rangle_{BC} \rightarrow  (R-T) |\psi\psi\rangle_{BC}, \qquad 
|\psi\psi_\perp\rangle \rightarrow R|\psi\psi_\perp\rangle-T |\psi_\perp \psi\rangle.
\end{equation}
It follows that the unbalanced beam splitter applies the filter
$\Pi^{-}+a \Pi^{+}$ with $a=R-T$. A schematic setup of the proposed
asymmetric cloning experiment is shown in Fig. \ref{figasymmetryfilip1}. Optimal symmetric cloning is
accomplished by stimulated parametric downconversion as discussed in detail in Sec.
\ref{secUCOdownconversion}. At the output, the two clones are separated on an auxiliary balanced beam splitter
and one of the clones is combined with the anti-clone on an unbalanced beam splitter. 
Successful asymmetric cloning is heralded by a coincident observation of a single photon
in each mode A, B, and C.

\input #progressfig10

The second  scheme for optimal asymmetric cloning (\cite{Filip2004b}) resembles 
very closely the scheme for teleportation of polarization states of photons, 
see Fig. \ref{figasymmetryfilip2}. The only difference is that the balanced beam 
splitter which is used in teleportation to perform a Bell analysis is 
replaced with an unbalanced beam splitter that conditionally applies 
the filter $\Pi_{-}+a\Pi_{+}$. The cloning succeeds if a single photon 
is detected in each mode A, B, and C. The initial state in the scheme 
shown in Fig. \ref{figasymmetryfilip2} is
$|\psi\rangle_B |\Psi^{-}\rangle_{AC}$ and after the interference on a beam splitter and
postselection we get 
\begin{equation}
|\tilde{\Psi}_{\mathrm{proj}}\rangle=I_A \otimes(a \Pi_{BC}^{+}+\Pi_{BC}^{-}) |\psi\rangle_B |\Psi^{-}\rangle_{AC} 
\end{equation}
After some algebra we arrive at
\begin{eqnarray}
|\tilde{\Psi}_{\mathrm{proj}}\rangle & \propto & 
\frac{1}{\sqrt{2(1+3a^2)}} \left[
(1+a)|\psi\rangle_A|\psi\rangle_B|\psi_\perp\rangle_C-
(1-a)|\psi\rangle_A|\psi_\perp\rangle_B|\psi\rangle_C  \right. \nonumber \\
& & \left. -2a|\psi_\perp\rangle_A|\psi\rangle_B|\psi\rangle_C \right].
\end{eqnarray}
This is again the output state of the optimal asymmetric cloning machine with
$p=(1-a)/(1+a)$ so the asymmetric cloning can be implemented by means of a partial
teleportation. An interesting feature of this scheme is that one of the clones is
teleported from Alice to Bob so we can speak about cloning at a distance.

The universal asymmetric cloning of polarization states of single
photons has been demonstrated experimentally by \cite{Zhao2005}
following the scheme illustrated in Fig. \ref{figasymmetryfilip2}.
In that experiment, a Mach-Zehnder interferometer  acted as an effective 
unbalanced beam splitter whose transmittance could be controlled by changing the
relative path difference between the two arms of the interferometer.
In this way it was possible to demonstrate the whole class of the
asymmetric  $1 \rightarrow 2$ cloning machines.

\subsection{Cloning of orthogonally-polarized photons}
\label{secCloningpairorthogonal}

It was shown in Section \ref{secUCoptimalmap} that the optimal universal quantum cloning and optimal 
quantum state estimation are closely related and that in the limit of infinite number of
clones the fidelity of cloning is equal to the fidelity of optimal state estimation.
In this context, a very interesting and surprising observation
was made by \cite{Gisin99}, who found that the state of a single qubit can be  better
estimated from  the state $|\psi\rangle|\psi_\perp\rangle$ than from the state
$|\psi\rangle|\psi\rangle$. Picturing the qubits as spin-$\frac{1}{2}$ particles, we can say that
the information about the direction  is better encoded in two anti-parallel
spins than in two parallel ones. The fidelity of the estimation of $|\psi\rangle$ from a
single copy of the two-qubit state $|\psi\rangle|\psi_\perp\rangle$ reads
(\cite{Gisin99,Massar2000}), 
\begin{equation}
F_{\perp}=\frac{1}{2}\left(1+\frac{1}{\sqrt{3}}\right) \approx 0.789,
\label{OCestimationfidelity}
\end{equation}
which is slightly higher than the fidelity of optimal estimation from
$|\psi\rangle|\psi\rangle$, $F=3/4$.
Motivated by this observation we may expect that this advantage of two anti-parallel
spins over two parallel ones extends also to the cloning. This 
is indeed the case, provided that the number of clones is large enough. We shall now 
describe the optimal universal cloning transformation which produces $M$ approximate clones 
of the state $|\psi\rangle$ from a single replica of $|\psi\rangle|\psi_\perp\rangle$
and maximizes the single-clone fidelity. 

Making the natural assumption 
that the output Hilbert space is the symmetric subspace of $M$ qubits, the optimal cloning 
CP map $\mathcal{S}$ can be determined analytically for any $M$
(\cite{Fiurasek2002}). The mean single-clone 
fidelity can  be expressed as  $F=\mathrm{Tr}[SR]$, where  $R$ has a rather complicated form
and can be found in (\cite{Fiurasek2003a}). In contrast to the universal cloning with 
input state $|\psi\rangle^{\otimes N}$, the maximum fidelity cannot be determined 
from the  maximum eigenvalue of $R$, and one has to 
solve the extremal Eqs. (\ref{Soptimality}) and prove the optimality by checking 
that the inequality (\ref{lambdaoptimality})  is satisfied. 

Since the input state of the cloner can be obtained as an orbit of the group SU(2),
$|\psi\rangle|\psi_{\perp}\rangle=U\otimes U |0\rangle|1\rangle$, the optimal cloner is
covariant and can be expressed as follows,
\begin{equation}
|\psi,\psi_\perp\rangle \rightarrow
\sum_{j=0}^M \alpha_{j,M} |(M-j)\psi,j\psi_\perp\rangle
\otimes |(M-j)\psi_\perp,j\psi\rangle,
\label{OCcloner}
\end{equation}
where the coefficients $\alpha_{j,M}$ are given by
\begin{equation}
\alpha_{j,M}=(-1)^j \left[\frac{1}{\sqrt{2(M+1)}}
                    +\frac{\sqrt{3}(M-2j)}{\sqrt{2M(M+1)(M+2)}}\right].
\label{alfa}
\end{equation}
The cloning machine (\ref{OCcloner}) is symmetric with respect to the interchange 
of $|\psi\rangle$ and $|\psi_\perp\rangle$. The cloner requires an ancilla 
whose size is the same as the size of the output Hilbert space, i.e. the ancilla Hilbert
space is also a symmetric subspace of $M$ qubits. The ancilla contains $M$ approximate
copies of the state $|\psi_\perp\rangle$ and the fidelity of these anti-clones 
is the same as the fidelity of the clones. The single-clone fidelity can be calculated as
weighted average of the coefficients $\alpha_{j,M}^2$,
\begin{equation}
F_\perp(M)=\sum_{j=0}^M \frac{M-j}{M}\alpha_{j,M}^2,
\end{equation}
and after a simple algebra we arrive at
\begin{equation}
F_{\perp}(M)= \frac{1}{2}\left(1+\sqrt{\frac{M+2}{3M}}\right).
\label{OCfidelity}
\end{equation}
The fidelity monotonically decreases with the increasing number of clones $M$ and in the
limit $M\rightarrow \infty$ we recover the fidelity (\ref{OCestimationfidelity}) of the optimal state estimation 
from $|\psi\rangle|\psi_\perp\rangle$. 
Upon comparing the fidelity $F_\perp(M)$ to the fidelity of the optimal 
cloner for a pair of identical qubits, $F_{||}(M)=(3M+2)/(4M)$, we see that
$F_{||}(M) \geq F_{\perp}(M)$ for $M \leq 6$, while
$F_{\perp}(M) > F_{||}(M)$ for $M > 6$ and the cloner (\ref{OCcloner})
outperforms the standard  universal cloner.

\input #progressfig11

We have seen in Section \ref{secUCOdownconversion} that the optimal universal cloning of polarization states of
photons can be realized by means of stimulated parametric down-conversion. It turns out
that also the optimal cloning with a pair of orthogonal qubits as the input can be
performed in the same way, if the photons in states $|\psi\rangle$ and $|\psi_\perp\rangle$
are fed to the input signal and idler ports of the amplifier, respectively, as
schematically illustrated in figure \ref{figorthogonalqubits}. 
We assume that the parametric amplification in
the nonlinear crystal C$_3$ is governed by the singlet-type Hamiltonian
(\ref{UCsingletHamiltonian}), which is
invariant under the simultaneous rotation of the signal and idler qubits, 
$(U\otimes U) H(U^\dagger \otimes U^\dagger)=H$.  
Assuming the input state $|1\rangle_{aV}|0\rangle_{aH}|0\rangle_{bV}|1\rangle_{bH}$,
the output state after the amplification in the crystal C$_3$ reads,
\begin{equation}
|\Psi_{\mathrm{out}}\rangle=\sum_{M=0}^\infty \lambda^{M-1}(1-\Gamma^2) |\Psi_{\perp,M}\rangle,
\label{OCpsiout}
\end{equation}
where the state with $M$ clones and $M$ anti clones is given by
\begin{equation}
|\Psi_{\perp,M}\rangle= \sum_{j=0}^M (-1)^{j} \left[(M-j)(1-\lambda^2)-\lambda^2\right]
|M-j\rangle_{aV} |j\rangle_{aH} |j\rangle_{bV} |M-j\rangle_{bH},
\label{OCpsiperpM}
\end{equation}
$\lambda=\tanh(\kappa t)$ and $t$ is the effective interaction time. 

In contrast to universal cloning with $N$ identical photons at the input, the state
$|\Psi_{\perp,M}\rangle$ depends on the strength of the parametric amplification
$\lambda$. The cloner that produces $M$ copies is obtained by  post-selecting only the
events when exactly $M$ photons are detected in signal and idler spatial modes, which
corresponds to the selection of the state $|\Psi_{\perp,M}\rangle$ from the superposition
(\ref{OCpsiout}). The fidelity of the cloner depends on  $\lambda$,
\begin{equation}
F_{\perp}(M,y)=\frac{3 y^2-2y(2M+1)+\frac{3}{2}M(M+1)}{6y^2-6My+M(2M+1)},
\label{OCdcfidelity}
\end{equation}
where $y=\lambda^2/(1-\lambda^2)=\sinh^2(\kappa t)$. The optimal parametric gain which
 maximizes the fidelity (\ref{OCdcfidelity}) can be found by solving the equation 
$\frac{\partial F_{\perp}(M,y)}{\partial y}=0$ which yields
\begin{equation}
y_{\rm opt}=\frac{M}{2}-\frac{1}{2}\sqrt{\frac{M(M+2)}{3}}.
\label{yopt}
\end{equation}
On inserting the optimal $y$ into Eq. (\ref{OCdcfidelity}) we recover the fidelity
(\ref{OCfidelity}) hence
the optimal cloning of a pair of orthogonal qubits can be achieved by means
of stimulated parametric down-conversion with properly chosen gain.

\section{Phase-covariant cloning of photons}
\label{section-phase-covariant}

Up to now we have studied universal cloning machines that clone equally
well all states. In many situations, however, one deals only with a
subset of the states. An archetypal example is the Bennett-Brassard 1984
protocol for quantum key distribution (\cite{BB84}), which utilizes four
non-orthogonal states $|0\rangle$, $|1\rangle$,  $|0\rangle +|1\rangle$,
and $|0\rangle-|1\rangle$. If we restrict the range of the admissible input
states of the cloning machine, then we can expect that the machine will
exhibit better performance than the universal cloner and will reach
higher fidelity. In this section we shall study the phase-covariant cloning
machines which optimally clone all states that are balanced
superpositions of the computational basis states,
\begin{equation}
|\psi\rangle=\frac{1}{\sqrt{d}} \sum_{j=0}^{d-1} e^{i\phi_j} |j\rangle,
\label{PCqudit}
\end{equation}
where the phases $\phi_j$ can be arbitrary.

\subsection{Phase-covariant cloning of qubits}
\label{secPCqubits}

The simplest and perhaps most important example is the $1\rightarrow 2$
 phase-covariant cloning machine which can be used as the optimal
individual eavesdropping attack on the BB84 protocol (\cite{Fuchs97}). In contrast to the
universal cloners, the optimal cloning transformation depends on whether
the single-clone fidelity or the global fidelity are taken as a figure
of merit that should be maximized. In the context of eavesdropping on
quantum key distribution protocol, it is natural to consider the
single-clone fidelities, since they quantify the amount of information
transmitted to the receiver and gained by the eavesdropper.

The optimal symmetric $1\rightarrow 2$ cloning transformation for qubits
that maximizes the single-clone fidelity  has the following form 
(\cite{Bruss2000b,Bruss2001b,Fan2002b}),
\begin{eqnarray}
|0\rangle|A_{\mathrm{in}}\rangle \rightarrow
\frac{1}{\sqrt{2}}|0\rangle_A|0\rangle_B|0\rangle_C +
\frac{1}{2}(|0\rangle_A|1\rangle_B+|1\rangle_A|0\rangle_B)|1\rangle_C,
\nonumber \\
|1\rangle|A_{\mathrm{in}}\rangle \rightarrow
\frac{1}{\sqrt{2}}|1\rangle_A|1\rangle_B|1\rangle_C +
\frac{1}{2}(|0\rangle_A|1\rangle_B+|1\rangle_A|0\rangle_B)|0\rangle_C.
\label{PCcloningmapqubits12}
\end{eqnarray}
The two clones are stored in qubits $A$ and $B$ and
the fidelity of each clone reads $F_{1\rightarrow 2}^{\mathrm pc}=(1+1/\sqrt{2})/2 \approx 0.855$ which
is indeed slightly higher than the fidelity of the optimal universal $1\rightarrow
2$ cloner for qubits, $F_{1\rightarrow 2}^{\mathrm univ}=5/6 \approx 0.833$. Note that besides a blank
copy qubit, the transformation also requires another  ancilla qubit C.
However, in contrast to universal cloning, this ancilla is not necessary
and one can design a simplified cloning transformation which achieves
the same fidelity and requires only two qubits: the input and a blank
copy (\cite{Niu99,Durt2004}). This is very important from the experimental point of view since
it is much easier to realize a two-qubit transformation than three qubit
transformation. The economic phase covariant
cloner can be obtained by projecting the ancilla C on the basis
state $|0\rangle$ (or $|1\rangle$). If we project on $|0\rangle$, then
we get
\begin{equation}
|0\rangle_A|0\rangle_B \rightarrow |0\rangle_A|0\rangle_B,
\qquad
|1\rangle_A|0\rangle_B \rightarrow
\frac{1}{\sqrt{2}}(|0\rangle_A|1\rangle_B+|1\rangle_A|0\rangle_B).
\label{PCeconomic12}
\end{equation}
An alternative economic cloning transformation can be obtained from
(\ref{PCeconomic12}) by exchanging $0$ and $1$. 
Interestingly, the cloning machine  (\ref{PCeconomic12}) is optimal
not only for the states on the equator of the Poincar\'{e} sphere but also
for all the states on the northern hemisphere, i.e. all states
$\cos(\theta/2)|0\rangle+ e^{i\phi}\sin(\theta/2)|1\rangle$ with
$\theta \leq \pi/2$ (\cite{Fiurasek2003b}). The optimal asymmetric cloning machine which
produces two clones with different fidelities $F_A$ and $F_B$ is obtained
by breaking the symmetry in the  output superposition of $|10\rangle$ and
$|01\rangle$,
\begin{equation}
|0\rangle_A|0\rangle_B \rightarrow |0\rangle_A|0\rangle_B,
\qquad
|1\rangle_A|0\rangle_B \rightarrow
\cos\vartheta|0\rangle_A|1\rangle_B+\sin\vartheta|1\rangle_A|0\rangle_B,
\label{niu-griff-asymm}
\end{equation}
and the two fidelities can be expressed as follows,
\begin{equation}
F_A=\frac{1}{2}(1+\sin\vartheta), \qquad
F_B=\frac{1}{2}(1+\cos\vartheta), \qquad \quad \vartheta \in
\left[0,\frac{\pi}{2}\right].
\end{equation}

The phase-covariant cloning machine that maximizes the global two-qubit
fidelity has qualitatively similar structure as the cloner (\ref{PCcloningmapqubits12}),
\begin{eqnarray}
|0\rangle|A_{\mathrm{in}}\rangle \rightarrow
\frac{1}{\sqrt{3}}[|0\rangle_A|0\rangle_B|0\rangle_C +
(|0\rangle_A|1\rangle_B+|1\rangle_A|0\rangle_B)|1\rangle_C],
\nonumber \\
|1\rangle|A_{\mathrm{in}}\rangle \rightarrow
\frac{1}{\sqrt{3}}[|1\rangle_A|1\rangle_B|1\rangle_C +
(|0\rangle_A|1\rangle_B+|1\rangle_A|0\rangle_B)|0\rangle_C],
\end{eqnarray}
and it reaches fidelity $F_{1 \rightarrow 2}^{\mathrm pc,G}=3/4$ which is again higher
than the global fidelity of the universal cloner, $F_{1\rightarrow 2}^{\mathrm univ,G}=2/3$.

The phase-covariant cloning can be extended to the case when we posses
$N$ copies of the state and would like to prepare $M$ clones, $M>N$. The
optimal $1 \rightarrow M$ phase-covariant cloning machine
was determined  by \cite{Fan2001b}, who considered the single-clone
fidelity as the figure of merit. The structure of the cloning
transformation depends of the parity of $M$.
If $M$ is even, there exist two independent cloning transformations,
\begin{equation}
|0\rangle \rightarrow|M,M/2-1\rangle, \qquad
|1\rangle \rightarrow |M,M/2\rangle,
\label{PCcloningevenA}
\end{equation}
and
\begin{equation}
|0\rangle \rightarrow |M,M/2\rangle, \qquad
|1\rangle \rightarrow |M,M/2+1\rangle.
\label{PCcloningevenB}
\end{equation}
Note that there are in fact infinitely many cloning transformations
since any convex mixture of the operations  (\ref{PCcloningevenA}) 
and (\ref{PCcloningevenB}) is also optimal. On the other hand, 
if $M$ is odd then we get only one optimal transformation
\begin{equation}
|0\rangle \rightarrow |M,(M-1)/2\rangle, \qquad
|1\rangle \rightarrow |M,(M+1)/2\rangle.
\label{PCcloningodd}
\end{equation}
The resulting fidelity is
\begin{equation}
F= \left\{
\begin{array}{lcl} \displaystyle
\frac{1}{2} +\frac{\sqrt{M(2+M)}}{4M}, & & M~\mathrm{even}, \\[3mm]
\displaystyle \frac{1}{2} +\frac{M+1}{4M}, & & M~\mathrm{odd}.
\end{array}
\right.
\end{equation}
The optimality of the cloning transformations (\ref{PCcloningevenA}), 
(\ref{PCcloningevenB}) and (\ref{PCcloningodd}) can be proved
using the same method that was employed in Sec. \ref{secUCproof} to prove the
optimality of the $1 \rightarrow M$ universal cloning machine for
qubits. In particular, the single-clone fidelity can be expressed as
$F=\mathrm{Tr}[SR]$, where $S$ is the operator isomorphic to the cloning
CP map and
\begin{equation}
R=\frac{1}{4}I\otimes \Pi_{+,M} + \frac{1}{4}
(|0\rangle\langle 1| +|1\rangle\langle 0|)\otimes
\sum_{k=0}^{M-1} D_{M,k}(|M,k+1\rangle\langle M,k|
+|M,k\rangle\langle M,k+1|),
\end{equation}
where $D_{M,k}=\sqrt{(M-k)(k+1)}/M$.
The fidelity is upper bounded by  the maximum eigenvalue
$r_{\mathrm{max}}$ of $R$, $F \leq 2 r_{\mathrm{max}}$,
and this bound is saturated by the above phase covariant cloners.

\cite{Fan2001b} also conjectured the structure of the  general optimal
$N\rightarrow M$ phase covariant cloning transformation for  qubits.
The proposed generalization is very straightforward, namely,  every input
symmetric $N$-qubit state $|N,k\rangle$ is transformed to an $M$ qubit
symmetric state $|M,k+j\rangle$  with the constant $j$ adjusted such
that the fidelity is maximized.
If $N$ and $M$ have the same parity, $M=N+2L$,
then the suggested cloning map is $|N,j\rangle \rightarrow |M,j+L\rangle,$
and the corresponding fidelity is
\begin{equation}
F_{N\rightarrow M}^{\mathrm pc}= \frac{1}{2}+\frac{1}{M 2^N}
\sum_{j=0}^{N-1} \sqrt{{N \choose j}{N \choose j+1}}
\sqrt{\left(N+L-j\right)\left(L+j+1\right)}.
\label{PCcloningNMevenfidelity}
\end{equation}
When the parity of $M$ and $N$ differs, $M=N+2L+1$, then the two
possible cloning transformations  are either
$|N,j\rangle \rightarrow |M,j+L\rangle,$
or $|N,j\rangle \rightarrow |M,j+L+1\rangle,$
and the corresponding fidelity is
\begin{eqnarray}
F_{N \rightarrow M}^{\mathrm pc}&=& \frac{1}{2}+\frac{1}{M 2^{N+1}}
\sum_{j=0}^{N-1} \sqrt{{N \choose j}{N \choose j+1}}
\nonumber \\
&&\times [\sqrt{(N+L-j+1)(L+j+1)}+\sqrt{(L+j+2)(N+L-j)}].
\nonumber \\
\label{PCcloningNModdfidelity}
\end{eqnarray}
The optimality of the fidelity (\ref{PCcloningNMevenfidelity}) was proved 
in \cite{D'Ariano2003} by exploiting the generic theory of covariant cloning
machines, see \cite{D'Ariano2001b}. In contrast, 
if $N$ and $M$ have different parities,
the optimal phase-covariant cloning transformation that was found in
\cite{D'Ariano2003} differs from Eq.~(\ref{PCcloningNModdfidelity}).

\subsection{Phase-covariant cloning of qudits}
\label{secPCqudits}

Going beyond the cloning of qubits, the $1\rightarrow 2$ phase covariant
cloning of qudits (\ref{PCqudit}) has been investigated 
(\cite{Fan2003a,Lamoureux2005,Rezakhani2005}). 
It can be shown that the optimal cloning
transformation for qudits (\ref{PCqudit}) has the structure
\begin{equation}
|j\rangle \rightarrow \alpha |jj\rangle_{AB}|j\rangle_C
+\frac{\beta}{\sqrt{2(d-1)}}\sum_{l\neq j}^{d-1}
(|jl\rangle_{AB}+|lj\rangle_{AB})|l\rangle_C,
\label{PCquditmap}
\end{equation}
where $\alpha^2+\beta^2=1$. The two clones are contained in qudits A and B
while the qudit C serves as an ancilla. Note that Eq. (\ref{PCquditmap})
is a direct extension of the cloning transformation for qubits (\ref{PCeconomic12}).
The coefficients $\alpha$ and $\beta$ have to be optimized such that the
cloning fidelity is maximized. After some algebra one arrives at
\begin{equation}
\alpha=\left(\frac{1}{2}-\frac{d-2}{2\sqrt{d^2+4d-4}}\right)^{1/2}, \qquad
\beta=\left(\frac{1}{2}+\frac{d-2}{2\sqrt{d^2+4d-4}}\right)^{1/2},
\end{equation}
and the fidelity reads
\begin{equation}
F=\frac{1}{4}+\frac{1}{2d}+\frac{\sqrt{d^2+4d-4}}{4d}.
\end{equation}
In contrast to the phase covariant cloning of qubits, we cannot get rid
of the ancilla $C$ because if we project ancilla on the computational
basis state $|k\rangle$  then the conditional  map is not unitary.
So for $d>2$ it seems impossible to implement the optimal phase covariant
$1\rightarrow 2$ cloning in an economic way, without ancilla.


\subsection{Optical phase-covariant cloning}
\label{secPCoptical}

In contrast to the universal cloning, the optical experimental implementation of phase
covariant cloning machines has received much less attention. This may come as a surprise
in view of the apparent simplicity of the optimal cloning transformation 
(\ref{PCeconomic12}). However, the
phase-covariant cloning exhibits much less symmetry than the universal copying and the
methods such as stimulated amplification or symmetrization cannot be readily extended to
implement the $1\rightarrow 2$ phase covariant cloning machine.

\input #progressfig12

It is nevertheless possible to conditionally realize the $1\rightarrow 2$ phase
covariant cloning of photonic qubits with linear optics (\cite{Fiurasek2003b}). As usual, the
qubits are encoded into polarization states of single photons,
and the state to be cloned is a balanced superposition of vertical and horizontal
polarization, $|\psi\rangle=(|V\rangle+e^{i\phi}|H\rangle)/\sqrt{2}$.
Besides the input state, the cloning  requires also a second photon, the blank copy
which we assume to be initially prepared in the state $|V\rangle$.
Written in the basis of polarization states, the cloning transformation (\ref{PCeconomic12}) becomes
\begin{equation}
|VV\rangle \rightarrow |VV\rangle,   \qquad
|HV\rangle \rightarrow \frac{1}{\sqrt{2}}(|HV\rangle+|VH\rangle).
\label{PCopticalmap}
\end{equation}
The scheme of the cloning machine is illustrated in Fig. \ref{figphasecovariant}(a). The
input photon and the blank copy are combined on an unbalanced beam
splitter whose transmittance $t_j$ and reflectance $r_j$ for the vertical ($j=V$) and
horizontal ($j=H$) polarizations are different. Only the events when the two 
photons leave the beam splitter in different output ports are post-selected. 
The principle of operation of the cloner is easy to grasp. If the input $|\psi\rangle$ 
is in state  $|V\rangle$, the two photons at the output must be in state  $|VV\rangle$
since the blank copy is initially in the state $|V\rangle$. On the other hand, if the
input to be cloned would be in the state $|H\rangle$ then the beam splitter would
produce a superposition of $|HV\rangle$ and $|VH\rangle$. By 
properly choosing $r_j$ this superposition can be made balanced and the conditional map 
becomes exactly the unitary (\ref{PCopticalmap}). 

The mixing of the modes on a beam splitter is governed by the 
linear canonical transformations:
\begin{eqnarray}
a_{j,\rm out}^\dagger= r_j a_{j}^\dagger + t_j b_{j}^\dagger, \qquad
b_{j,\rm out}^\dagger= r_j b_{j}^\dagger  - t_j a_{j}^\dagger,
\end{eqnarray}
$j=V,H$ and $r_j^2+t_j^2=1$. 
The conditional transformation  corresponding to selecting only the events when there is 
one photon in the left output arm  (mode $a$) and one photon  in the right output 
arm (mode $b$) reads
\begin{equation}
|VV\rangle \rightarrow (r_V^2-t_V^2) |VV\rangle, \qquad
|HV\rangle \rightarrow r_H r_V |HV\rangle - t_H t_V |VH\rangle.
\end{equation}
This transformation becomes fully equivalent to Eq. (\ref{PCopticalmap}) if the following conditions are
satisfied,
\begin{equation}
r_V^2-t_V^2=\sqrt{2}r_H r_V=-\sqrt{2} t_H t_V.
\end{equation}
These constraints imply that $r_H=t_V$, $t_H=-r_V$ and 
$(r_V^2-t_V^2)=\sqrt{2} r_V t_V.$ On combining this equation
with the normalization $r_V^2+t_V^2=1$ we can determine $r_V$.
After simple algebra we obtain
\begin{equation}
r_V^2 = \frac{1}{2}\left(1+\frac{1}{\sqrt{3}}\right).
\label{RV}
\end{equation}
The probability of successful realization of the phase-covariant cloner is given by
\begin{equation}
P=(r_V^2-t_V^2)^2=\frac{1}{3}.
\end{equation}

The required beam splitter with different transmittances for  vertical and horizontal
polarizations can be simulated by a Mach-Zehnder interferometer with the polarization
dependent phase shifters in its arms, such as Soleil-Babinet compensators, so that the
phase shift and, consequently, the splitting ratio could be controlled independently for
vertical and horizontal polarizations. The setup could be also modified to work with 
a beam splitter whose reflectance is the same for both vertical and horizontal
polarizations. This alternative configuration is depicted in Fig. \ref{figphasecovariant}(b).
The signal and blank copy photons are first combined on a polarizing beam
splitter PBS that reflects vertically polarized photons and transmits
horizontally polarized photons. The two beams are then re-combined on a
beam splitter with reflectance $r$. If the signal photon
is initially vertically polarized, then a vertically polarized photon
enters each input port of  BS. If the signal photon is
polarized horizontally, then it is switched to the right arm and two
photons in orthogonal polarization states impinge on the right input port of
BS. The polarizing beam splitter ensures that the role of the transmittance and
reflectance for the horizontally polarized photon is interchanged with
respect to the scheme shown in Fig. \ref{figphasecovariant}(a).
It can be easily shown that this setup leads to the
cloning transformation (\ref{PCopticalmap}) provided that the
reflectance of the beam splitter is equal to $r^2=(1+1/\sqrt{3})/2$.

In the experiment, it may not be easy to precisely control the transmittance. 
It is therefore important to investigate how the performance of the setup shown 
in Fig. \ref{figphasecovariant}(b) depends on the reflectance $r$ of the beam splitter.
The cloning transformation remains phase-covariant  and the
cloning fidelity $F$ is the same for both clones and
does not depend on $\phi$. However, $F$  becomes a function of
$r$. After some algebra one  arrives at the formula for the fidelity of cloning of
equatorial qubits,
\begin{equation}
F=\frac{1}{2} \left[ 1+\frac{2r(2r^2-1)\sqrt{(1-r^2)}}{2r^4-2r^2+1}\right].
\label{FR}
\end{equation}
It turns out that  the cloning is rather robust with respect to the variations 
of the reflectance of the beam splitter and cloning fidelity $F>0.8$ can be achieved 
for a broad range of beam splitter  reflectances $0.7\leq r^2 \leq 0.9$.

In the experiment, the required pair of photons can be produced in spontaneous
Type-I parametric down-conversion and the desired initial states of the photons can be
prepared with the use of wave plates. After cloning, the states of the two clones can be
analyzed by a sequence of wave plates, polarizing beam splitters, and single photon
detectors, similarly as in the experiments on universal cloning. 

\subsection{Experimental 1-to-3 phase-covariant cloning}

Remarkably, while the optimal $1 \rightarrow 2$ phase covariant cloning
transformation (\ref{PCcloningmapqubits12}) has not yet been implemented for optical qubits, the optimal
$1\rightarrow 3$ phase-covariant cloning of polarization state of a single photon has
been demonstrated experimentally by  \cite{Sciarrino2004c}. The set of cloned
states included all linear polarization states $\cos\theta|V\rangle+\sin\theta|H\rangle$.
The first step in the copying process
 consisted of the optimal $1 \rightarrow 2$ universal cloner described in
Section \ref{secUCOdownconversion} which produced two clones and one anti-clone. In the next step, the
anti-clone was converted into a clone by applying a unitary transformation
$\sigma_y$ with the help of a half-wave plate. The final step was
to symmetrize the state of the three clones by combining the two cones and the
anti-clone on a balanced beam splitter and selecting only the events where all
three photons ended up in the same output spatial mode. In this way
three copies of equal fidelity were produced. 

The experimentally observed fidelities were $F_{1\rightarrow
3}^{\mathrm pc}(|+\rangle)=0.76$ for the state 
$|+\rangle=2^{-1/2}(|V\rangle+|H\rangle)$ and 
$F_{1\rightarrow3}^{\mathrm pc}(|H\rangle)=0.80$. This should be compared with the
theoretical maximum $F_{1\rightarrow 3}^{\mathrm pc}=5/6 \approx 0.833.$ It is also
instructive to make a comparison with the fidelity of the optimal universal
$1\rightarrow 3$ cloner, $F_{1\rightarrow 3}^{\mathrm univ}=7/9 \approx 0.778$.
One can conclude that the experimental phase-covariant cloning machine operates
very close to its theoretical limit and for certain inputs it achieves better
fidelity than what would be possible with universal cloning machine.

\section{Cloning of optical continuous variables}
\label{section-continuous-variables}

So far we have considered cloning of quantum states in finite dimensional Hilbert spaces. 
During recent years, however, quantum information processing in systems with infinite
dimensional Hilbert space, such as modes of the electromagnetic field, has attracted great
deal of attention, see, e.g, \cite{Braunstein2005}. 
In this approach, the quantum information is usually encoded into two
noncommuting quadrature operators $x$ and $p$ which satisfy canonical commutation
relations $[x,p]=i$. Since these operators have continuous spectra, one speaks about
quantum information processing with continuous variables.

The universal cloning machine for states belonging to infinite dimensional Hilbert space
can be formally obtained as a limit of the universal cloning machine for qudits when
$d\rightarrow \infty$. One finds that the single-clone fidelity  of the universal
$1\rightarrow 2$ cloner is $1/2$ which means that the optimal cloning can be achieved be a
very simple strategy where the input state is sent with probability $1/2$ to the first 
or second output, while the other output is prepared in maximally mixed state. 
Besides being rather trivial, this universal cloner is not of great practical interest
because most of the quantum information protocols with continuous variables involve only the 
so-called Gaussian states. These state have Gaussian Wigner function and their great
advantage is that they can be relatively easily generated and manipulated in the lab 
with the help of linear optical interferometers and optical parametric amplifiers which
produce squeezed and entangled Gaussian states.

\subsection{Cloning of coherent states}
\label{secCVcoherent}

Among the Gaussian states, the coherent state is perhaps the best known example. The
coherent state $|\alpha\rangle$ can be defined as a displaced vacuum state
$D(\alpha)|0\rangle$, where $D(\alpha)=\exp(\alpha a^\dagger -\alpha^\ast a)$ is the
displacement operator. Coherent state is the eigenstate of the annihilation operator, $a
|\alpha\rangle= \alpha |\alpha\rangle$ and it is also a minimum uncertainty state. The
variance of all rotated quadratures $x_{\theta}=x\cos \theta + p\sin\theta $ is the same
and equal to $1/2$. The Glauber P-distribution of the coherent state is a Dirac delta
function. In this sense, the coherent states are not usually considered as non-classical
states in the quantum-optical sense. Still, these are pure quantum states and they carry
quantum noise. This makes these states suitable for applications such as quantum key
distribution. It has been shown theoretically 
(\cite{Grosshans2002,Grosshans2003b,Grosshans2004,Iblisdir2004}) and demonstrated experimentally
(\cite{Grosshans2003a}) that secure key distribution can be achieved with coherent states 
and balanced homodyne detection.

Let us first consider the optimal Gaussian cloning of coherent states 
introduced in \cite{Cerf2000c,Lindblad2000}. In this scenario,
the class of admissible cloning transformations is restricted to Gaussian operations,
which preserve the Gaussian shape of the Wigner function. Intuitively, one could expect
that the Gaussian cloning should be optimal. This is indeed true if the figure of merit is
the global $M$-clone fidelity or if the quality of the clones is quantified in terms of
the noise added to the two quadratures $x$ and $p$
(\cite{Cerf2000d}). However, it has been realized
recently that, remarkably, the single-clone fidelity of the $1\rightarrow 2$ cloning of
coherent states is maximized by a non-Gaussian cloner (\cite{Cerf2005}).

Let us begin with the Gaussian $1\rightarrow 2$ cloning. We require that the mean values
of the quadratures of the two clones $A$ and $B$ are equal to the mean values of the
quadratures $x_{\mathrm{in}}$ and $p_{\mathrm{in}}$ of the input coherent state $\alpha$. This guarantees
that the cloning transformation is invariant with respect to the displacements and the
cloning fidelity does not depend on the amplitude $\alpha$. In the Heisenberg picture, 
the most general Gaussian cloning transformation can be written in the form,
\begin{eqnarray}
&&x_{A}=x_{\mathrm{in}}+\tilde{x}_A, \qquad x_{B}=x_{\mathrm{in}}+\tilde{x}_B, \\
&&p_{A}=p_{\mathrm{in}}+\tilde{p}_A, \qquad p_{B}=p_{\mathrm{in}}+\tilde{p}_B. 
\label{CVoutputquadratures}
\end{eqnarray}
The operators $\tilde{x}_{A}$, $\tilde{p}_{A}$, $\tilde{x}_{B}$, $\tilde{p}_{B}$ represent
the noise that is added to the two copies during the cloning process and they all commute
with $x_{\mathrm{in}}$ and $p_{\mathrm{in}}$. The quadrature operators must satisfy the canonical
commutation relations, which implies
\begin{equation}
[\tilde{x}_A,\tilde{p}_B]=-i, \qquad  [\tilde{x}_B,\tilde{p}_A]=-i.
\end{equation}
The Heisenberg uncertainty relation gives a lower bound on the products of the variances 
of the noise operators,
\begin{equation}
\langle (\Delta \tilde{x}_{A})^2\rangle \langle (\Delta \tilde{p}_{B})^2\rangle
\geq \frac{1}{4}, \qquad
\langle (\Delta \tilde{x}_{B})^2\rangle \langle (\Delta \tilde{p}_{A})^2\rangle
\geq \frac{1}{4}.
\label{CVuncertainty}
\end{equation}
The cloning should add noise isotropically, the variance of the $x$ and $p$ quadratures 
of each clone should be the same. Since the noise operators are not correlated with 
$x_{\mathrm{in}}$ and $p_{\mathrm{in}}$, the variance of the quadratures 
(\ref{CVoutputquadratures}) of the two clones is sum of two variances,
and the isotropy condition is satisfied if 
\begin{equation}
\langle(\Delta \tilde{x}_A)^2\rangle=\langle(\Delta \tilde{p}_A)^2\rangle=\bar{n}_{A},
\qquad
\langle(\Delta \tilde{x}_B)^2\rangle=\langle(\Delta \tilde{p}_B)^2\rangle=\bar{n}_{B}.
\end{equation}
The two uncertainty relations (\ref{CVuncertainty}) the boil down to a single constraint
\begin{equation}
\bar{n}_{A} \bar{n}_{B} \geq \frac{1}{4}.
\label{CVnbarproduct}
\end{equation}
The state of each clone is a mixed Gaussian state, namely a coherent state with
added thermal noise with mean number of thermal photons equal to $\bar{n}_j$, $j=A,B$. 
The fidelity of cloning can be most easily calculated from the Husimi Q-function, 
which is defined as the overlap of the density matrix with coherent state, 
$Q(\beta)=\pi^{-1}\langle \beta|\rho|\beta\rangle$. The Q function of the $j$-th 
clone reads (\cite{Fiurasek2001a}),
\begin{equation}
Q_j(\beta)= \frac{1}{\pi (1+\bar{n}_j)}
\exp\left[-\frac{(\beta-\alpha)^2}{1+\bar{n}_j}\right].
\label{CVQfunction}
\end{equation}
 The fidelity can be calculated as $F_j(\alpha)= \pi Q_j(\alpha)=1/(1+\bar{n}_j)$.
 The best trade-off between the fidelities of the two clones is obtained when the equality
 holds in Eq. (\ref{CVnbarproduct}), and we get
 \begin{equation}
 F_A=\frac{2}{2+e^{2\gamma}},  \qquad F_B= \frac{2}{2+e^{-2\gamma}},
 \end{equation}
where $\gamma$ is a parameter which controls the asymmetry of the cloning. The fidelity of
the optimal $1\rightarrow 2$ symmetric ($\gamma=0$) Gaussian cloner is $F=2/3$.

\subsection{Cloning by phase-insensitive amplification}
\label{secCVamplification}

If the coherent states are carried by optical modes, then the cloning  
can be realized with the use of a phase insensitive amplification of light
(\cite{Braunstein2001b,Fiurasek2001a}). This is very
natural and intuitive result, because the idealized perfect cloning amounts to noiseless
amplification of the coherent state, $|\alpha\rangle \rightarrow |\sqrt{2}\alpha\rangle$.
The optimal amplification that adds the minimum amount of noise can be performed, e.g.,
 in nondegenerate optical parametric amplifier (NOPA), which transforms the input
annihilation operator $a$ as
$a_{\mathrm{out}}=\sqrt{G}a_{\mathrm{in}}+\sqrt{G-1}c_{\mathrm{in}}^\dagger$, where  
$c$ is the annihilation operator of the idler mode in NOPA. The setup for asymmetric cloning
of coherent states is shown in Fig. \ref{figcoherentstates}(a). It consists of a Mach-Zehnder interferometer with
an amplifier in one of its arms. The signal is initially divided into two beams and one
beam is amplified such that the total mean intensity is twice the input intensity. The two
clones are obtained by  re-combining the two beams on the second beam splitter.
The splitting ratios of the unbalanced beam splitters $BS_1$ and $BS_2$ and 
the intensity gain G of the amplifier can be expressed in terms of the asymmetry 
parameter $\gamma$ as follows,
\begin{equation}
t_1=-\frac{\sqrt{2}\sinh\gamma}{\sqrt{1+2\sinh^2\gamma}}, \qquad
G=1+\cosh(2\gamma), \qquad t_2=\frac{e^{2\gamma}}{\sqrt{1+e^{4\gamma}}}.
\end{equation}
The setup becomes particularly simple for symmetric cloner. In this case the first beam
splitter disappears and the whole input signal is amplified with gain $G=2$ and then
divided into two modes on a balanced beam splitter, see Fig. \ref{figcoherentstates}(b).

\input #progressfig13

The procedure for symmetric cloning can be readily extended to the optimal symmetric
$N\rightarrow M$ Gaussian cloning of coherent states. The cloning consists of three steps.
First, the whole signal is collected in a single mode using an array of $N-1$ beam
splitters with properly chosen transmittances, $|\alpha\rangle^{\otimes N} \rightarrow
|\sqrt{N}\alpha\rangle$. Next, the collected signal is amplified with a gain $G=M/N$.
Finally, the amplified signal is distributed among the $M$ modes with the help of another
array of $M-1$ unbalanced beam splitters such that the mean complex amplitude in each 
mode is the same and equal to $\alpha$. The fidelity of this cloner does not depend on
$\alpha$ and each clone is in a coherent state with thermal noise described by the Husimi
function (\ref{CVQfunction}). The total mean number of thermal photons in all modes is $G-1=M/N-1$ and 
the noise is equally divided into $M$ modes hence the thermal noise in each clone is
$\bar{n}=1/N-1/M$. On inserting this into the expression for the fidelity, 
$F=1/(1+\bar{n})$, we obtain
\begin{equation}
F= \frac{MN}{MN+M-N}.
\label{CVFidelity}
\end{equation}
In the limit of infinite number of copies, $M\rightarrow \infty$, we  get $F=N/(N+1)$
which is the fidelity of optimal estimation of a coherent state from $N$ copies. 
Similarly as in the case of universal cloning of qubits (\cite{Bruss98b}), the connection
between optimal cloning and optimal state estimation can be exploited to prove that 
(\ref{CVFidelity}) is the maximal fidelity of the Gaussian $N\rightarrow M$ 
cloning of coherent states (\cite{Cerf2000d}). As shown by \cite{vanLoock2001},
it is also possible to clone coherent states via an extended 
continuous-variable teleportation. 
The telecloning requires a specific multimode entangled Gaussian state that can 
be generated by mixing single-mode squeezed vacuum states on an array of unbalanced beam
splitters (\cite{vanLoock2000}).
 
As already noted before, the Gaussian machine depicted in  Fig. \ref{figcoherentstates} is not the optimal one
if the single-clone fidelity is taken as a figure of merit (\cite{Cerf2005}). 
The optimal non-Gaussian cloner can achieve a fidelity $F_{\mathrm{max}}=0.6826$, 
which is slightly higher than the
maximum fidelity achievable by Gaussian transformations, $F=2/3\approx 0.6667$.
Interestingly, the optimal non-Gaussian cloner can be obtained from the setup shown in
Fig. \ref{figcoherentstates}(b) if the input ports of the idler mode of the amplifier  
and the auxiliary  mode of the beam splitter are fed with a specific non-Gaussian 
entangled state $|\psi\rangle_{BC}=\sum_{n=0}^\infty c_{n}|2n\rangle_B|2n\rangle_{C}$. 
The coefficients $c_n$ can be optimized in order to maximize the cloning fidelity 
which results in the above given value $F_{\mathrm max}=0.6826$. It should be stressed 
that while the non-Gaussian cloner maximizes the fidelity, the variance of 
the quadratures of the clones is higher than for the optimal Gaussian cloner. 
In applications such as quantum key distribution with coherent states and balanced 
homodyning, where the quantum channel between Alice and Bob is characterized in terms 
of the first and second moments of the transmitted quadratures, the aim of 
the eavesdropper is to minimize the quadrature variance instead of the fidelity 
and the Gaussian cloning (or its variant entangling Gaussian cloner in case
of reverse reconciliation protocol) can be the most dangerous individual eavesdropping 
attack.

\subsection{Experimental cloning of coherent states}

\input #progressfig14

Recently, optimal Gaussian $1\rightarrow 2$ cloning of coherent states
has been experimentally demonstrated by \cite{Andersen2005}. The
distinct feature of this experiment is that it does not require an
amplifier which is replaced by a clever combination of measurement and
feedback. A simplified  scheme of the experimental setup is depicted in
Fig. \ref{figandersenfigprogress}. The mode A contains the coherent state to be cloned. The
beam  is split into two parts on a balanced beam splitter whose
auxiliary input port $\nu_1$ is in vacuum state. The output annihilation
operators thus read
\[
a^\prime=\frac{1}{\sqrt{2}}(a_{\mathrm in}+\nu_{1,\mathrm in}), \qquad
\nu_1^\prime=\frac{1}{\sqrt{2}}(a_{\mathrm in}-\nu_{1,\mathrm in}).
\]

The output beam $\nu_1^\prime$ is sent to an eight-port homodyne
detector, which consists of a balanced beam splitter followed
by two balanced homodyne detectors. This detector  effectively measures
the operator $\lambda=\nu_1^\prime+\nu_2^\dagger$, where $\nu_2^\dagger$ is the creation
operator of an auxiliary vacuum mode. After the measurement, the mode
$a^\prime$ is displaced by the amount $\lambda$ which is in practice
achieved by mixing this beam with a strong coherent beam with amplitude
$\lambda/\sqrt{1-T}$ on a highly unbalanced beam splitter with transmittance
$T \approx 99$ \%. The resulting displaced beam is effectively the
amplified input,
\[
a_{\mathrm disp}= \sqrt{2}a_{\mathrm in}+\nu_2^\dagger.
\]
The cloning is finished by dividing the amplified beam into two parts
with the help of another balanced beam splitter, thereby preparing the
two clones of the input coherent state. The fidelity observed in the
experiment was about $65$\% which is very close to the optimal value
$2/3 \approx 0.667$.

\subsection{Gaussian distribution with finite width}
\label{secCVfinite}

Up to now, we have assumed that the distribution of the coherent states 
that should be cloned is uniform over the whole phase space. 
However, this is clearly an idealization, since the mean energy of the input 
state would be infinite. A more realistic scenario, considered
by \cite{Cochrane2004}, is that the coherent states are drawn from a Gaussian 
distribution with width $\sigma$ and centered on vacuum, so that the  
\emph{a-priori} probability that the cloned state is $|\alpha\rangle$ is given by
\begin{equation}
P(\alpha)=\frac{1}{2\pi \sigma^2} \exp\left(-\frac{|\alpha|^2}{2\sigma^2}\right).
\label{CVinputdistribution}
\end{equation}
This occurs for instance  in the quantum key distribution with coherent states
(\cite{Grosshans2003a}).

If the width $\sigma$ of the distribution (\ref{CVinputdistribution}) is finite, 
then we possess some information that can be explored in order to increase 
the average cloning fidelity. Also, the probability (\ref{CVinputdistribution}) 
is not invariant with respect to the displacements, so there is no
reason to search for covariant cloner. The fidelity of the cloner may depend on the
input state and the figure of merit that should be maximized is  the average fidelity,
\begin{equation}
{\mathcal{F}}=\int P(\alpha) F(\alpha) d^2\alpha.
\end{equation}

It turns out that the optimal symmetric $1\rightarrow 2$ Gaussian cloning transformation 
is still the amplification followed by the beam splitting on a balanced beam splitter.
However, the gain $G$ depends on $\sigma$. After the amplification and beam splitting, the
coherent amplitude in each mode is $\alpha\sqrt{G/2}$ and the mean number of chaotic
photons in each mode is $\bar{n}=(G-1)/2$.  The fidelity of cloning a particular coherent
state $|\alpha\rangle$ reads
\begin{equation}
F(\alpha)=\frac{2}{G+1} \exp\left[-\frac{2(1-\sqrt{G/2})^2}{G+1} |\alpha|^2\right].
\end{equation}
After the averaging over the Gaussian distribution (\ref{CVinputdistribution}) we arrive at the expression for the
mean fidelity,
\begin{equation}
{\mathcal{F}}= \frac{2}{G+1+2\sigma^2(2+G-2\sqrt{2G})}.
\end{equation}
We have to find the maximum of $\mathcal{F}$ under the constraint $G \geq 1$. 
It turns out that there are two different solutions in dependence on the value 
of $\sigma$. If $\sigma^2> \sigma_{\mathrm{th}}^2=(1+\sqrt{2})/2$, then it is optimal 
to amplify the signal and the optimal gain is
\begin{equation}
G= \frac{8\sigma^4}{(1+2\sigma^2)^2}.
\end{equation}
On the other hand, if $\sigma^2< \sigma_{\mathrm{th}}^2$ then it is optimal 
to simply divide the input signal into two beams without any amplification, and $G=1$.
The resulting cloning fidelity is 
\begin{equation}
{\mathcal{F}}=\left\{
\begin{array}{lcl} 
\displaystyle \frac{4\sigma^2+2}{6\sigma^2+1}, & & \sigma^2 \geq \sigma_{\mathrm{th}}^2 ,\\[4mm]
\displaystyle \frac{1}{1+(3-2\sqrt{2})\sigma^2}, & & \sigma^2 < \sigma_{\mathrm{th}}^2.
\end{array}
\right.
\end{equation}
The average fidelity monotonically increases with the decreasing width of the 
distribution (\ref{CVinputdistribution}) and in the limit $\sigma \rightarrow 0$ we get 
${\mathcal{F}}=1$, as expected.

\subsection{Cloning of conjugate coherent states}
\label{secCVconjugate}

In Section \ref{secCloningpairorthogonal} we have discussed a cloning machine 
for a pair of orthogonal qubits. This device possesses a natural and very 
interesting continuous-variable analogue, 
namely, one can consider a cloning machine
for coherent states $|\alpha\rangle$ whose input consists of $N$ copies
of the state $|\alpha\rangle$ and of $N^\prime$ copies of the
complex conjugate coherent state $|\alpha^\ast\rangle$. This problem was
analyzed in detail in even more general setting in  \cite{Cerf2001a}.

Without any loss of generality, we can assume that a pair of arrays of
beam splitters is used to collect all signal into two modes and 
 the input state of the cloning machine thus reads
$|\sqrt{N}\alpha\rangle_A|\sqrt{N^\prime} \alpha^\ast\rangle_B$.
The goal of cloning is to produce $M$ copies of $|\alpha\rangle$ with
minimum added noise. This could be again accomplished with the help of
the non-degenerate  parametric amplifier.
While the mode $A$ represents the signal input similarly as before, the
mode $B$ is sent to the idler input port of the amplifier. Assuming
amplification with intensity gain G, the output annihilation operator
of the signal mode is given by,
$a_{\mathrm out}=\sqrt{G} a_{\mathrm in}+\sqrt{G-1}\,b_{\mathrm in}^\dagger.$
Note that both terms $a_{\mathrm in}$ and $b^\dagger$ in
the above formula contribute to the total coherent signal in $a_{\mathrm out}$.
If the cloning should be performed with unity gain, then $G$ must
satisfy
\begin{equation}
\sqrt{M}=\sqrt{G}\sqrt{N}+\sqrt{G-1}\sqrt{N^\prime},
\end{equation}
and $G$ can be easily determined by solving the above quadratic
equation.

A careful analysis reveals that for certain values of $N$, $N^\prime$ and
$M$ the cloning with conjugate inputs could be more efficient than the
standard cloning of coherent states. To be fair,
we should compare the cloning fidelities for inputs  consisting either
of  $N+N^\prime$ copies of $|\alpha\rangle$ or including  N copies of
$|\alpha\rangle$ and $N^\prime$ copies of $|\alpha^\ast\rangle$.
The advantage of dealing with complex conjugate inputs could be most
easily illustrated in the limit of infinite number of clones, $M
\rightarrow \infty$, where the optimal cloning becomes equivalent with
optimal state estimation. It has been shown in \cite{Cerf2001b} that
when possessing a single copy of $|\alpha\rangle|\alpha^\ast\rangle$ we
can estimate $|\alpha\rangle$ with fidelity $F_{c.c.}=4/5$ which is strictly
higher than the estimation fidelity $F=2/3$ corresponding to the input
state $|\alpha\rangle^{\otimes 2}$. In the former case the optimal 
detection strategy is the nonlocal continuous-variable Bell measurement
where the quadratures $x_A+x_B$ and $p_{A}-p_{B}$ are measured
simultaneously.

\section{Conclusions}
\label{section-conclusions}

The quantum no-cloning theorem is a crucial aspect of modern
quantum mechanics and one of the cornerstones of quantum information theory. 
Besides its fundamental interest for the foundations of quantum physics, the impossibility of exactly copying an unknown quantum state is crucial for the security of the quantum key distribution protocols. 
Going beyond the no-cloning theorem, it is possible to design approximate quantum cloning machines, which enable the copying of quantum information
in an optimal -- albeit imperfect -- way, an issue which has
attracted a considerable attention over the last decade.

This review paper aims at providing an exhaustive view of the
various quantum cloning machines that have been introduced since 
this concept was put forward by Bu\v{z}ek and Hillery in 1996. 
The mathematical description of quantum cloning machines based on the
isomorphism between maps and operators is developed in details.
A special attention is also paid to the experimental optical implementations 
of these machines. The cloning of single photons has been
accomplished by several groups by now, and these experiments represent 
a very valuable contribution to the toolbox 
of available optical methods for quantum information processing.

In the course of years, quantum cloning has grown 
into a genuine subfield of quantum information sciences, 
which is still currently very active both on the theoretical and
experimental sides. The advanced methods of preparation, manipulation, 
and measurement of quantum states of light, whose development has been 
stimulated to a large extent by the perspectives
of quantum information processing, have recently 
enabled the demonstration of even more complex cloning machines. 
In the years to come, we anticipate many new achievements and breakthroughs
in quantum information sciences, and there is no doubt that quantum cloning
will play an important role in these future developments.

\section{Acknowledgments}

We acknowledge financial support from the EU under projects 
COVAQIAL (FP6-511004) and RESQ (IST-2001-37559), from  
the Communaut\'e Fran\c{c}aise de Belgique under grant ARC 00/05-251, 
and from the IUAP programme of the Belgian
government under grant V-18. JF also acknowledges support under the Research 
project Measurement and Information in Optics MSM 6198959213 and  
from the grant  202/05/0498 of the Grant Agency of Czech Republic. 




\end{document}